\documentclass[%
 preprint,% linenumbers,
 amsmath,amssymb,
 aps, physrev,
]{revtex4-2}

\usepackage{graphicx}% Include figure files
\usepackage{dcolumn}% Align table columns on decimal point
\usepackage{bm}% bold math
\usepackage{subcaption}
\usepackage[T1]{fontenc}
\usepackage[utf8]{inputenc}
\begin{document}

\preprint{APS/123-QED}

\title{\textbf{Comparative study and critical assessment of phase-field lattice Boltzmann models for laminar and turbulent two-phase flow simulations}
}%

\author{Xuming Li}
\affiliation{%
 Key Laboratory of High Efficiency and Clean Mechanical Manufacture, Ministry of Education, School of Mechanical Engineering, Shandong University, Jinan,
 250061, Shandong, China
}%

\author{Chunhua Zhang}%
\affiliation{%
 School of Energy and Power Engineering, North University of China, Taiyuan
 030051, Shanxi, China
}%

\author{Xinnan Wu}%
\affiliation{%
 Shandong Scitech Innovation Group Co., Ltd., Jinan,
 250101, Shandong, China
}%

\author{Cheng Peng}%
 \email{Contact author: pengcheng@sdu.edu.cn}
\affiliation{%
 Key Laboratory of High Efficiency and Clean Mechanical Manufacture, Ministry of Education, School of Mechanical Engineering, Shandong University, Jinan,
 250061, Shandong, China
}%

\author{Wenrui Wang}%
\affiliation{%
 GuanYun Technology, Co., Ltd., Jinan,
 250003, Shandong, China
}%

\date{\today}% It is always \today, today,

\begin{abstract}
Phase-field lattice Boltzmann (LB) models have undergone continuous development, resulting in multiple variants widely used for simulating multiphase flows. However, direct performance comparisons remain limited, especially for three-dimensional cases. In this study, we present a systematic comparative analysis of several recent and representative phase-field LB models, covering four major categories: conservative Allen–Cahn, nonlocal Allen–Cahn, hybrid Allen–Cahn, and Cahn–Hilliard models. Their accuracy, numerical stability, and mass/volume conservation are assessed through a series of canonical two-phase flow problems. Beyond the commonly tested two-dimensional laminar cases, we extend the evaluation to three-dimensional droplet-laden turbulent flows, which expose more critical limitations of the existing models. The results show that while all models perform satisfactorily in two dimensions, they still suffer from substantial droplet volume loss in turbulence, particularly at high Weber numbers. Overall, conservative Allen–Cahn-based LB models exhibit the most favorable balance of numerical stability, accuracy, and computational efficiency.
\end{abstract}

\maketitle

%\tableofcontents

\section{\label{sec:introduction}Introduction}

Multiphase flows are ubiquitous in both natural and engineering contexts, such as breaking ocean waves, immiscible groundwater mixing, fuel injection, emulsion polymerization, and bubble column reactors. Their interfacial dynamics span a wide range of spatial and temporal scales, making accurate modeling a formidable challenge. The lattice Boltzmann (LB) method  has emerged as a promising alternative to conventional computational fluid dynamics (CFD), owing to its algorithmic simplicity, inherent parallelism, and flexible boundary treatment~\cite{chen1998lattice,aidun2010lattice}. Over the past three decades, LB has been extensively advanced and applied to multiphase flow problems.

Several major classes of multiphase LB models have been developed~\cite{huang2015multiphase}. The color-gradient model represents phases with distinct particle populations and enforces interfacial dynamics via local color gradients~\cite{gunstensen1991lattice,zahid2025review}. The pseudopotential model introduces interparticle potentials to capture surface tension and wettability~\cite{shan1993lattice,shan1994simulation,peng2021thermodynamically}. The free-energy model employs a nonideal pressure tensor derived from a free-energy functional to account for surface tension effects~\cite{swift1995lattice,swift1996lattice,guo2021well}. More recently, phase-field lattice Boltzmann (PF-LB) models based on the Cahn–Hilliard (CH) equation \cite{cahn1958free,cahn1959free}, Allen–Cahn (AC) equation \cite{sun2007sharp,allen1979microscopic}, or their variations, have attracted increasing attention. These PF-LB formulations can simulate multiphase flows with large density ratios and have shown satisfactory accuracy in many cases~\cite{he1999lattice,bao2024phase}.

The CH equation has long been employed in PF-LB modeling to describe interface dynamics. The first CH-based LB model, proposed by He et al.~\cite{he1999lattice}, introduced an order parameter to track fluid interfaces, but it could not recover the correct CH equation through Chapman–Enskog expansion. To address this, Zheng et al.~\cite{zheng2006lattice} and Zu et al.~\cite{zu2013phase} modified the standard LB equation by incorporating spatial differences of adjacent distribution functions, while Liang et al.~\cite{liang2014phase} redefined the discrete force term and equilibrium distribution within the standard LB framework. These CH-LB models all achieve second-order temporal accuracy. More recently, Zhang et al.~\cite{zhang2019high} analyzed the third-order error terms of the CH-LB model using a fourth-order Chapman–Enskog expansion and improved both accuracy and stability of interface capturing by eliminating the leading error terms.

A well-known drawback of CH-LB models is the spontaneous shrinkage of droplets during simulations. Several theoretical studies have analyzed this phenomenon~\cite{yue2007spontaneous,zhang2019spontaneous}. To mitigate this issue, Bao and Guo~\cite{bao2024phase} proposed an improved CH PF-LB model with singular mobility. The singularity was designed to suppress droplet shrinkage—a long-standing challenge in CH-based formulations. By tuning free parameters linked to chemical potential gradients and local shear rates, their model reduced numerical dissipation and improved the physical fidelity of long-term simulations.

In addition to the CH equation, the AC equation has also gained significant attention in the LB community due to its better volume conservation and its reliance on only second derivatives. The original AC equation does not conserve mass. Two main strategies have been developed to overcome this limitation. The first introduces a Lagrange multiplier to enforce global conservation, leading to the nonlocal AC equation~\cite{rubinstein1992nonlocal}. Although physically consistent, this formulation suffers from coarsening, whereby small droplets are absorbed by larger ones—an undesirable behavior in multiphase flow simulations. The second strategy adds a local counter term, yielding the conservative or local AC equation~\cite{sun2007sharp,chiu2011conservative}, which achieves mass conservation without coarsening but is prone to numerical dispersion and bulk-phase fluctuations~\cite{chai2018comparative}.

Building on the conservative AC equation, several AC-LB models have been developed. Geier et al.~\cite{geier2015conservative} first proposed a conservative AC PF-LB model by treating the source term analogously to the convective term. Fakhari et al.~\cite{fakhari2017improved} refined the discrete force term to recover the conservative AC equation while introducing additional velocity derivative terms. Following similar strategies to those used for CH equations, other improved AC-LB models have been proposed~\cite{ren2016improved,wang2016comparative,liang2018phase}. When coupled with LB models for incompressible or quasi-incompressible Navier–Stokes equations, these AC-LB formulations have been successfully applied to multiphase flows with density ratios up to $1000$.

To reconcile the advantages of both conservative and nonlocal AC formulations, Hu et al.~\cite{hu2019hybrid} introduced a hybrid AC PF-LB model that combined the two through a single global weight. This approach reduced numerical dissipation and produced sharper interfaces, but it struggled to preserve fine-scale interfacial structures. Kang and Yun~\cite{kang2022local} extended this concept by introducing a spatially varying weight, which simultaneously suppressed dissipation and coarsening; however, accurately resolving three-dimensional interfaces remained difficult. Most recently, Liu et al.~\cite{liu2023improved} developed an improved hybrid AC PF-LB model with variable weights, which effectively reduced dispersion and preserved small-scale interface features, marking an important step toward robust simulations of complex multiphase flows.

The diversity of PF-LB models underscores the need for systematic comparative studies, and several efforts have been made in this direction. Kim et al.~\cite{kim2012phase} compared the AC and CH equations using finite-difference discretization with a nonlinear multigrid solver, focusing primarily on the phase-field formulations themselves. Wang et al.~\cite{wang2016comparative} examined the conservative AC and CH PF-LB models for interface tracking through several simple two- and three-dimensional cases. Chai et al.~\cite{chai2018comparative} investigated conservative and nonlocal AC PF-LB models, highlighting the recovery of correct PF equations via Chapman–Enskog analysis, as well as their mass conservation and numerical stability. More recently, Xu et al.~\cite{xu2023high} evaluated multiple conservative AC models through high-order analysis and explored the role of relaxation times.

Despite these contributions, most comparative studies remain confined to two-dimensional laminar flows. The performance of existing PF-LB models in complex three-dimensional flows—particularly turbulent flows—remains largely unclear. To address this gap, we conduct a systematic comparison of representative PF-LB models, extending the assessment from conventional two-dimensional laminar flows to three-dimensional turbulent flows. This study provides practical guidance for selecting models suitable for increasingly complex flow conditions. To the best of our knowledge, this is the first comparative study of PF-LB models in three-dimensional two-phase turbulence.

The remainder of this paper is organized as follows. Section~\ref{sec:models} introduces the governing equations and summarizes the key features of the selected PF-LB models. Section~\ref{sec:laminar} validates the models against laminar benchmarks and extends the analysis to three-dimensional turbulent cases. Section~\ref{sec:turbulence} investigates droplet-laden turbulence in detail, analyzing the effects of both physical and numerical parameters. Finally, Section~\ref{sec:summary} concludes with recommendations for model selection and potential applications.

\section{Phase-field lattice Boltzmann models}\label{sec:models}

In PF models, different phases are distinguished by an order parameter $\phi$, whose evolution is governed by either the CH or AC equation. The values of $\phi$ for the heavy and light phases are denoted by $\phi_H$ and $\phi_L$, with the interface defined at $\phi_0 = (\phi_H + \phi_L)/2$. The terms “\emph{heavy}” and “\emph{light}” are used only as labels and do not necessarily imply a density contrast. The densities, dynamic viscosities, and kinematic viscosities of the two phases are denoted by $\rho_H$ and $\rho_L$, $\mu_H$ and $\mu_L$, and $\nu_H$ and $\nu_L$, respectively.

The choice of $\phi_H$ and $\phi_L$ influences numerical stability. For high-density-ratio flows, setting $\phi_H = 1$ and $\phi_L = 0$ generally yields more robust results, while for unity density ratio, using $\phi_H = 0.5$ and $\phi_L = -0.5$ provides better stability. Unless otherwise specified, we adopt $\phi_H = 1$ and $\phi_L = 0$ for all models in this work \cite{fakhari2017improved}.

In the following, we first present the governing PF equations (AC and CH), and the Navier–Stokes (NS) equations for fluid motion. We then outline the LB framework for PF-LB methods and briefly introduce the representative PF-LB models selected for comparison.

\subsection{Hydrodynamic equations}\label{subsection:Allen-Cahn equation}
In the phase-field theory, the total free energy of an isothermal system can be written as
\begin{equation}
\mathit{F}=\int{\left(\psi(\phi)+ \frac{\kappa}{2} \left| \boldsymbol{\nabla}\phi \right|^2  \right)}dV
\end{equation}
where $\psi$ is the bulk free energy density, and the second term represents the interfacial free energy density. In most cases, the bulk energy is approximated by
\begin{equation}
\psi(\phi) =
\beta (\phi_H - \phi)^2 (\phi - \phi_L)^2,
\label{eq:bulk_energy}
\end{equation}
with the coefficient
\begin{equation}
\beta = \frac{12\sigma}{\xi |{\phi_H} - {\phi_L}|^4},
\label{eq:coefficient1}
\end{equation}
where $\sigma$ is the surface tension and $\xi$ is the interface thickness~\cite{ren2016improved}.

The original AC equation is expressed as
\begin{equation}
\frac{\partial \phi}{\partial t} + \boldsymbol{\nabla} \cdot (\phi \bm{u})
= M_\phi \left[ \nabla^2 \phi - \frac{4\beta}{\kappa} (\phi - \phi_H)(\phi - \phi_L)
\left(\phi - \phi_0\right) \right]
\label{eq:originalAC}
\end{equation}
where $\bm u$ is the macroscopic velocity vector, $M_\phi$ is the phase-field mobility, and $\kappa$ is defined as
\begin{equation}
\kappa = \frac{3\sigma \xi}{2|{\phi_H} - {\phi_L}|^2}.
\label{eq:coefficients}
\end{equation}
For the common choice ${\phi_H} = 1$ and ${\phi_L} = 0$, the parameters reduce to $\beta = 12\sigma/\xi$ and $\kappa = 3\sigma \xi/2$, which are adopted in this work unless otherwise specified.

The original AC equation does not conserve total mass under the boundary conditions $\bm{n}\cdot\bm{u}|_{\partial\Omega}=0$ and $\bm{n}\cdot\boldsymbol{\nabla}\phi|_{\partial\Omega} = 0$, where $\bm{n} = \boldsymbol{\nabla} \phi/\vert\boldsymbol{\nabla} \phi\vert$ is the unit normal at the domain boundary $\partial\Omega$. To restore global conservation, Rubinstein and Sternberg \cite{rubinstein1992nonlocal} introduced a nonlocal Lagrange multiplier $\gamma(t)$, leading to the nonlocal AC equation
\begin{equation}
\frac{\partial \phi}{\partial t} + \boldsymbol{\nabla}\cdot(\phi \bm{u})
= M_\phi \Biggl[ \nabla^2\phi - \frac{4\beta}{\kappa}(\phi-\phi_H)(\phi-\phi_L)
(\phi - \phi_0) + \gamma(t) (\phi_H - \phi)(\phi - \phi_L) \Biggr],
\label{eq:nonlocalAC}
\end{equation}
with
\begin{equation}
\gamma(t) = \frac{\displaystyle \int_{\Omega}  \frac{4\beta}{\kappa} (\phi - \phi_{H}) (\phi - \phi_{L}) \left( \phi - \phi_0 \right) d\bm{x}}
{\displaystyle \int_{\Omega} (\phi_{H} - \phi) (\phi - \phi_{L}) \, d\bm{x}}.
\label{eq:lagrange_multiplier}
\end{equation}
LB models derived from Eq.~\eqref{eq:nonlocalAC} are referred to as nonlocal AC models.

An alternative approach introduces a local counter term associated with interface curvature, yielding the conservative (or local) AC equation~\cite{sun2007sharp,chiu2011conservative}
\begin{equation}\label{eq:local_ACE_2}
\frac{\partial \phi}{\partial t} + \boldsymbol{\nabla} \cdot (\phi \bm{u})
= M_\phi \boldsymbol{\nabla} \cdot \left[ \boldsymbol{\nabla} \phi - \sqrt{\frac{2\beta}{\kappa}} (\phi_H - \phi)(\phi - \phi_L){\bm n} \right].
\end{equation}
LB models based on this formulation are termed conservative AC models.

A third category, hybrid AC models, combines the conservative and nonlocal forms through a weighting factor $\lambda$, leading to
\begin{equation}
\begin{aligned}
\frac{\partial\phi}{\partial t} + \boldsymbol{\nabla}\cdot(\phi\mathbf{u})
=\, M_\phi\Bigg\{ &\nabla^2\phi
- \boldsymbol{\nabla}\cdot\Bigg[ \lambda\sqrt{\frac{2\beta}{\kappa}}
\frac{(\phi_H-\phi)(\phi-\phi_L)}{|\boldsymbol{\nabla}\phi|}\,\boldsymbol{\nabla}\phi \Bigg] \\
&+ (1-\lambda)\Bigg[ -\frac{4\beta}{\kappa}(\phi-\phi_H)(\phi-\phi_L)
(\phi-\phi_0) + \gamma(t)\,(\phi_H-\phi)(\phi-\phi_L) \Bigg] \Bigg\}.
\end{aligned}
\label{eq:hybrid_ACE}
\end{equation}

On the other hand, the CH equation reads~\cite{kendon2001inertial,badalassi2003computation,jacqmin1999calculation}:
\begin{equation}
\frac{\partial \phi}{\partial t} + \boldsymbol{\nabla} \cdot (\phi \bm{u})
= \boldsymbol{\nabla} \cdot \big( M_\phi \boldsymbol{\nabla} \mu_\phi \big),
\label{eq:CH}
\end{equation}
with the chemical potential
\begin{equation}
\mu_{\phi} = 4\beta (\phi-\phi_{H})(\phi-\phi_{L})(\phi-\phi_{0})
- \kappa \nabla^{2}\phi,
\label{eq:chemical_potential}
\end{equation}

It should be noted that although $M_\phi$ in the CH equation is also referred to as mobility, its physical dimension and underlying essence differ fundamentally from those in the AC equation. This distinction makes it difficult to conduct fully fair comparisons between the two sets of LB models, as pointed out in previous studies~\cite{ren2016improved}. Nevertheless, in practice, the mobility in PF-LB equations is often treated as a numerical parameter, whose value should not substantially affect the physical results provided it ensures numerical stability and does not alter the scale separation in the Chapman–Enskog analysis. In our comparative study, we adopt this perspective and assign the same moderate value of mobility across both AC and CH PF-LB models to maintain numerical stability as much as possible, similar to some previous studies~\cite{wang2016comparative}. We emphasize, however, that this compromise may influence the quantitative aspects of the comparison, and therefore the results should be interpreted with appropriate caution.

A well-known drawback of the CH formulation is spontaneous droplet shrinkage, where droplets smaller than a critical radius vanish over time \cite{wagner1997effect,yue2007spontaneous,zhang2019spontaneous}. To overcome this, Bao et al. \cite{bao2024phase}proposed a singular mobility formulation
\begin{equation}
M_\phi = \frac{4}{(\phi_H - \phi_L)^2} M_0 \, |\phi - \phi_H| \, |\phi - \phi_L|,
\label{eq:bao_singular_mobility}
\end{equation}
where $M_0$ is the mobility at the interface ($\phi=\phi_0$). This ensures vanishing mobility in the bulk phases and alleviates droplet shrinkage.

The fluid motion is governed by the incompressible NS equations modified with a surface tension force\cite{ding2007diffuse,li2012additional}
\begin{subequations}
\begin{align}
&\boldsymbol{\nabla}\cdot\bm{u} = 0, \label{eq:continuity} \\[1mm]
&\frac{\partial(\rho\bm{u})}{\partial t} + \boldsymbol{\nabla}\cdot(\rho\bm{u}\bm{u}) = -\boldsymbol{\nabla} p + \boldsymbol{\nabla}\cdot\left[\mu(\boldsymbol{\nabla}\bm{u}+\boldsymbol{\nabla}\bm{u}^T)\right] + \bm{F}_s + \bm{F}_b. \label{eq:momentum}
\end{align}
\label{eq:NS}
\end{subequations}
where $\rho$ is the density, $p$ is the pressure, $\mu$ is the viscosity, ${\bm F}_s$ is the interfacial force, and ${\bm F}_b$ is the body force.

Alternatively, the quasi-incompressible NS equations are also employed to simulate the multiphase flow. When coupled with the CH equation, the divergence of the velocity should satisfy the following equality \cite{shen2013mass}:
\begin{equation}
\boldsymbol{\nabla} \cdot \bm{u}
= - \theta \boldsymbol{\nabla} \cdot \left( M_\phi \boldsymbol{\nabla} \mu_\phi \right),
\qquad \theta = \frac{\tfrac{\rho_{H}}{\rho_{L}} - 1}{\phi_{H} - \phi_{L} \tfrac{\rho_{H}}{\rho_{L}}}.
\label{eq:quasi_incomp_system}
\end{equation}

\subsection{Lattice Boltzmann equations}
The lattice Boltzmann method solves the above hydrodynamic equations by evolving distribution functions associated with discrete velocity sets through the so-called lattice Boltzmann equations. In PF-LB models, two sets of distribution functions are typically employed: one for the PF equation and the other for the NS equations. For consistency and fairness in comparison, we recast all selected models into the single-relaxation-time (BGK) form, even if their original formulations adopted the multi-relaxation-time (MRT) operator. This allows the essential differences between the models to be highlighted without the influence of different collision operators.

All models considered here share the following general lattice Boltzmann equation:
\begin{subequations}
\begin{align}
& g_\alpha (\bm{x} + \bm{e}_\alpha \delta t, t + \delta t)
= g_\alpha (\bm{x},t)
- \frac{g_\alpha (\bm{x},t) - g_\alpha^{eq}(\bm{x},t)}{\tau_\phi}
+ \left(1 - \frac{1}{2\tau_\phi}\right) \delta t \, G_\alpha (\bm{x},t),
\label{eq:fakhari_g_simplified} \\[2mm]
& f_\beta (\bm{x} + \bm{e}_\beta \delta t, t + \delta t)
= f_\beta (\bm{x},t)
- \frac{f_\beta (\bm{x},t) - f_\beta^{eq}(\bm{x},t)}{\tau}
+ \left(1 - \frac{1}{2\tau}\right) \delta t \, F_\beta (\bm{x},t).
\label{eq:fakhari_f_simplified}
\end{align}
\label{eq:fakhari_model_simplified}
\end{subequations}
where $\bm{x}$ and $t$ denote space and time, respectively. The distribution function $g_\alpha$ corresponds to the PF equation with discrete velocity ${\bm e}_\alpha$, while
${f_\beta }$ corresponds to the NS equations with discrete velocity ${\bm e}_\beta$. To emphasize that the velocity sets for the PF and NS solvers need not be identical, different subscripts ($\alpha$ and $\beta$) are used. The relaxation times are defined as
\begin{equation}
\tau_\phi = \frac{M_\phi}{c_s^2} + \frac{1}{2},
\qquad \tau = \frac{\mu}{\rho c_s^2} + \frac{1}{2},
\label{eq:fakhari_tau_model}
\end{equation}
where $c_s$ is the lattice sound speed, whose specific value is determined by the lattice velocity set.

Because viscosity can vary across phases, $\mu$ must be updated locally. Three interpolation strategies are commonly adopted in the literature\cite{lai2025analytical}: inverse-linear, linear, and logarithmic:
\begin{subequations}
\begin{align}
&{\rm }\mu = \frac{\mu_H \mu_L (\phi_H - \phi_L)}{(\phi - \phi_L) \mu_L + (\phi_H - \phi) \mu_H},
\label{eq:mu_inverse} \\[2mm]
&\mu = \mu_H \frac{\phi - \phi_L}{\phi_H - \phi_L} + \mu_L \frac{\phi - \phi_H}{\phi_L - \phi_H},
\label{eq:mu_linear} \\[2mm]
&\mu = \mu_H^{\frac{\phi - \phi_L}{\phi_H - \phi_L}} \, \mu_L^{\frac{\phi - \phi_H}{\phi_L - \phi_H}},
\label{eq:mu_log}
\end{align}
\end{subequations}

Nevertheless, the choice of viscosity interpolation can strongly influence certain flows, such as stratified Poiseuille flow. Therefore, the update scheme should be selected according to the specific physical problem under consideration. A more detailed discussion of this issue can be found in the recent work of Lai et al.~\cite{lai2025analytical}. Unless otherwise noted, the linear form is adopted in this study.

The main differences among PF-LB models lie in the definitions of the equilibrium distribution functions ($g_\alpha^{eq}$, $f_\beta^{eq}$) and the corresponding source terms ($G_\alpha$, $F_\beta$). These will be detailed for each model in the following subsections.

\subsection{AC based LB models}\label{subsection:AC-PH-LB-model}

\subsubsection{Fakhari et al.'s conservative AC model}\label{subsubsection:Fakhari}

The first model examined in this work is the conservative AC model proposed by Fakhari et al.\cite{fakhari2017improved} (hereafter referred to as Fakhari2017).
In this formulation, the two equilibrium distribution functions are defined as
\begin{subequations}
\begin{align}
& g_\alpha^{eq}(\bm{x},t) = w_\alpha \phi \left[ 1 + \frac{\bm{e}_\alpha \cdot \bm{u}}{c_s^2}
+ \frac{(\bm{e}_\alpha \cdot \bm{u})^2}{2 c_s^4} - \frac{\bm{u} \cdot \bm{u}}{2 c_s^2} \right],
\label{eq:fakhari_geq} \\[2mm]
& f_\beta^{eq}(\bm{x},t) = \frac{p}{\rho c_s^2} w_\beta
+ w_\beta \left[ \frac{\bm{e}_\beta \cdot \bm{u}}{c_s^2}
+ \frac{(\bm{e}_\beta \cdot \bm{u})^2}{2 c_s^4} - \frac{\bm{u} \cdot \bm{u}}{2 c_s^2} \right],
\label{eq:fakhari_feq}
\end{align}
\label{eq:fakhari_eqs}
\end{subequations}
where $w_\alpha$ and $w_\beta$ are the lattice weights determined by the discrete velocity sets.
The associated source terms are given by
\begin{subequations}
\begin{align}
& G_\alpha(\bm{x},t)
= w_\alpha \, \bm{e}_\alpha \cdot
\left( \sqrt{\frac{2\beta}{\kappa}} \;
\frac{(\phi_H - \phi)(\phi - \phi_L)}{|\boldsymbol{\nabla} \phi|} \, \boldsymbol{\nabla} \phi \right),
\label{eq:fakhari_G_modified} \\[2mm]
& F_\beta(\bm{x},t)
= w_\beta \, \frac{\bm{e}_\beta \cdot \bm{F}}{\rho c_s^2}.
\label{eq:fakhari_F}
\end{align}
\label{eq:fakhari_forces_modified}
\end{subequations}

In Fakhari2017, the total force is expressed as
\begin{equation}
   {\bm F} = {\bm F}_s + {\bm F}_b + {\bm F}_p + {\bm F}_\mu
\end{equation}
Here, ${\bm F}_s = \mu_\phi \boldsymbol{\nabla}\phi$ is the surface tension force,
${\bm F}_b = \rho{\bm g}$ is the external body force,
${\bm F}_p = - \tfrac{p}{\rho} \, \boldsymbol{\nabla} \rho$ is the pressure force,
and ${\bm F}_\mu = \nu \, [\boldsymbol{\nabla} {\bm u} + (\boldsymbol{\nabla} {\bm u})^T] \cdot \boldsymbol{\nabla} \rho$
is the viscous force, with $\nu$ denoting the kinematic viscosity.

The hydrodynamic quantities are updated as
\begin{subequations}
\begin{align}
& \phi = \sum_\alpha g_\alpha,
\label{fakhari_phi_update} \\
& \rho = \rho_H \frac{\phi - \phi_L}{\phi_H - \phi_L} + \rho_L \frac{\phi - \phi_H}{\phi_L - \phi_H},
\label{fakhari_rho_update} \\
& \nu = \nu_H \frac{\phi - \phi_L}{\phi_H - \phi_L} + \nu_L \frac{\phi - \phi_H}{\phi_L - \phi_H},
\label{fakhari_mu_update} \\
& p = \rho c_s^2 \sum_\beta f_\beta,
\label{fakhari_p_update} \\
& \bm{u} = \sum_\beta f_\beta \bm{e}_\beta + \frac{\bm{F}}{2\rho} \, \delta t
\label{fakhari_velocity_update}
\end{align}
\label{fakhari_update_all}
\end{subequations}

The updates of $\phi$, $\rho$, and $\mu$ (or $\nu$) are consistently applied across all PF LB models considered in this work, unless explicitly noted otherwise.
In contrast, the updates for pressure and velocity may differ between models.
In subsequent sections, any deviations from the above procedure will be explicitly stated; otherwise, it is assumed that the same update rules are followed.

\subsubsection{Liang et al's conservative AC model}\label{subsubsection:Liang}
In Fakhari2017, the equilibrium distribution functions for the conservative AC equation are retained up to second-order terms in velocity, $O(u^2)$. This results in more complex time and spatial derivatives in the recovered AC equation. Liang et al. simplified the equilibrium distribution functions by truncating $O(u^2)$ terms in the equilibrium distribution functions~\cite{liang2018phase}. This simplified model (hereafter referred to as Liang2018) modifies both the equilibrium distribution functions and the source terms as follows:
\begin{subequations}
\begin{align}
g_\alpha^{eq}(\bm{x},t) &= w_\alpha \, \phi \left( 1 + \frac{\bm{e}_\alpha \cdot \bm{u}}{c_s^2} \right),
\label{eq:liang_geq} \\[2mm]
f_\beta^{eq}(\bm{x},t) &=
\begin{cases}
\dfrac{p}{c_s^2} (w_\beta - 1) + \rho \, w_\beta \left[ \dfrac{\bm{e}_\beta \cdot \bm{u}}{c_s^2}
+ \dfrac{(\bm{e}_\beta \cdot \bm{u})^2}{2 c_s^4} - \dfrac{\bm{u} \cdot \bm{u}}{2 c_s^2} \right], & \beta = 0, \\[1mm]
\dfrac{p}{c_s^2} w_\beta + \rho \, w_\beta \left[ \dfrac{\bm{e}_\beta \cdot \bm{u}}{c_s^2}
+ \dfrac{(\bm{e}_\beta \cdot \bm{u})^2}{2 c_s^4} - \dfrac{\bm{u} \cdot \bm{u}}{2 c_s^2} \right], & \beta \neq 0,
\end{cases}
\label{eq:liang_feq}\\[2mm]
G_\alpha(\bm{x},t)
&= \omega_\alpha \, \bm{e}_\alpha \cdot \frac{\partial_t (\phi \bm{u})}{c_s^2}
+ \omega_\alpha \, \bm{e}_\alpha \cdot
\left( \sqrt{\frac{2\beta}{\kappa}} \; \frac{(\phi_H - \phi)(\phi - \phi_L)}{|\boldsymbol{\nabla} \phi|} \, \boldsymbol{\nabla} \phi \right),
\label{eq:liang_G_final} \\[2mm]
F_\beta(\bm{x},t) &= \omega_\beta \left[ \bm{u} \cdot \boldsymbol{\nabla} \rho + \frac{\bm{e}_\beta \cdot \bm{F}}{c_s^2} + \frac{\bm{u}\boldsymbol{\nabla} \rho : (\bm{e}_\beta \bm{e}_\beta - c_s^2 \mathbf{I})}{c_s^2} \right],
\label{eq:liang_F_final}
\end{align}
\label{eq:liang_eqs}
\end{subequations}
where ${\bf I}$ denotes the identity tensor, and $\bm F= {\bm F}_s + {\bm F}_b$.

It is worth noting that Fakhari2017 recovers the non-conservative form of the NS equations, whereas Liang2018 recovers the conservative form. Although these two forms are mathematically equivalent in the continuous sense, they can exhibit numerical differences in discrete errors, particularly at high density ratios.

The hydrodynamic quantities are updated as
\begin{subequations}
\begin{align}
& \rho \bm{u} = \sum_\beta \bm{e}_\beta f_\beta + \frac{1}{2} \bm{F}\delta t
\label{liang_velocity_update} \\[1mm]
& p = \frac{c_s^2}{1-\omega_0} \left[ \sum_{\beta \neq 0} f_\beta + \frac{\delta_t}{2} \bm{u} \cdot \boldsymbol{\nabla} \rho
+ \rho  w_0 \left( - \frac{\bm{u} \cdot \bm{u}}{2 c_s^2} \right) \right]
\label{liang_p_update}
\end{align}
\label{liang_update_all}
\end{subequations}

\subsubsection{Chai et al.'s nonlocal AC model}\label{subsubsection:Chai}
In addition to conservative AC models, several models have been developed based on the nonlocal AC equation. A representative example is the model proposed by Chai et al.~\cite{chai2018comparative} (hereafter Chai2018), whose equilibrium distribution functions and source terms are defined as
\begin{subequations}
\begin{align}
g_\alpha^{eq}(\bm{x},t) &= w_\alpha \, \phi \left( 1 + \frac{\bm{e}_\alpha \cdot \bm{u}}{c_s^2} \right),
\label{eq:Chai_geq} \\[2mm]
f_\beta^{eq}(\bm{x},t) &= w_\beta \left[ \frac{p}{\rho} + \frac{\bm{e}_\beta \cdot \bm{u}}{c_s^2}
+ \frac{(\bm{e}_\beta \cdot \bm{u})^2}{2 c_s^4} - \frac{\bm{u} \cdot \bm{u}}{2 c_s^2} \right].
\label{eq:Chai_feq} \\[2mm]
G_{\alpha}(\bm{x},t) &= \omega_\alpha M_\phi \Big[ - \frac{4\beta}{\kappa} (\phi - \phi_H)(\phi - \phi_L)
\left( \phi - \phi_0 \right) + \gamma(t) (\phi_H - \phi)(\phi - \phi_L) \Big],
\label{eq:chai_G} \\[2mm]
F_\beta(\bm{x},t) &= \omega_\beta \left[ \frac{\bm{e}_\beta \cdot \bm{F}}{c_s^2}
+ \frac{\bm{u} \bm{F} : (\bm{e}_\beta \bm{e}_\beta - c_s^2 \mathbf{I})}{c_s^2} \right],
\label{eq:chai_F}
\end{align}
\label{eq:chai_source}
\end{subequations}
where $\gamma(t)$ follows the definition in Eq.~(\ref{eq:lagrange_multiplier}), which involves a global integration step to enforce mass conservation.

The total force $\bm{F}$ is defined in the same way as in Fakhari2017, including surface tension, body, pressure, and viscous contributions.
The hydrodynamic quantities are updated as
\begin{subequations}
\begin{align}
& \phi = \sum_\alpha g_\alpha + \frac{G_\alpha}{2 \, \omega_\alpha} \, \delta t,
\label{chai2018_phi_update} \\[1mm]
& \bm{u} = \sum_\beta f_\beta \bm{e}_\beta + \frac{\bm{F}}{2} \, \delta t,
\label{chai2018_velocity_update} \\[1mm]
& p = \rho  \sum_\beta f_\beta
\label{chai2018_p_update}
\end{align}
\label{chai2018_update}
\end{subequations}

\subsubsection{Hu et al.'s hybrid AC model}\label{subsubsection:Hu}

To combine the advantages of both the conservative and nonlocal AC formulations, Hu et al.~\cite{hu2019hybrid} (hereafter Hu2019) proposed a hybrid AC model. In this approach, a global weight factor $\lambda$ is introduced to linearly blend the conservative AC equation with the nonlocal AC equation.
The equilibrium distribution functions are identical to those used in Chai2018, while the source terms are expressed as a weighted combination of the Fakhari2017 and Chai2018 formulations:
\begin{subequations}
\begin{align}
G_\alpha(\bm{x},t) &= \lambda \,
\omega_\alpha \, \bm{e}_\alpha \cdot
\left( \sqrt{\frac{2\beta}{\kappa}} \;
\frac{(\phi_H - \phi)(\phi - \phi_L)}{|\boldsymbol{\nabla} \phi|} \, \boldsymbol{\nabla} \phi \right)  \nonumber \\[2mm]
&\quad + (1-\lambda)\,
\omega_\alpha M_\phi \Big[ - \frac{4\beta}{\kappa} (\phi - \phi_H)(\phi - \phi_L)
\left( \phi - \phi_0 \right)
+ \gamma(t) (\phi_H - \phi)(\phi - \phi_L) \Big],
\label{eq:Hu_G} \\[3mm]
F_\beta(\bm{x},t) &= \omega_\beta \left[ \frac{\bm{e}_\beta \cdot \bm{F}}{c_s^2}
+ \frac{\bm{u} \bm{F} : (\bm{e}_\beta \bm{e}_\beta - c_s^2 \mathbf{I})}{c_s^2} \right].
\label{eq:Hu_F}
\end{align}
\label{eq:Hu_source}
\end{subequations}
The force term $\bm{F}$ in Eq.~\eqref{eq:Hu_F} follows the same definition as in Chai2018, and the hydrodynamic quantities are updated in the same manner (see Eq.~\eqref{chai2018_update}).

\subsubsection{Liu et al.'s hybrid AC model}\label{subsubsection:Liu}
Unlike the Hu2019 model, which employs a globally constant weighting factor $\lambda$, Liu et al.~\cite{liu2023improved} (hereafter Liu2023) introduced a spatially varying weighting coefficient $\lambda(\bm{x})$. This coefficient distinguishes between the bulk and interface regions according to the signed distance function $\psi(\bm{x})$:
\begin{equation}
\lambda(\bm{x}) =
\begin{cases}
0, & |\psi(\bm{x})| \geq d_{\min}, \\
1, & |\psi(\bm{x})| < d_{\min},
\end{cases}
\end{equation}
where $d_{\min} = 3\xi$ is the threshold distance to the interface. The signed distance function $\psi(\bm{x})$ is reconstructed using a predictor–corrector procedure.

In the predictor step, $\psi(\bm{x})$ is defined as
\begin{equation}
\psi(\bm{x}) =
\begin{cases}
-\dfrac{\xi}{4} \ln \left( \dfrac{\phi_H - \phi_L}{\phi - \phi_L} - 1 \right),
& \phi_L + 0.01 \leq \phi(\bm{x}) \leq \phi_H - 0.01, \\[8pt]
-\dfrac{\xi}{4} \ln \left( \dfrac{\phi_H - \phi_L}{\phi_H - 0.01 - \phi_L} - 1 \right),
& \phi(\bm{x}) > \phi_H - 0.01, \\[8pt]
-\dfrac{\xi}{4} \ln \left( \dfrac{\phi_H - \phi_L}{\phi_L + 0.01 - \phi_L} - 1 \right),
& \phi(\bm{x}) < \phi_L + 0.01,
\end{cases}
\end{equation}
and in the corrector step, a reinitialization equation is solved to enforce the signed-distance property:
\begin{equation}
\partial_{t'} \psi + S(\psi_0)\, (|\boldsymbol{\nabla} \psi| - 1) = 0,
\end{equation}
where $t'$ is a pseudo-time, $\psi_0 = \psi(\bm{x}, t'=0)$,
and the smeared-out sign function is defined as
\begin{equation}
S(\psi_0) = \dfrac{\psi_0}{\sqrt{\psi_0^2 + \Delta x^2}}.
\end{equation}

This design ensures that the conservative AC formulation (active where $\lambda = 1$) governs the interface region—allowing accurate capture of small-scale features—while the nonlocal AC formulation (active where $\lambda = 0$) applies in the bulk, thereby reducing numerical dispersion of the order parameter.

The equilibrium distribution functions follow Liang2018, while the source terms are given by:
\begin{subequations}
\begin{align}
G_\alpha(\bm{x},t) &= \lambda \, \omega_\alpha \, \bm{e}_\alpha \cdot
\left[ \frac{\partial_t (\phi \bm{u})}{c_s^2}
+ \sqrt{\frac{2\beta}{\kappa}} \; \frac{(\phi_H - \phi)(\phi - \phi_L)}{|\boldsymbol{\nabla} \phi|} \, \boldsymbol{\nabla} \phi \right] \nonumber \\[1mm]
&\quad + (1-\lambda)\,
\omega_\alpha M_\phi \Big[ - \frac{4\beta}{\kappa} (\phi - \phi_H)(\phi - \phi_L)
\left( \phi - \phi_0 \right)
+ \gamma(t) (\phi_H - \phi)(\phi - \phi_L) \Big],
\label{eq:Liu_G} \\[2mm]
F_\beta(\bm{x},t) &= \omega_\beta \left[ \bm{u} \cdot \boldsymbol{\nabla} \rho + \frac{\bm{e}_\beta \cdot \bm{F}}{c_s^2}
+ \frac{\bm{u} \boldsymbol{\nabla} \rho : (\bm{e}_\beta \bm{e}_\beta - c_s^2 \mathbf{I})}{c_s^2} \right].
\label{eq:Liu_F}
\end{align}
\label{eq:Liu_source}
\end{subequations}
It is worth noting that, unlike Hu2019, which combines Fakhari2017 with Chai2018, Liu2023 replaces Fakhari2017 with Liang2018, thereby introducing an additional time-derivative term. The force term is also defined following Liang2018, and the hydrodynamic quantities are updated in the same manner (see Eq.~\eqref{liang_update_all}).

\subsection{CH based LB models}\label{subsection:CH-PH-LB-model}
\subsubsection{Liang et al.'s CH model with incompressible NS}\label{subsubsection:LiangCHE}
While the CH equation-based LB models and the AC equation-based LB models discussed above recover different PF equations, they share the same LB equation. The essential differences remain in the definitions of the equilibrium distribution functions and the source terms.

The first CH-based model considered here is that of Liang et al.~\cite{liang2014phase} (hereafter Liang2014). In this model, the fluid flow is governed by the incompressible NS equations (Eq.~\eqref{eq:NS}).
The equilibrium distribution functions are defined as:
\begin{subequations}
\begin{align}
g_\alpha^{eq}(\bm{x},t) &=
\begin{cases}
\phi + (w_\alpha - 1) \, \eta \, \mu_\phi, & \alpha = 0, \\[1pt]
w_\alpha \, \eta \, \mu_\phi + w_\alpha \frac{\bm{e}_\alpha \cdot (\phi \bm{u})}{c_s^2}, & \alpha \ne 0,
\end{cases}
\label{eq:liang2_geq} \\[2mm]
f_\beta^{eq}(\bm{x},t) &=
\begin{cases}
\dfrac{p}{c_s^2} (w_\beta - 1) + \rho w_\beta \left[ \dfrac{\bm{e}_\beta \cdot \bm{u}}{c_s^2} + \dfrac{(\bm{e}_\beta \cdot \bm{u})^2}{2c_s^4} - \dfrac{\bm{u} \cdot \bm{u}}{2c_s^2} \right], & \beta = 0, \\[1pt]
\dfrac{p}{c_s^2} w_\beta + \rho w_\beta \left[ \dfrac{\bm{e}_\beta \cdot \bm{u}}{c_s^2} + \dfrac{(\bm{e}_\beta \cdot \bm{u})^2}{2c_s^4} - \dfrac{\bm{u} \cdot \bm{u}}{2c_s^2} \right], & \beta \ne 0.
\end{cases}
\label{eq:liang2_feq}
\end{align}
\label{eq:liang2_eqs}
\end{subequations}
where $\eta$ is an adjustable parameter controlling the mobility. Correspondingly, the relation between the relaxation time $\tau_\phi$ and the mobility is modified to
\begin{equation}
    {\tau _\phi } = \frac{M_\phi}{\eta {c_s}^2} + 0.5.
\end{equation}

The source terms are expressed as:
\begin{subequations}
\begin{align}
G_\alpha(\bm{x},t) &= \frac{\omega_\alpha \, \bm{e}_\alpha \cdot \partial_t (\phi \bm{u})}{c_s^2},
\label{eq:liang2_G} \\[2mm]
F_\beta(\bm{x},t) &= \frac{(\bm{e}_\beta - \bm{u})}{c_s^2} \cdot \left[
\Gamma_\beta(\bm{u})  \boldsymbol{\nabla} (\rho c_s^2)
+ (\bm{F}_s + \bm{F}_a + \bm{F}_b)
s_{\beta}\left(\bm u\right)\right],
\label{eq:liang2_F}
\end{align}
\label{eq:Liang2_source}
\end{subequations}
with
\begin{equation}
   \Gamma_\beta(\bm{u}) = w_\beta \left(
\dfrac{\bm{e}_\beta \cdot \bm{u}}{c_s^2}
+ \dfrac{(\bm{e}_\beta \cdot \bm{u})^2}{2c_s^4}
- \dfrac{\bm{u} \cdot \bm{u}}{2c_s^2} \right)\quad s_\beta(\bm{u}) = w_\beta \left(
1+\dfrac{\bm{e}_\beta \cdot \bm{u}}{c_s^2}
+ \dfrac{(\bm{e}_\beta \cdot \bm{u})^2}{2c_s^4}
- \dfrac{\bm{u} \cdot \bm{u}}{2c_s^2} \right).
\label{eq:Gamma_beta}
\end{equation}

Here, the forces are defined as $\bm{F}_s = \mu_\phi \, \boldsymbol{\nabla} \phi$ and $\bm{F}_a = \frac{\rho_H - \rho_L}{\phi_H - \phi_L} \, \boldsymbol{\nabla} \cdot \left( M_\phi \, \boldsymbol{\nabla} \mu_\phi \right) \, \bm{u}$.

The velocity is updated as
\begin{equation}
\bm{u} = \frac{\sum_\beta \bm{e}_\beta f_\beta + 0.5 \delta t  (\bm{F}_s + \bm{F}_b)}
{\rho - 0.5  \delta t  \frac{\rho_H - \rho_L}{\phi_H - \phi_L} \boldsymbol{\nabla} \cdot M_\phi  \boldsymbol{\nabla} \mu_\phi}.
\label{liang2014_velocity_update}
\end{equation}

The pressure is updated in the same manner as in Liang2018 (Eq.~\eqref{liang_p_update}).

\subsubsection{Yang et al.'s CH model with quasi-incompressible NS}\label{subsubsection:YangCHE}
Due to the density difference between the two fluids, the incompressible NS equations cannot conserve mass locally\cite{shen2013mass}. To better account for compressibility effects in multiphase flows, Yang et al.~\cite{yang2016lattice} (hereafter Yang2016) proposed a quasi-incompressible CH model. In this model, the CH equation itself is unchanged, while the fluid flow is governed by the quasi-incompressible NS equations (Eq.~\ref{eq:quasi_incomp_system}).

The equilibrium distribution functions are defined as
\begin{subequations}
\begin{align}
g_\alpha^{eq}(\bm{x},t) &=
\begin{cases}
\phi - (1 - w_\alpha) \, \eta \, \mu_\phi + w_\alpha \, \phi \left( \frac{\bm{e}_\alpha \cdot \bm{u}}{c_s^2} + \frac{(\bm{e}_\alpha \cdot \bm{u})^2}{2 c_s^4} - \frac{\bm{u} \cdot \bm{u}}{2 c_s^2} \right), & \alpha = 0, \\[1pt]
w_\alpha \, \eta \, \mu_\phi + w_\alpha \, \phi \left( \frac{\bm{e}_\alpha \cdot \bm{u}}{c_s^2} + \frac{(\bm{e}_\alpha \cdot \bm{u})^2}{2 c_s^4} - \frac{\bm{u} \cdot \bm{u}}{2 c_s^2} \right), & \alpha \ne 0,
\end{cases}
\label{eq:Yang_g_eq} \\[2mm]
f_\beta^{eq}(\bm{x},t) &= w_\beta \left[ p + c_s^2 \, \rho \left( \frac{\bm{e}_\beta \cdot \bm{u}}{c_s^2} + \frac{(\bm{e}_\beta \cdot \bm{u})^2}{2 c_s^4} - \frac{\bm{u} \cdot \bm{u}}{2 c_s^2} \right) \right].
\label{eq:Yang_f_eq}
\end{align}
\label{eq:Yang_eq}
\end{subequations}

The corresponding source terms are
\begin{subequations}
\begin{align}
&G_\alpha(\bm{x},t) = - \frac{\phi}{c_s^2 \rho} (\bm{e}_\alpha - \bm{u}) \cdot (\boldsymbol{\nabla} p - \bm{F}) \,  \, s_\alpha(\mathbf{u}),
\label{eq:Yang_G} \\[2mm]
&F_\beta(\bm{x},t) = (\bm{e}_\beta - \bm{u}) \cdot \left[  \bm{F} \, s_\beta(\bm{u}) +  \Gamma_\beta(\bm{u}) \, c_s^2 \boldsymbol{\nabla} \rho \right] - \omega_\beta c_s^2 \rho \, \gamma \, \boldsymbol{\nabla} \cdot (M_\phi \boldsymbol{\nabla} \mu_\phi),
\label{eq:Yang_F} \\[2mm]
&s_\alpha(\bm{u}) = w_\alpha\left[1 + \frac{\bm{e}_\alpha \cdot \bm{u}}{c_s^2} + \frac{(\bm{e}_\alpha \cdot \bm{u})^2}{2 c_s^4} - \frac{\bm{u} \cdot \bm{u}}{2 c_s^2}\right],
\label{eq:Yang_Gamma}
\end{align}
\label{eq:Yang_source}
\end{subequations}
Here, $\Gamma_\beta(\bm{u})$ and $s_\beta(\bm{u})$ are consistent with the definitions in Eq.~\eqref{eq:liang2_F}, and the total force is given by $\bm{F} = \bm{F}_s + \bm{F}_b$, with $\bm{F}_s$ defined as in Liang2014.

The hydrodynamic quantities are updated as
\begin{subequations}
\begin{align}
&p = \sum_{\beta} f_{\beta}
    + \frac{1}{2} \delta_t c_s^2 \Big[ \bm{u} \cdot \boldsymbol{\nabla} \rho
    - \gamma \rho \boldsymbol{\nabla} \cdot (M_\phi  \boldsymbol{\nabla} \mu_\phi) \Big],
    \label{eq:yang_p_update} \\[1mm]
&\rho  c_s^2  \bm{u} = \sum_\beta \bm{e}_\beta f_\beta
    + \frac{1}{2}  \delta_t  c_s^2  \bm{F},
    \label{eq:yang_u_update}
\end{align}
\label{eq:yang_update}
\end{subequations}

\subsubsection{Bao et al.'s singular mobility CH model with quasi-incompressible NS}\label{subsubsection:BaoCHE}
The final model included in this comparative study is the recent CH equation-based model proposed by Bao et al.~\cite{bao2024phase} with singular mobility (hereafter Bao2024). This model was developed to address the spontaneous disappearance of small droplets below a critical size, which can occur in traditional CH models. The targeted hydrodynamic equations are introduced in Sec.~\ref{subsection:AC-PH-LB-model}.

The key idea of this model is to set the mobility to zero in the bulk phase, as discussed in Sec.~\ref{subsection:Allen-Cahn equation}. However, since the mobility is linearly related to the relaxation time $\tau_\phi$, zero mobility corresponds to $\tau_\phi = 0.5$, which would lead to numerical instability in standard PF-LB models. To overcome this limitation, the model adopts a lattice kinetic scheme~\cite{inamuro2002lattice}, which introduces gradient-related terms into the equilibrium distributions, thereby decoupling the linear relationship between diffusivity coefficients (mobility and viscosity) and the relaxation times.

The equilibrium distribution functions are modified as
\begin{subequations}
\begin{align}
g_\alpha^{eq}(\bm{x},t) &=
\begin{cases}
\phi - (1-w_\alpha)\eta \mu_\phi + \phi \, {\Gamma}_\alpha(\bm{u}) + w_\alpha A \delta t \, \eta \, \bm{e}_\alpha \cdot \boldsymbol{\nabla} \mu_\phi, & \alpha = 0, \\[1pt]
w_\alpha \eta \mu_\phi +  \phi \, {\Gamma}_\alpha(\bm{u}) + w_\alpha A \delta t \, \eta \, \bm{e}_\alpha \cdot \boldsymbol{\nabla} \mu_\phi, & \alpha \ne 0,
\end{cases}
\label{eq:bao_g_eq} \\[2mm]
f_\beta^{eq}(\bm{x},t) &= w_\beta p + c_s^2 \rho \, {\Gamma}_\beta(\bm{u})
+ \frac{1}{2} \rho w_\beta B \delta t \, \mathbf{S} : \left( \bm{e}_\beta \bm{e}_\beta - c_s^2 \mathbf{I} \right),
\label{eq:bao_f_eq}
\end{align}
\label{eq:bao_eq}
\end{subequations}
where $\mathbf{S} = \boldsymbol{\nabla}\bm{u} + (\boldsymbol{\nabla}\bm{u})^T$ is the shear stress tensor, and $A$ and $B$ are free parameters that locally adjust the relationship between the relaxation times and the mobility and kinematic viscosity:
\begin{subequations}
\begin{align}
M_\phi &= c_s^2 \, \eta \, \delta t \, (\tau_\phi - 0.5 - A),
\label{eq:bao_Mphi} \\[2mm]
\nu &= c_s^2 \, \delta t \, (\tau - 0.5 - B),
\label{eq:bao_v}
\end{align}
\label{eq:bao_tau}
\end{subequations}

In practice, $\tau_\phi$ and $\tau$ are kept constant to ensure numerical stability, while $A$ and $B$ are varied locally to match the required mobility and viscosity. The source terms $G_\alpha$ and $F_\beta$ are defined in the same manner as in Yang2016~\cite{yang2016lattice}, and the hydrodynamic quantities are updated according to Eq.~\eqref{eq:yang_update}.

In summary, this section reviewed eight representative PF-LB models developed in recent years:
\begin{itemize}
    \item Two conservative AC models: Fakhari2017 (non-conservative NS) and Liang2018 (conservative NS);
    \item One nonlocal AC model: Chai2018;
    \item Two hybrid AC models: Hu2019 (uniform weight) and Liu2023 (spatial variant weight);
    \item One CH model with incompressible NS: Liang2014;
    \item Two CH models with quasi-incompressible NS: Yang2016 and Bao2024 (sigular mobility).
\end{itemize}

In the following sections, the performance of these models will be systematically evaluated and compared through a series of laminar and turbulent flow cases.

\section{Two-dimensional laminar flow tests}\label{sec:laminar}

In this section, we evaluate the performance of the eight PF-LB models introduced in the previous section through a series of two-dimensional laminar test cases. For the Hu2019 model, the global weighting coefficient is set to $\lambda = 0.9$, following the recommendation in the original work to accurately capture small-scale features.

For all two-dimensional cases considered here, the D2Q9 lattice model is employed for both the PF equation and NS equations. Unless otherwise specified, the BGK collision operator is used. In the simulations, the second-order isotropic central schemes are used for the discretizations of the gradient and the Laplacian operator, taking the order parameter as an example,
\begin{subequations}
\begin{align}
\boldsymbol{\nabla}\phi(\bm x) &=\frac{1}{c_s^2\delta t}\sum_\alpha \omega_\alpha \bm e_\alpha \phi(\bm x+\bm e_\alpha\delta t), \\
\nabla^2 \phi(\bm x) &=\frac{2}{c_s^2\delta t^2}\sum_\alpha \omega_\alpha \left[\phi(\bm x+\bm e_\alpha\delta t)-\phi(\bm x)\right].
\end{align}
\end{subequations}

\subsection{Diagonal translation of a circular droplet}\label{subsec:diag-translation-circle}

The first case considered is the diagonal translation of a circular droplet, which primarily tests the ability of different models to maintain the interface~\cite{liu2023improved}. A circular droplet of radius $R = L_0/5$ is initially placed at the center of a square domain of size $L_0 \times L_0$. The initial order parameter distribution is
\begin{equation}
\phi(x,y,0) = \frac{\phi_H + \phi_L}{2}
+ \frac{\phi_H - \phi_L}{2} \,
\tanh \left\{
\frac{2 \left[ R - \sqrt{(x - 0.5 L_0)^2 + (y - 0.5 L_0)^2} \right]}{\xi}
\right\}.
\label{eq:init-phi-circle}
\end{equation}

The velocity field is specified as ${\bm{U}} = \left( {U_0, U_0} \right)$, which causes the droplet to migrate along the diagonal and return to its initial position after one or integer numbers of period $T_0 = L_0 / U_0$, ideally without deformation. Simulation parameters are ${L_0} = 200$, ${U_0} = 0.02$, ${\rho_H} = {\rho_L} = 1$, $\sigma = 0.01$, ${M_\phi} = 0.0013$, and $\xi = 0.02 L_0$. The initial pressure is uniform and set to 0.

\begin{figure}
    \centering
    \includegraphics[width=1.0\textwidth]{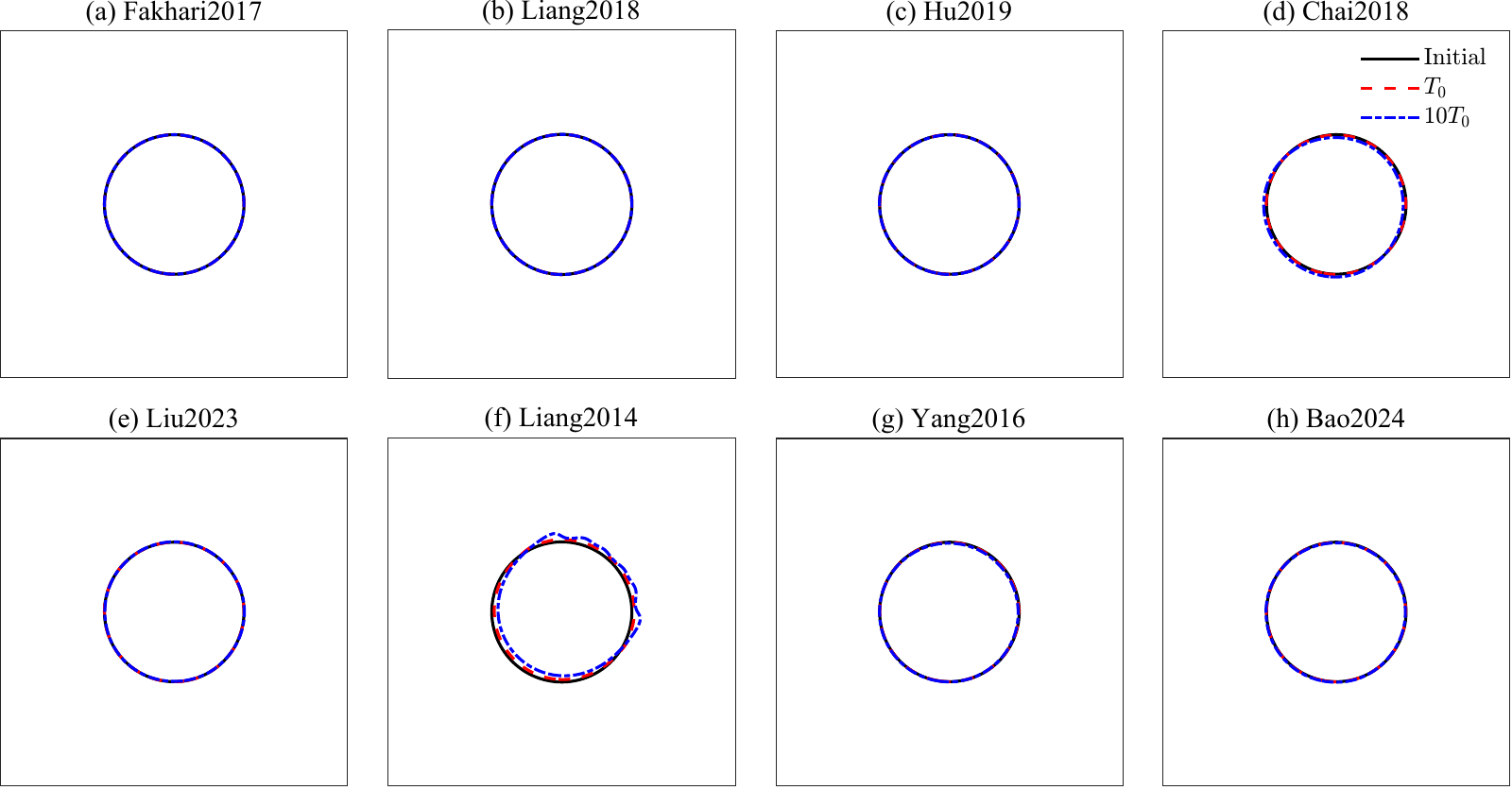}
    \caption{Contours of the droplet interface at $t = 0$, $t = T_0$, and $t = 10T_0$.}
    \label{fig:circular_interface_1and10T0}
\end{figure}

Figures~\ref{fig:circular_interface_1and10T0} show the droplet interface (contour line $\phi = 0.5$) predicted by each model after one and ten periods. After one period, the incompressible CH model Liang2014 shows slight deviation from the initial interface, which becomes more pronounced after ten periods. The nonlocal AC model Chai2018 also shows deviation after ten periods, but to a lesser extent than Liang2014. These results indicate that Liang2014 has relatively poor interface-capturing ability, whereas Chai2018 performs better. These differences have been noted in previous studies \citep{bao2024phase,liu2023improved}. For the other six models, no visible differences between the initial and predicted interfaces are observed.

To quantitatively assess model performance, the following four quantities are computed:
\begin{enumerate}
    \item Numerical dispersion of the order parameter:
\begin{equation}
\delta \phi_{\max} = \max \left\{ |a|, |b| \right\}, \qquad a = \max \phi(\bm{x}) - \phi_H, \qquad
b = \min \phi(\bm{x}) - \phi_L.
\end{equation}

\item L2 error of the order parameter:
\begin{equation}
 \mathrm{L2\ Err}(\phi) =
\sqrt{
\frac{
\sum_{\bm x} \left[ \phi({\bm x},t) - \phi({\bm x},0) \right]^2
}{
\sum_{\bm x} \left[ \phi({\bm x},0) \right]^2
}
}.
\label{eq:phi-error}
\end{equation}

\item Migration of the droplet centroid:
\begin{equation}
\text{Centroid Offset} = \sqrt{
\left(X_{\rm final} - X_0\right)^2 +
\left(Y_{\rm final} - Y_0\right)^2
},
\label{eq:centroid_offset}
\end{equation}
where $(X_0, Y_0)$ denote the initial coordinates of the droplet centroid, while $(X_{\rm final}, Y_{\rm final})$ represent the centroid coordinates at the final time, e.g., after one or several periods.

\item Roundness of the interface:
\begin{equation}
C_r = \frac{\text{Perimeter of a circle with the same area as the droplet}}{\text{Actual perimeter of the droplet interface}}.
\label{eq:roundness}
\end{equation}

A value of $C_r = 1$ corresponds to a perfect circle, with smaller values indicating increasing deviation from circularity.

\end{enumerate}

Table~\ref{tab:diagonal_translation_combined_8cols} presents the four quantities computed from the simulations using different models after one ($T_0$) and ten ($10T_0$) periods. Except for the roundness, whose ideal value is 1, the other three quantities should be as small as possible for high accuracy.
After ten periods, Fakhari2017 and Liang2018 maintain both roundness and centroid position very well, indicating that conservative AC models clearly excel in interface preservation.

%The inclusion of the time-derivative term in Liang2018, while enabling the exact recovery of the conservative AC equation after Chapman-Enskog analysis, does not produce a noticeable impact on the numerical results, consistent with the findings of Fakhari et al.~\cite{fakhari2017improved}.
%This ineffectiveness can be understood by examining the order of magnitude of the added time derivative under diffusive scaling corresponding to incompressible N-S equations: $\delta t = h^2$, $\delta x = h$, ${\bm e}_\alpha\sim\delta h/\delta t\sim O(1/h)$, $c_s^2= a/h^2\sim \vert{{\bm e}_\alpha}\vert^2 $, and all hydrodynamic quantities are $O(1)$, where $h$ is a small parameter.

%Under this scaling, the source term in Liang2018 (Eq.~\ref{eq:liang_G_final}) can be expressed as
%\begin{equation}
%\begin{aligned}
%    G_\alpha(\bm{x},t) =& \omega_\alpha \, \bm{e}_\alpha \cdot \left[\frac{\partial_t (\phi \bm{u})}{c_s^2}
%+
%\sqrt{\frac{2\beta}{\kappa}} \; \frac{(\phi_H - \phi)(\phi - \phi_L)}{|\boldsymbol{\nabla} \phi|} \, \boldsymbol{\nabla} \phi \right],\\
%=& \omega_\alpha \, \frac{\overline{\bm{e}}_\alpha}{h} \cdot \left[\frac{\partial_t (\phi \bm{u})h^2}{a}
%+
%\sqrt{\frac{2\beta}{\kappa}} \; \frac{(\phi_H - \phi)(\phi - \phi_L)}{|\boldsymbol{\nabla} \phi|} \, \boldsymbol{\nabla} \phi \right].
%\end{aligned}
%\end{equation}
%It is evident that the added time-derivative term is two orders of magnitude smaller than the main contribution in the source term, which explains why it does not have a significant effect on the numerical results.

Hu2019 and Liu2023, as hybrid AC models, exhibit some centroid offset due to their reliance on the nonlocal AC equation. Moreover, since Hu2019 employs a globally uniform weight $\lambda = 0.9$, whereas Liu2023 uses a spatially varying weight with the full nonlocal AC equation in the bulk region, the centroid offset in Hu2019 is smaller. This indicates that the nonlocal AC model is less effective at interface preservation compared to the conservative AC models. Among the CH models, all quasi-incompressible variants maintain roundness well after ten periods, in contrast to the incompressible CH model of Liang2014. This highlights the importance of properly accounting for compressibility when selecting a CH-based model.

\begin{table}
\centering
\caption{Comparison of different models for the diagonal translation of a circular interface at times $t = T_0$ and $t = 10T_0$}
\label{tab:diagonal_translation_combined_8cols}
\resizebox{\textwidth}{!}{%
\begin{tabular}{c c c c c c c c c}
\hline
Quantity & Fakhari2017 & Liang2018 & Hu2019 & Chai2018 & Liu2023 & Liang2014 & Yang2016 & Bao2024 \\
\hline
$\delta \phi_{\rm max}$ (T$_0$) & $2.5700\times 10^{-4}$ & $2.5800\times 10^{-4}$ & $2.2800\times 10^{-4}$ & $6.8000\times 10^{-5}$ & $1.7900\times 10^{-4}$ & $2.3686\times 10^{-1}$ & $6.1300\times 10^{-3}$ & $1.4936\times 10^{-2}$ \\
$\delta \phi_{\rm max}$ (10T$_0$) & $2.6400\times 10^{-4}$ & $2.6400\times 10^{-4}$ & $2.3100\times 10^{-4}$ & $7.5000\times 10^{-5}$ & $1.8300\times 10^{-4}$ & $2.9286\times 10^{-1}$ & $6.7900\times 10^{-3}$ & $5.6308\times 10^{-2}$ \\
Centroid Offset (T$_0$) & 0.0000 & 0.0000 & 0.0013 & 0.0252 & 0.0000 & 0.2540 & 0.0132 & 0.0226 \\
Centroid Offset (10T$_0$) & 0.0000 & 0.0000 & 0.0122 & 0.2694 & 0.1066 & 0.6033 & 0.0699 & 0.0573 \\
Roundness (T$_0$) & 1.0000 & 1.0000 & 1.0000 & 1.0000 & 1.0000 & 0.9994 & 0.9999 & 0.9999 \\
Roundness (10T$_0$) & 1.0000 & 1.0000 & 1.0000 & 0.9998 & 0.9999 & 0.9876 & 0.9999 & 0.9999 \\
L2 Err($\phi$) (T$_0$) & 0.0032 & 0.0031 & 0.0036 & 0.0140 & 0.0020 & 0.1486 & 0.0125 & 0.0145 \\
L2 Err($\phi$) (10T$_0$) & 0.0031 & 0.0031 & 0.0076 & 0.1212 & 0.0053 & 0.2898 & 0.0382 & 0.0370 \\
\hline
\end{tabular}%
}
\end{table}

In summary, the conservative AC models, hybrid AC models, and CH models with quasi-incompressibility effectively maintain the droplet interface, whereas the standard CH model and nonlocal AC model exhibit larger errors. However, the nonlocal AC model demonstrates the best boundedness of the order parameter among all tested models, indicating that the enforcement of global mass conservation helps suppress numerical dispersion, consistent with the findings of Hu et al.~\cite{hu2019hybrid}.

\subsection{A rising bubble}\label{subsec:Bubbles rise}

\begin{figure}
    \centering
    \includegraphics[width=0.2\textwidth]{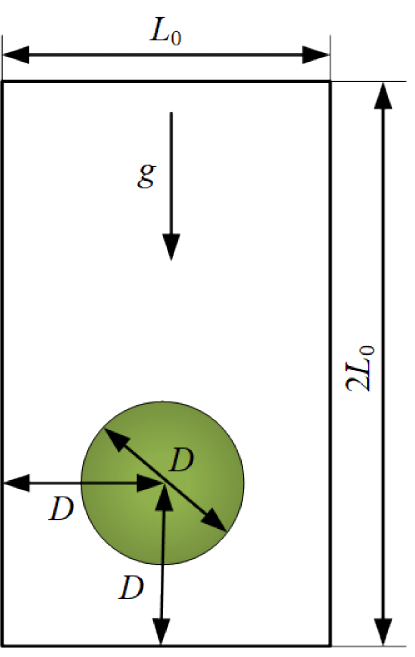} % 将宽度缩小到文本宽度的 20%
    \caption{A schematic diagram of bubble rising under gravity.}
    \label{fig:bubble_schematic}
\end{figure}

The second case considers a bubble rising under the action of gravity, used to assess the models' ability to simulate large topological deformations \cite{aland2012benchmark}. In this case, the density and viscosity ratios are 1000 and 100, respectively. As illustrated in Figure~\ref{fig:bubble_schematic}, a circular bubble with initial diameter $D$ is placed in the liquid, with its center at $(D, D)$ in a computational domain of size $L_0 \times 2L_0$, where $L_0 = 2D$. The motion of the rising bubble is characterized by the Reynolds and E\"otv\"os numbers, defined as
\begin{equation}
    \mathrm{Re} = \frac{\rho_H U_g D}{\mu_H}, \qquad
\mathrm{Eo} = \frac{\rho_H U_g^2 D}{\sigma},
\end{equation}
where $U_g = \sqrt{gD}$ is the characteristic velocity scale determined by gravity, with $g$ denoting the gravitational acceleration.

For the simulation, the parameters are set as $D = 120$, $U_g = 0.005$, $\xi = 5$, $M_\phi = 0.01$, $\rho_H = 100$, and $\rho_L = 0.1$, corresponding to $\mathrm{Re} = 35$ and $\mathrm{Eo} = 125$. Periodic boundary conditions are applied horizontally, while no-slip boundary conditions are enforced on the top and bottom walls. Time is nondimensionalized as $T_b = t U_g / D$. The initial distribution of the order parameter is given by
\begin{equation}
\phi(x,y,0) = \frac{\phi_H + \phi_L}{2}
+ \frac{\phi_H - \phi_L}{2} \,
\tanh \frac{2 \Big[ \sqrt{(x - D)^2 + (y - D)^2} - 0.5 D \Big]}{\xi}.
\label{eq:init-phi-bubbles}
\end{equation}

\begin{figure}
    \centering

    % 子图 a
    \begin{subfigure}[b]{0.48\textwidth}
        \centering
        \includegraphics[width=\textwidth]{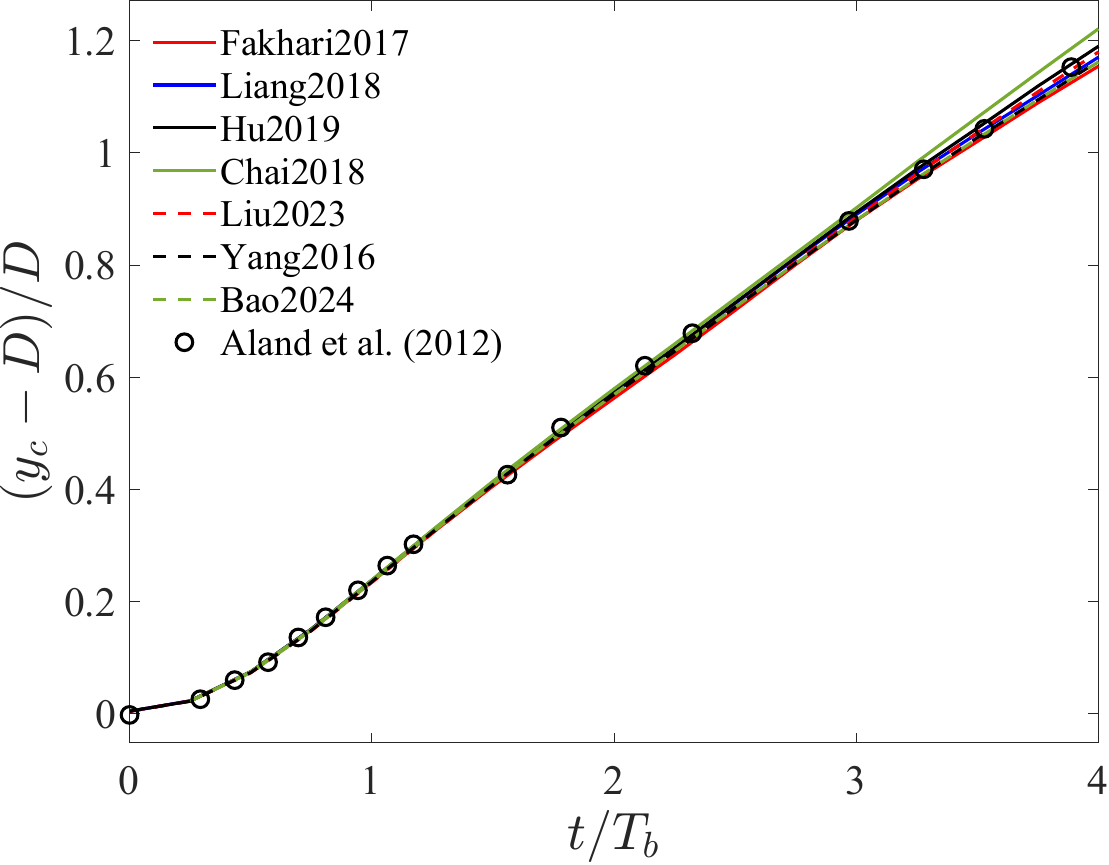}
        \caption{Comparison of bubble centroid positions}
        \label{fig:bubble_centroid}
    \end{subfigure}
    \hfill
    % 子图 b
    \begin{subfigure}[b]{0.48\textwidth}
        \centering
        \includegraphics[width=\textwidth]{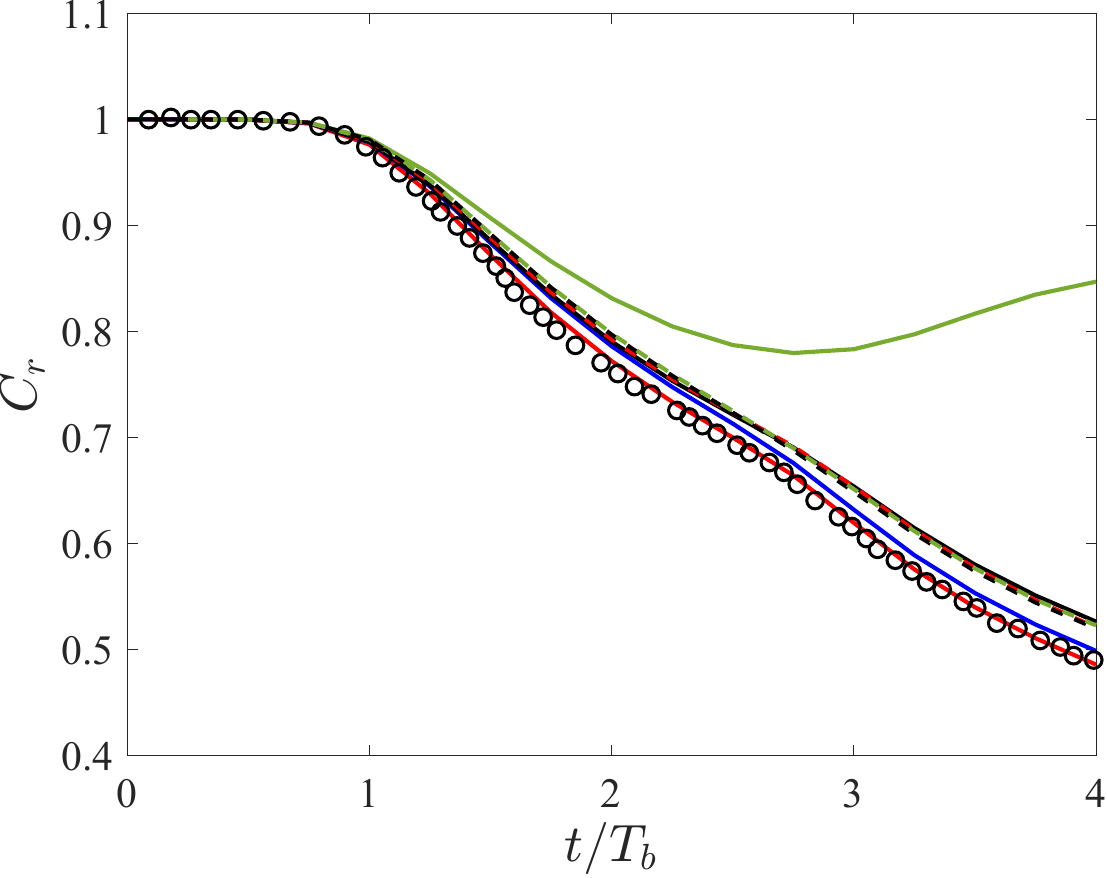}
        \caption{Comparison of bubble roundness}
        \label{fig:bubble_roundness}
    \end{subfigure}

    \caption{Comparison of bubble centroid positions and roundness with the results of Aland~\cite{aland2012benchmark}.}
    \label{fig:bubble_subfigs}
\end{figure}

It should be noted that the incompressible CH model of Liang2014 is excluded due to numerical instability when handling large density and viscosity ratios.
Figures~\ref{fig:bubble_centroid} and \ref{fig:bubble_roundness} show the time evolution of the normalized bubble centroid position $(y_c - D)/D$ and roundness (Eq.~\ref{eq:roundness}) predicted by the different models, compared with benchmark results from Aland et al.~\cite{aland2012benchmark}. These benchmarks were obtained using high-resolution finite element discretizations, including Eulerian level-set and arbitrary Lagrangian-Eulerian moving grid methods, providing a standard for quantitatively assessing interface evolution and bubble dynamics under high density and viscosity ratios.

Most models predict the bubble centroid location in good agreement with the benchmark, except the nonlocal AC model of Chai2018, which shows a slight deviation. This discrepancy is further amplified in the roundness, where the nonlocal AC model exhibits significant deviations.

\begin{figure}
    \centering
    \includegraphics[width=0.35\textwidth]{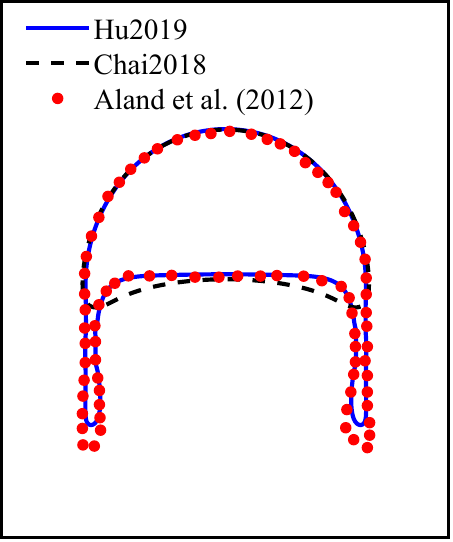}
    \caption{Contours of the nonlocal ACE model.}
    \label{fig:bubble_nonlocal_ACE}
\end{figure}

Interface contours at $T_b = 4$ (Figure~\ref{fig:bubble_nonlocal_ACE}) suggest that this deviation arises from the coarsening effect inherent to the nonlocal AC model, which captures small features poorly. Consequently, it fails to resolve the thin filament beneath the bubble, causing a notable error in roundness and a slightly elevated centroid position compared to the benchmark. The hybrid AC models, inheriting features of the nonlocal AC model, also produce shortened filaments, resulting in roundness deviations.

\begin{figure}
    \centering
    \includegraphics[width=0.8\textwidth]{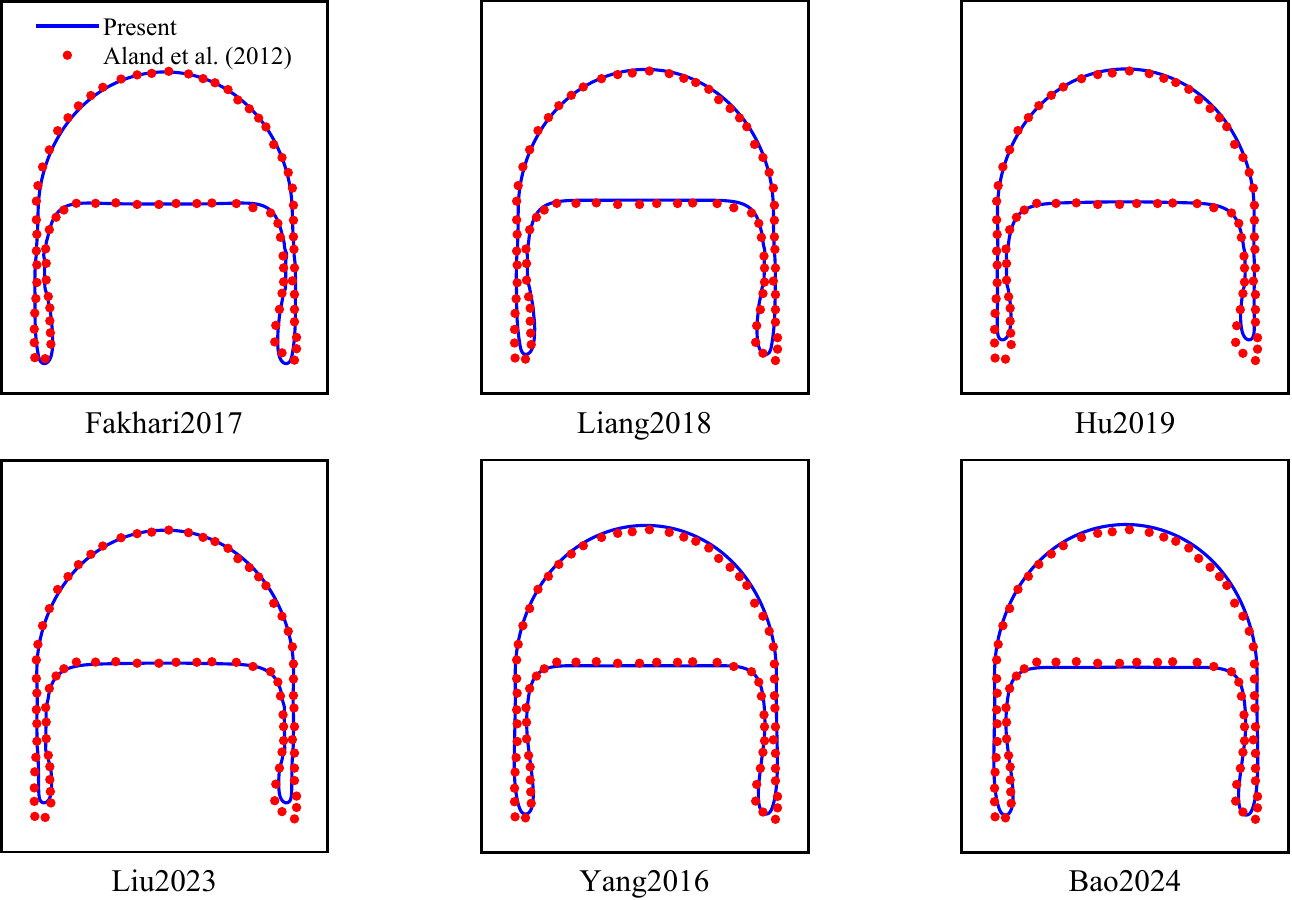}
    \caption{Comparison of bubble interface contours.}
    \label{fig:bubble_interface}
\end{figure}

For the remaining models, interface contours at $T_b = 4$ are shown in Figure~\ref{fig:bubble_interface}, all demonstrating reasonable agreement with the benchmark results. The conservative AC models and the two quasi-incompressible CH models show good agreement, with the conservative AC model of Fakhari2017 providing the most accurate prediction. Notably, the time-derivative correction slightly worsens the results in this case, likely due to errors in its numerical evaluation. In our simulations, this term is approximated using the backward Euler method:
\begin{equation}
\partial_t(\phi \bm{u}) \approx \frac{\phi(t)\bm{u}(t) - \phi(t-\delta t)\bm{u}(t-\delta t)}{\delta t},
\end{equation}
which is only first-order accurate. While higher-order schemes could improve accuracy, they would also increase algorithmic complexity and reduce computational efficiency.

\subsection{Rayleigh–Taylor instability}\label{subsec:Rayleigh–Taylor instability}

In the third case, we consider the classical Rayleigh-Taylor instability to assess the models’ ability to simulate flows with complex interface evolution \cite{liska2003comparison}. Initially, the heavy fluid is placed above the light fluid, with the interface located at $y = 2L_0$ in a computational domain of $L_0 \times 4L_0$, subject to periodic boundary conditions in the horizontal direction and no-slip conditions on the top and bottom walls. A sinusoidal perturbation is imposed on the interface:
${y_t} = 2{L_0} + 0.05{L_0}\cos \left( {2\pi x/{L_0}} \right)$, which defines the initial order parameter as
\begin{equation}
    \phi (x,y,0) = \frac{{{\phi _H} + {\phi _L}}}{2} + \frac{{{\phi _H} - {\phi _L}}}{2} \tanh \left[\frac{{2\left(y - {y_t}\right)}}{\xi }\right].
\end{equation}
The flow is characterized by the Reynolds number ($\mathrm{Re}$), Atwood number ($\mathrm{At}$), and Péclet number ($\mathrm{Pe}$) \cite{tryggvason2001front}:
\begin{equation}
{\rm Re} = \frac{L_0}{\nu} \sqrt{\frac{{\rm At} \, g \, L_0}{1 + {\rm At}}},\qquad {\rm At} = \frac{\rho_H - \rho_L}{\rho_H + \rho_L},\qquad {\rm Pe} = \frac{L_0 \sqrt{g L_0}}{M_\phi}.
\end{equation}
The parameters are set as $L_0 = 256$, $U_g = \sqrt{gL_0} = 0.04$, $\sigma = 5 \times 10^{-5}$, $\mathrm{Re} = 3000$, $\mathrm{At} = 0.1$, $\mathrm{Pe} = 1000$, $\rho_H = 11/9$, and $\rho_L = 1$. The time is normalized by $T_0 = \sqrt{L_0/({\rm At} g) }$.

During the early stage of the instability, the heavy fluid penetrates into the light fluid, forming a spike from the initial convex shape. Two counter-rotating vortices subsequently develop on either side of the spike. As time progresses, the tip stretches and breaks up into numerous small structures.

\begin{figure}
    \centering

    % 子图 a
    \begin{subfigure}[b]{\textwidth}
        \centering
        \includegraphics[width=0.8\textwidth]{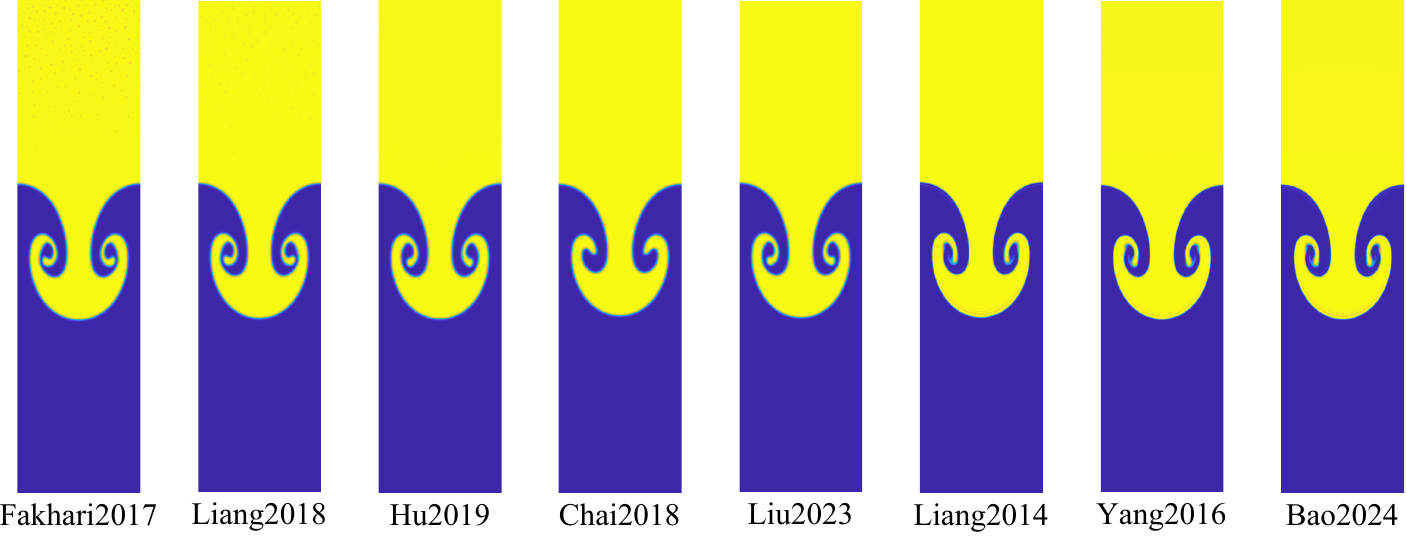}
        \caption{$t/T_0=2$}
        \label{fig:RT_snap_a}
    \end{subfigure}
    \vspace{0.5em}

    % 子图 b
    \begin{subfigure}[b]{\textwidth}
        \centering
        \includegraphics[width=0.8\textwidth]{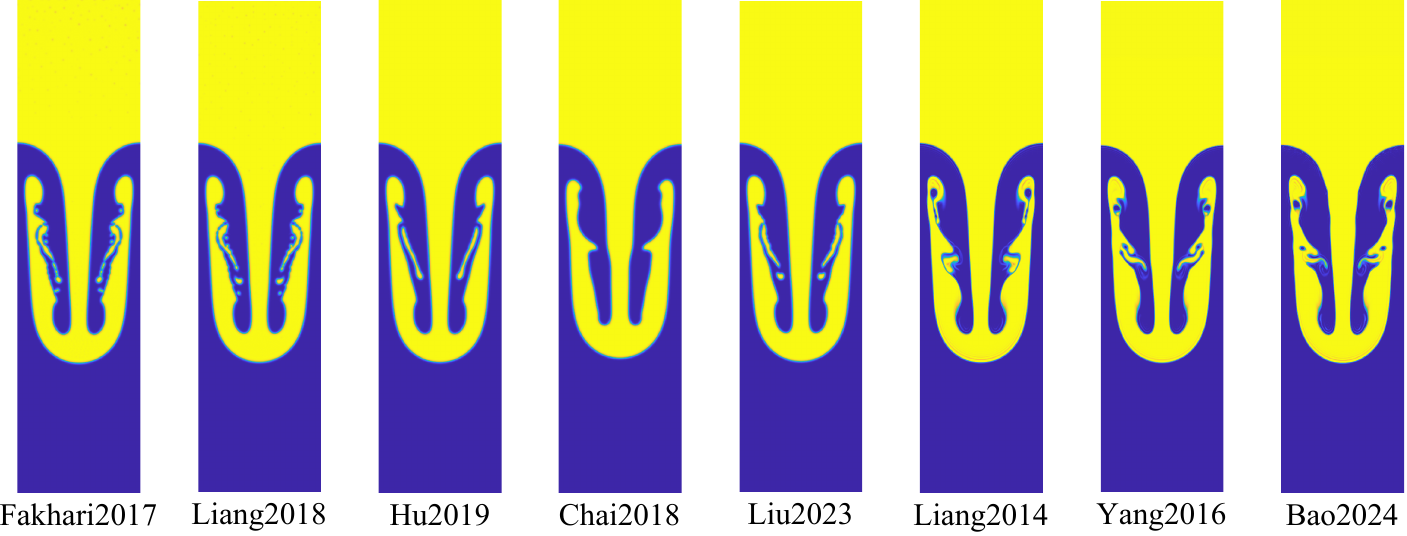}
        \caption{$t/T_0=3$}
        \label{fig:RT_snap_b}
    \end{subfigure}
    \vspace{0.5em}

    % 子图 c
    \begin{subfigure}[b]{\textwidth}
        \centering
        \includegraphics[width=0.8\textwidth]{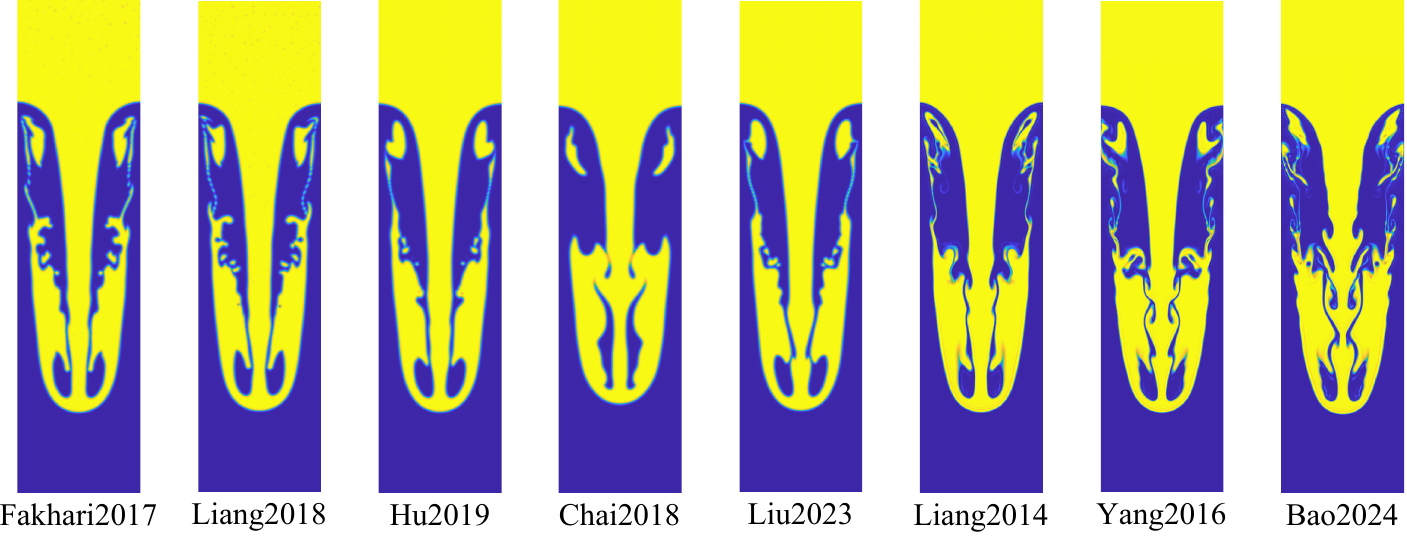}
        \caption{$t/T_0=3.7$}
        \label{fig:RT_snap_c}
    \end{subfigure}

    \caption{Snapshots of the two-dimensional Rayleigh--Taylor instability simulated by different methods.}
    \label{fig:RT_snapshots}
\end{figure}

Figures~\ref{fig:RT_snapshots}(a)--(c) show snapshots of the order parameter at $T_0 = 2$, $3$, and $3.7$, respectively. The nonlocal AC model of Chai2018 exhibits noticeable differences from the other models even during early vortex formation, due to its strong coarsening effect and weaker ability to capture fine-scale features. As the flow evolves, this limitation becomes more pronounced, and the model fails to resolve small vortex breakup. Among the test models, the CH-based models—particularly the quasi-incompressible models Yang2016 and Bao2024—capture the richest small-scale structures, followed by the conservative and hybrid AC models.

\begin{figure}
    \centering

    % 子图 (a) Bubble front
    \begin{subfigure}[b]{0.48\textwidth}
        \centering
        \includegraphics[width=\textwidth]{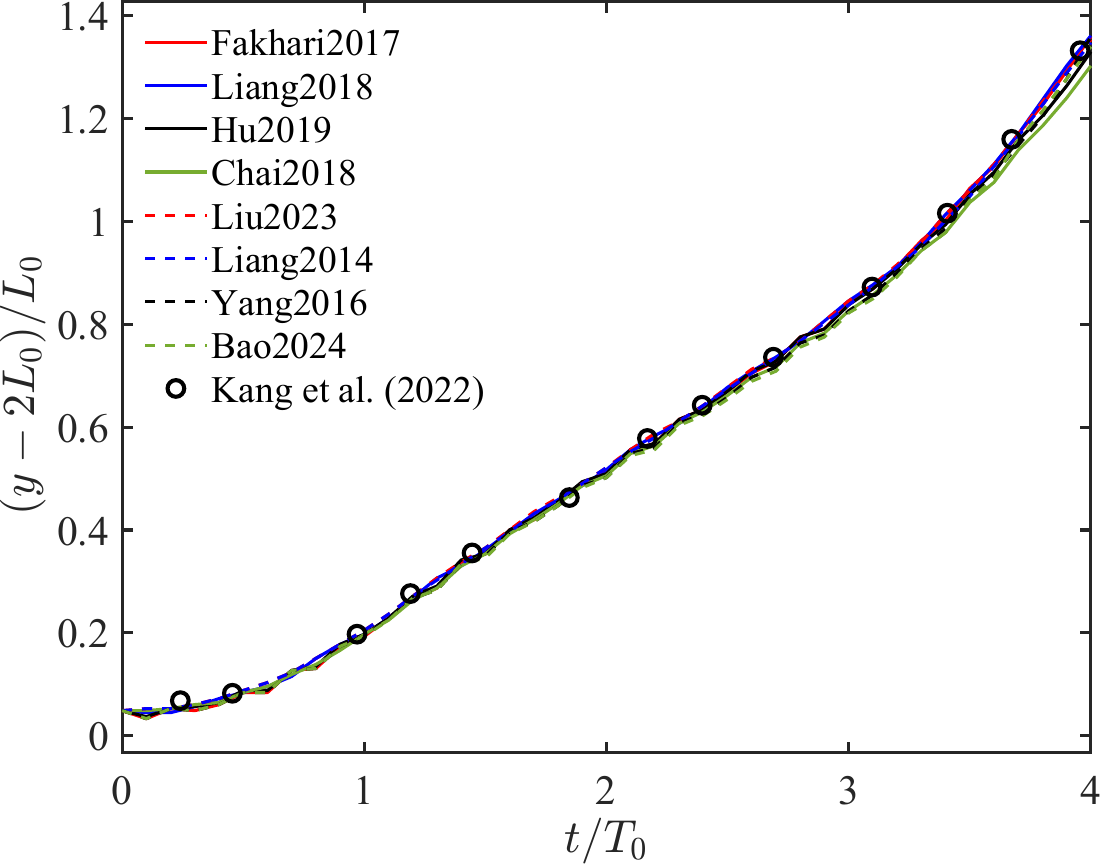}
        \caption{Bubble front}
        \label{fig:bubble_front}
    \end{subfigure}
    \hfill
    % 子图 (b) Spike tip
    \begin{subfigure}[b]{0.48\textwidth}
        \centering
        \includegraphics[width=\textwidth]{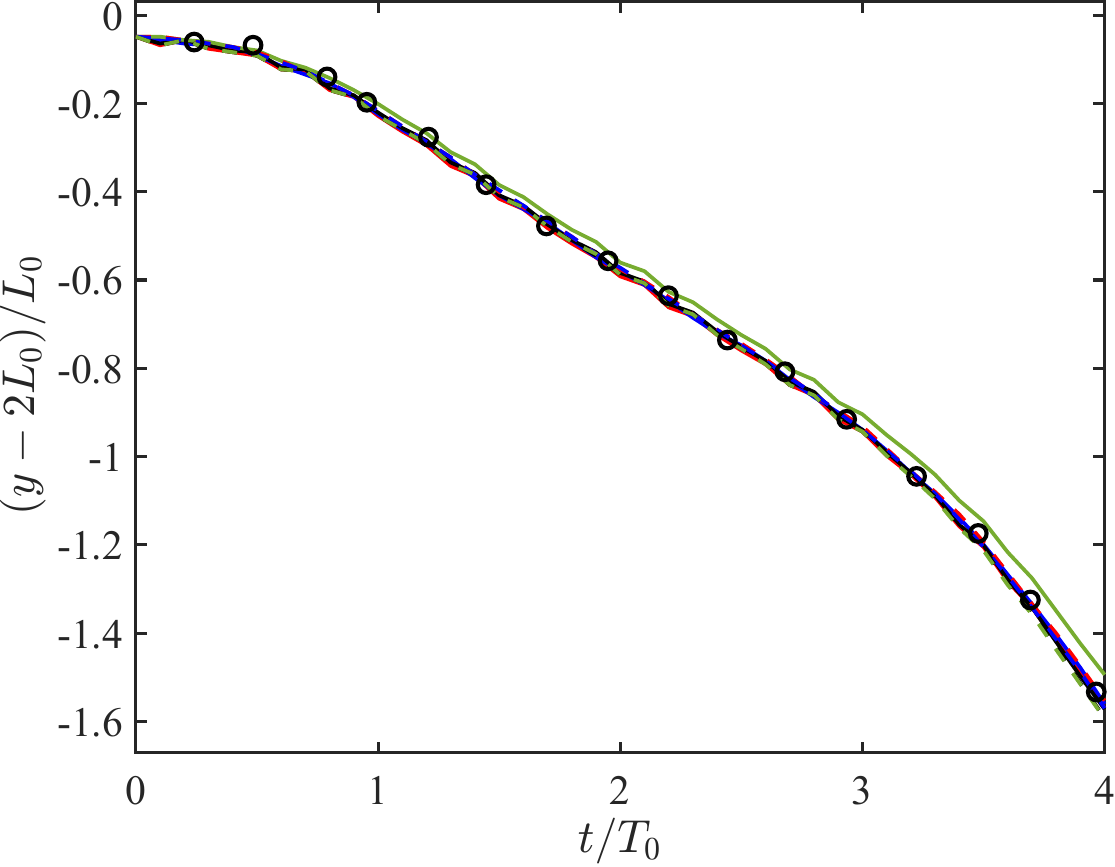}
        \caption{Spike tip}
        \label{fig:spike_tip}
    \end{subfigure}

\caption{Comparison with Kang et al.~\cite{kang2022local}: Dimensionless positions of (a) the bubble front and (b) the spike tip for different models as functions of non-dimensional time $t/T_0$.}
    \label{fig:bubble_spike_position}
\end{figure}

Figure~\ref{fig:bubble_spike_position} shows the time evolution of the dimensionless positions of the heavy-fluid front and spike tip, together with the benchmark results of Kang et al.~\cite{kang2022local}. All models yield consistent predictions, except for the nonlocal AC model (Chai2018), which exhibits slight deviations, consistent with previous observations~\cite{liu2023improved}. These results suggest that, although the models differ in their ability to resolve small-scale structures, they perform comparably in capturing the large-scale interface evolution.

\subsection{Two-phase Poiseuille flow}\label{subsec:Two-phase Poiseuille flow}

The two-phase stratified Poiseuille flow is a classical benchmark problem for assessing model accuracy. In earlier studies, PF-LB results were often compared with analytical solutions based on a sharp interface assumption, which inevitably introduced deviations due to the neglect of interface thickness. More recently, Lai et al.~\cite{lai2025analytical} derived analytical solutions that explicitly account for the diffuse-interface assumption, making them more appropriate for benchmarking diffuse-interface models. In this study, we therefore adopt these solutions as the reference.

The flow setup consists of two immiscible fluids stratified into horizontal layers, with the heavy fluid occupying the lower half-domain ($-H \le y \le 0$) and the light fluid occupying the upper half ($0 \le y \le H$). The flow is driven by a constant body force $\bm{G} = (G_x,0)$, where ${G_x} = {u_c}\left( {{\mu _H} + {\mu _L}} \right)/{H^2}$, $u_c$ is the velocity at the centerline. Periodic boundary conditions are applied in the streamwise direction, while no-slip conditions are imposed on the top and bottom walls.

The initial order parameter field is specified as
\begin{equation}
\phi (x,y,0) = \frac{\phi_H + \phi_L}{2} + \frac{\phi_H - \phi_L}{2} \, \tanh \frac{2(H - y)}{\xi}.
\end{equation}
The distribution of the order parameter remains unchanged under the applied driving force. To eliminate the influence of order-parameter fluctuations, its evolution is disabled in this case.

The simulation parameters are
$N_x = 10$,
$N_y = 100$,
$u_c = 0.001$,
$M_\phi = 0.1$,
$\xi = 4$,
$\rho_H / \rho_L = 1000$,
$\mu_H / \mu_L = 1000$,
and $\sigma = 0.001$.

At steady state, the analytical solution of the streamwise velocity reads
\begin{equation}
u_x = \frac{G_x}{\displaystyle \int_{-H}^{H} \frac{1}{\mu} \, dY} \cdot
\left\{
\int_{-H}^{H} \frac{1}{\mu} \, dY \int_{y}^{H} \frac{Y}{\mu} \, dY
+ \int_{-H}^{H} \frac{Y}{\mu} \, dY \int_{H}^{y} \frac{1}{\mu} \, dY
\right\},
\label{eq:analytical_solution}
\end{equation}
where the specific form depends on the viscosity interpolation across the interface. In line with the other benchmark cases, we test only the linear viscosity interpolation given in Eq.~(\ref{eq:mu_linear}).

\begin{figure}
    \centering
    \includegraphics[width=0.6\textwidth]{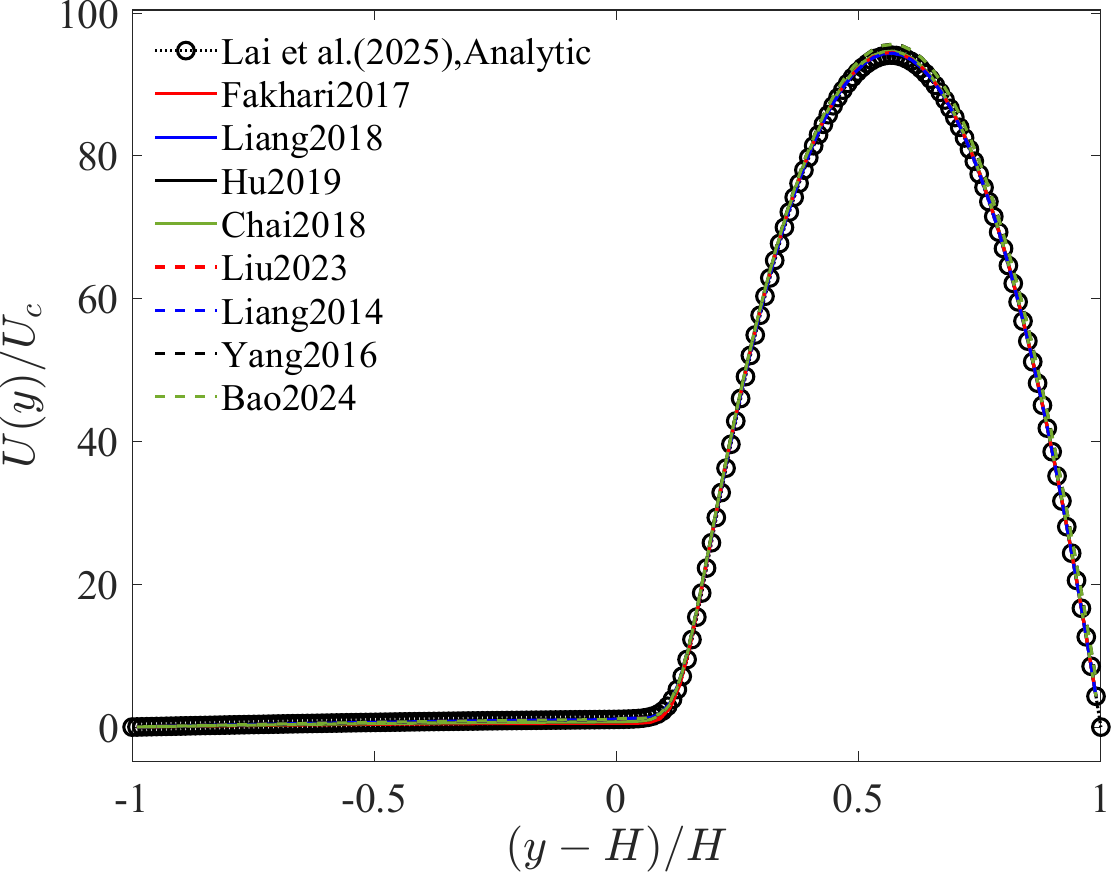}
    \caption{Velocity profile in the stratified Poiseuille flow.}
    \label{fig:poiseuille_velocity}
\end{figure}

Figure~\ref{fig:poiseuille_velocity} compares the velocity profiles in the stratified Poiseuille flow with the analytic solution derived by Lai et al.~\cite{lai2025analytical}. All models reproduce the analytical solution with satisfactory accuracy.

To quantify the model performance, we evaluate the $L_2$ error norm of the velocity field:
\begin{equation}
 \mathrm{L2\ Err}(u_x) =
\sqrt{
\frac{
\sum_y \left[ u_x(y,t) - u_x^a(y) \right]^2
}{
\sum_y \left[ u_x^a(y) \right]^2
}
},
\label{eq:ux-error}
\end{equation}
where $u_x^a$ is the analytical solution. The results, summarized in Table~\ref{tab:poiseuille_velocity}, show that the conservative AC model Liang2018 achieves the highest accuracy, followed by the incompressible CH model Liang2014, the hybrid AC model Liu2023, and the conservative AC model Fakhari2017. The hybrid AC model Hu2019 and the nonlocal AC model Chai2018 exhibit comparable performance. By contrast, the two quasi-incompressible CH models (Yang2016 and Bao2024) yield the largest errors. This suggests that while quasi-incompressible formulations enhance local mass conservation, they do so at the expense of reduced velocity accuracy in this stratified flow benchmark.

\begin{table}
    \centering
    \caption{Relative errors of velocity in stratified Poiseuille flow for different models}
    \label{tab:poiseuille_velocity}
  %  \resizebox{\textwidth}{!}{%
    \begin{tabular}{c c c c c c c c c}
        \hline
         & Fakhari2017 & Liang2018 & Hu2019 & Chai2018 & Liu2023 & Liang2014 & Yang2016 & Bao2024 \\
        \hline
         $\mathrm{L2\ Err}(u_x)$ & 0.008374 & 0.005284 & 0.009702 & 0.009702 & 0.006506 & 0.005772 & 0.023655 & 0.023323  \\
        \hline
    \end{tabular}%
%    }
\end{table}

\subsection{Two stationary bubbles with different radii}\label{subsec:Two stationary bubbles with different radii}

The final two-dimensional test involves two static bubbles of unequal radii coexisting in a periodic domain. As discussed earlier, the nonlocal AC model exhibits coarsening; that is, in the presence of two bubbles of different sizes, the smaller bubble gradually shrinks while its mass transfers to the larger one~\cite{chai2018comparative}. In contrast, the CH model suffers from spontaneous dissolution, where bubbles below a critical radius disappear into the surrounding fluid, even in single-bubble cases~\cite{zhang2010cahn,liu2003phase}. This test is therefore used to evaluate the models’ ability to preserve bubble volume.

The initial order parameter is prescribed as
\begin{equation}
\begin{split}
\phi (x,y,0) = \frac{\phi_H + \phi_L}{2}
- \frac{\phi_H - \phi_L}{2} \,
\left(
\tanh \left\{\frac{2\left[\sqrt{(x - x_1)^2 + (y - y_1)^2} - R_1\right]}{\xi}\right\}\right.\\
\left.+ \tanh \left\{\frac{2\left[\sqrt{(x - x_2)^2 + (y - y_2)^2} - R_2\right]}{\xi}\right\}
- 1
\right),
\end{split}
\end{equation}
where $\left( x_1,y_1 \right) = (50,50)$ and $\left( x_2,y_2 \right) = (120,120)$ are the bubble centers. The remaining parameters are set to ${L_0} = 200$, ${\rho _H} = 1.0$, ${\rho _L} = 0.1$, $\sigma  = 0.001$, ${M_\phi } = 0.1$, $\xi  = 3$, ${R_1} = 10$, and ${R_2} = 25$.

\begin{figure}
    \centering
    \includegraphics[width=0.8\textwidth]{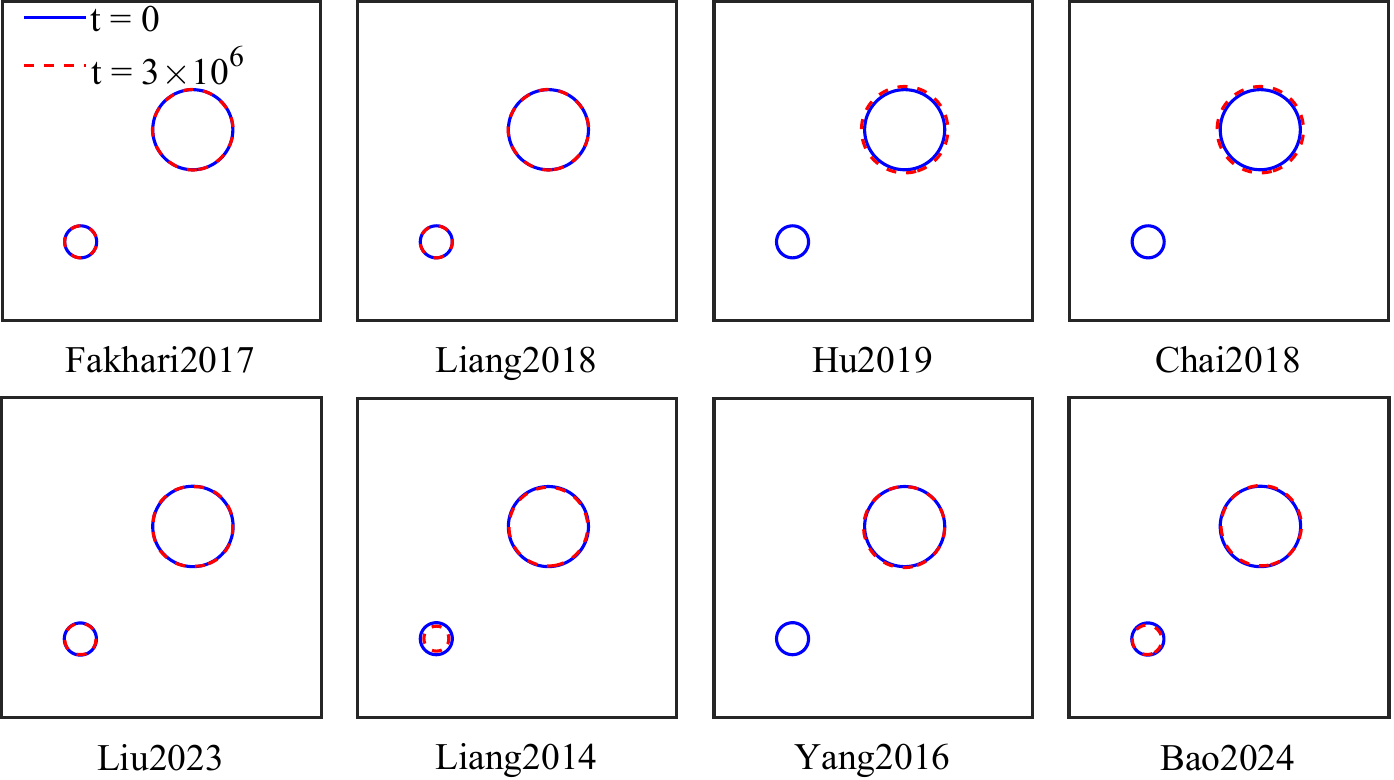}
    \caption{Comparison of order parameter contours for stationary bubbles with different radii at different simulation time. }
    \label{fig:stationary_bubble_contours}
\end{figure}

Figure~\ref{fig:stationary_bubble_contours} shows the interface contours ($\phi=0$) after $3\times 10^6$ time steps. With the nonlocal AC model of Chai2018, the small bubble vanishes completely, accompanied by a substantial increase in the large bubble volume—an unambiguous coarsening phenomenon. The hybrid AC model of Hu2019, which applies a globally constant weight, also inherits this behavior, leading to full disappearance of the small bubble. In contrast, the hybrid AC model of Liu2023, which employs a spatially varying weight, effectively suppresses coarsening and preserves the small bubble throughout the simulation, in agreement with Liu et al.~\cite{liu2023improved}. The two conservative AC models, Fakhari2017 and Liang2018, also preserve the small bubble volume well, further demonstrating their robustness in this regard.

The three CH models exhibit different outcomes. In Yang2016, the small bubble disappears completely, consistent with the observation by Bao et al.~\cite{bao2024phase}, but without a corresponding increase in the large bubble volume, confirming that this effect arises from spontaneous dissolution rather than coarsening. For CH models, the critical radius for spontaneous dissolution of a two-dimensional bubble is given by~\cite{yue2007spontaneous}
\begin{equation}
r_c = \left( \frac{\sqrt{3}}{16\pi}\,\xi V \right)^{\tfrac{1}{3}},
\label{eq:critical_radius}
\end{equation}
where $\xi$ is the interface thickness and $V$ the domain volume. With $\xi = 3$ and $V = 0.5L_0^2$ (since each bubble corresponds to half of the domain), the critical radius is $r_c = 12.74$. The small bubble ($R_1 = 10$) is thus below this threshold and expected to dissolve spontaneously. The incompressible CH model of Liang2014 and the singular-mobility CH model of Bao2014 both retain the small bubble, but visible shrinkage is still observed.

To quantify the volume evolution, the bubble volume is defined as
\begin{equation}
V(t) = \int_{\Omega\in\Omega_k} \frac{\phi(\bm{x},t)-\phi_L} {\phi_H-\phi_L}\, d\Omega \approx \sum_{i,j\in\Omega_k} \frac{\phi_{i,j}(t)-\phi_L}{\phi_H-\phi_L}\, \delta x^2 ,
\label{eq:volume_continuous}
\end{equation}
where $\Omega_k$ is the region of the $k$th bubble, and $i$ and $j$ indicate the node indices. We denote $V_S(t)$ and $V_L(t)$ for the small and large bubbles, normalized by their initial values.

\begin{figure}[htbp]
    \centering

    % 子图 (a) 小气泡体积随时间变化
    \begin{subfigure}[b]{0.48\textwidth}
        \centering
        \includegraphics[width=\textwidth]{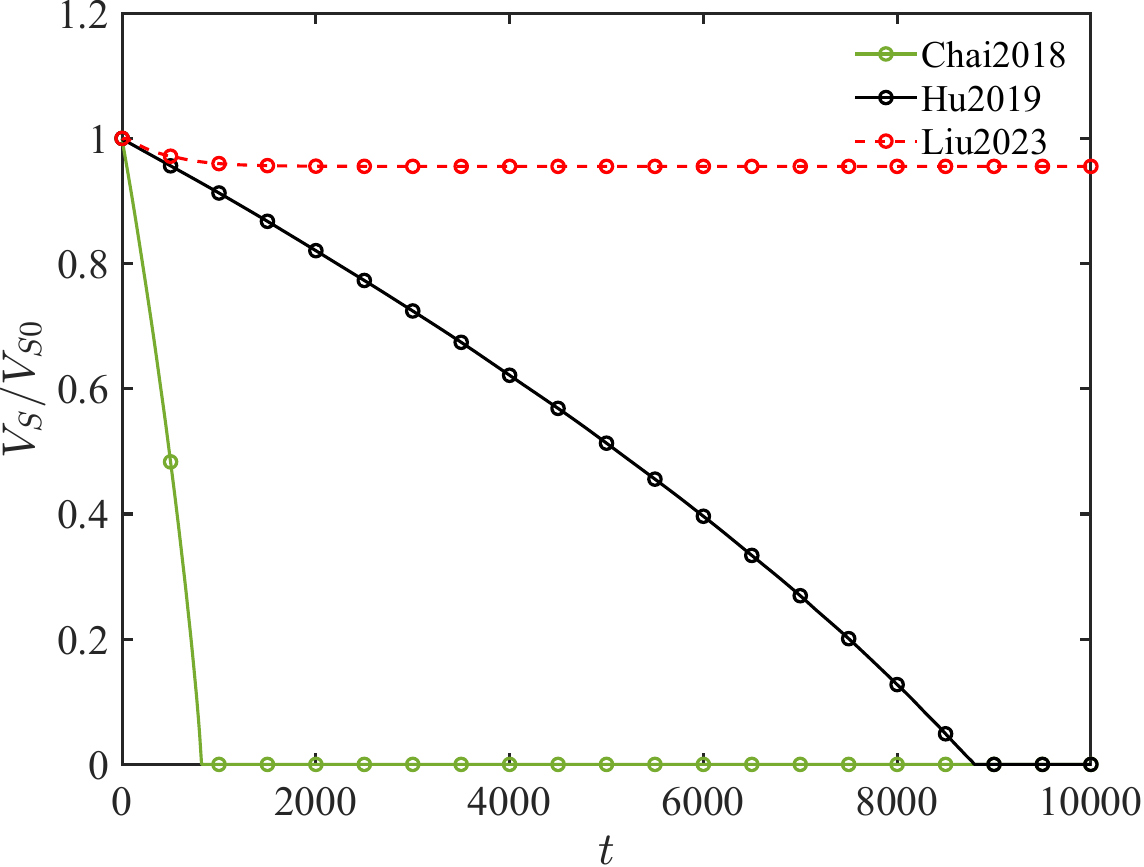}
        \caption{Small bubble volume over time}
        \label{fig:small_bubble_volume}
    \end{subfigure}
    \hfill
    % 子图 (b) 大气泡体积随时间变化
    \begin{subfigure}[b]{0.48\textwidth}
        \centering
        \includegraphics[width=\textwidth]{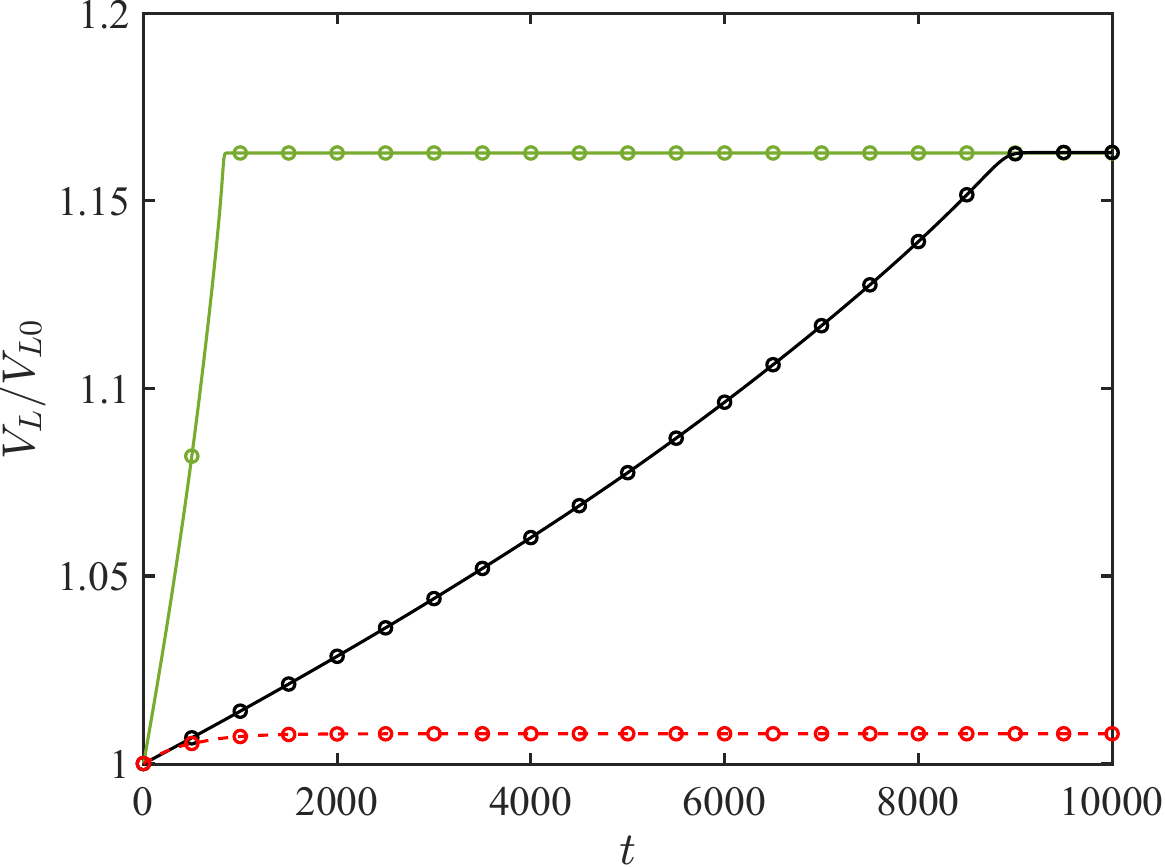}
        \caption{Large bubble volume over time}
        \label{fig:large_bubble_volume}
    \end{subfigure}

    \caption{Temporal evolution of small and large bubble volumes for different models.}
    \label{fig:bubble_volume_comparison_nonlocal}
\end{figure}

In Figure~\ref{fig:bubble_volume_comparison_nonlocal}, we present the time evolution of the bubble volumes for the nonlocal AC model (Chai2018) and the two hybrid AC models. In the nonlocal AC case, the small bubble vanishes rapidly (within $10^3$ time steps), accompanied by a corresponding increase in the large bubble volume, demonstrating a strong coarsening effect. The hybrid AC model of Hu2019, which employs a global weighting factor, reduces the influence of the nonlocal term and significantly delays the disappearance of the small bubble ($t = 820$ vs.\ $t = 8810$). However, it still fails to prevent the eventual loss of the small bubble. In contrast, the hybrid AC model of Liu2023 with a spatially varying weighting factor experiences some initial coarsening and associated volume errors but ultimately stabilizes the sizes of both bubbles at later times.

\begin{figure}[htbp]
    \centering

    % 子图 (a) 小气泡体积随时间变化
    \begin{subfigure}[b]{0.48\textwidth}
        \centering
        \includegraphics[width=\textwidth]{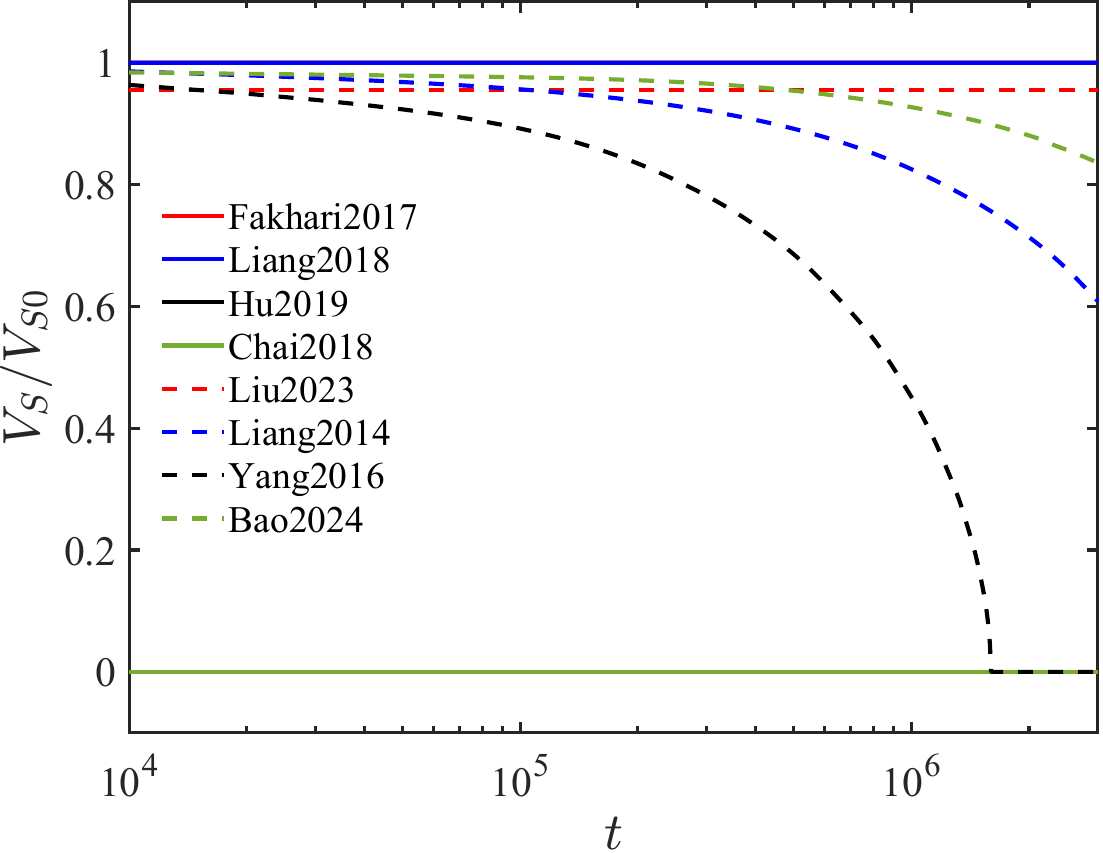}
        \caption{Small bubble volume over time}
        \label{fig:small_bubble_volume}
    \end{subfigure}
    \hfill
    % 子图 (b) 大气泡体积随时间变化
    \begin{subfigure}[b]{0.48\textwidth}
        \centering
        \includegraphics[width=\textwidth]{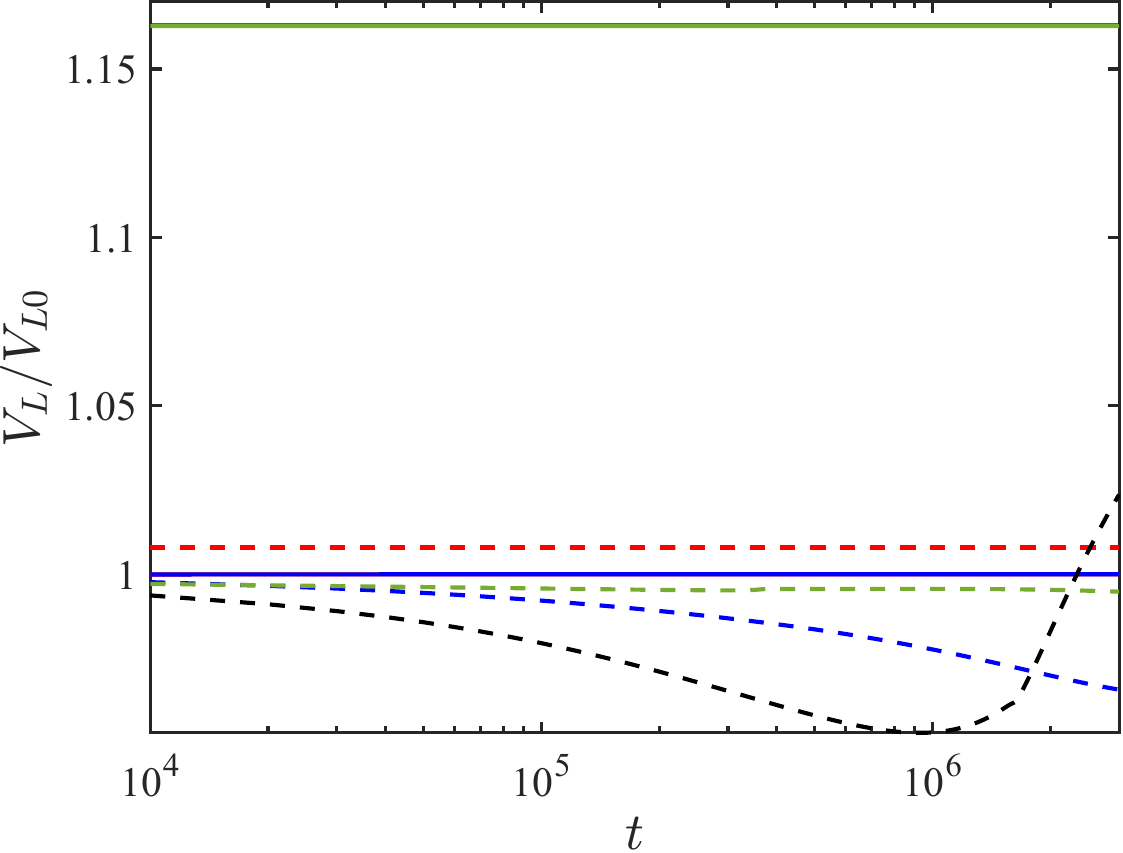}
        \caption{Large bubble volume over time}
        \label{fig:large_bubble_volume}
    \end{subfigure}

    \caption{Temporal evolution of bubble volumes for stationary bubbles. The result of Fakhari2017 and Liang2018 collapses fully, so are the results of Chai2018 and Hu2019.}
    \label{fig:bubble_volume_comparison}
\end{figure}

Figure~\ref{fig:bubble_volume_comparison} extends this analysis to later times. All three CH models show persistent dissolution of the small bubble, albeit at different rates, implying that complete disappearance would eventually occur on longer time scales. The singular-mobility model of Bao2024 significantly slows dissolution but does not eliminate it. In contrast, the hybrid AC model with spatially varying weight stabilizes the small bubble volume after an initial coarsening phase. Most notably, the two conservative AC models exhibit neither coarsening nor spontaneous dissolution, underscoring their superiority in preserving bubble volume.

\section{Three-dimensional turbulent flow tests}\label{sec:turbulence}

In the previous section, the PF LB models were tested in a series of two-dimensional laminar flow cases. While these tests allow for some preliminary conclusions, they do not fully challenge the models or reveal their limitations. This limitation also exists in most prior studies, where only a few cases were extended to three dimensions, leaving aside the more demanding turbulent scenarios. For instance, the conservative AC models, which perfectly preserved the droplet volume in the two-dimensional static droplet test, may still encounter difficulties in more complex flow environments. To address this gap, we now investigate the performance of different models in three-dimensional homogeneous isotropic turbulence with a droplet. To save computational resources, only a subset of models is included based on the results from the previous section.

Specifically, six out of the eight models examined in the last section are selected to represent the main categories of PF LB approaches: the conservative AC models of Fakhari2017 and Liang2018, the nonlocal AC model of Chai2018, the hybrid AC model of Liu2023, the incompressible CH model of Liang2014, and the quasi-incompressible CH model of Bao2024. Both conservative AC models are included because Fakhari2017 recovers the non-conservative form of the NS equations, whereas Liang2018 recovers the conservative form. Although these two models exhibited nearly identical performance in two-dimensional tests, they may display discernible differences in more complex three-dimensional turbulent cases. The hybrid AC model of Hu2019 is excluded because it generally performs worse than Liu2023; and the quasi-incompressible CH model of Yang2016 is excluded as an earlier version of Bao2024 with overall inferior performance.

The test case concerns the dynamic evolution of a large droplet immersed in sustained homogeneous isotropic turbulence. Prior to droplet insertion, statistically stationary turbulence was generated in the single-phase system using stochastic forcing, following the method described in our recent work~\cite{peng2023parameterization}. The turbulence intensity of the statistically stationary state is characterized by the Taylor-scale Reynolds number, ${\rm Re}_\lambda = u_{\rm rms}\lambda_T/\nu = 42.28$, where $u_{\rm rms}$ is the root-mean-square velocity and $\lambda_T$ is the Taylor microscale. Unless otherwise specified, the characteristic velocity and length used in defining the Weber number are $u_{\rm rms}$ and the initial droplet diameter, respectively: ${\rm We} = \rho_H u_{\rm rms}^2(2R)/\sigma$.

The droplet is initially placed at the center of the computational domain. Its order parameter $\phi$ is initialized using a three-dimensional hyperbolic tangent profile:
\begin{equation}
\phi(x,y,z) = \frac{\phi_H + \phi_L}{2}
+ \frac{\phi_H - \phi_L}{2} \,
\tanh \left\{\frac{2 \Big[ \sqrt{(x - 0.5 L)^2 + (y - 0.5 L)^2 + (z - 0.5 L)^2} - R \Big]}{\xi}\right\},
\label{eq:init-phi-droplet-3D}
\end{equation}
where the initial droplet radius is $R = 30$ lattice units and the cubic domain size is $L = 256$. Periodic boundary conditions are applied in all three directions. The initial pressure field is uniformly set to zero, and stochastic forcing is continuously applied after droplet insertion to sustain turbulence.

The NS equations are solved using the D3Q27 lattice with a multiple-relaxation-time (MRT) collision operator for enhanced stability\cite{peng2017lattice}, while the PF equations (AC or CH) are solved using a D3Q7 lattice\cite{liu2023improved}. In both lattice models, the speed of sound is set to $c_s^2 = 1/3$. The performance of the five selected models is systematically assessed under varying density ratios, Weber numbers, and interface thicknesses. The effects of the mobility coefficient and Reynolds number were also examined, but as the main observations remain consistent, these results are omitted here for conciseness. The evaluation focuses on three key aspects: numerical stability, droplet volume conservation, and the boundedness of the order parameter (numerical dispersion).

\subsection{Effect of density ratio}\label{subsubsection:DensityRatio}

As a first step, we examined the effect of density ratio by considering
$\rho^{*} = 2, 10, 50,$ and $100$, while keeping the remaining parameters fixed:
$\nu_H = \nu_L = 0.006$ and $We = 1.2$. The light-phase density was set to $\rho_L = 1.0$, and the heavy-phase density $\rho_H$ was determined from
$\rho^{*} = \rho_H / \rho_L$. To ensure stability at high density ratios, the interface thickness was set to $\xi = 6$. For a fair comparison, the mobility coefficient was fixed at $M_\phi = 0.01$ across all cases.

\begin{figure}
    \centering

    % 子图 (a) ρ*=2
    \begin{subfigure}[b]{0.48\textwidth}
        \centering
        \includegraphics[width=\textwidth]{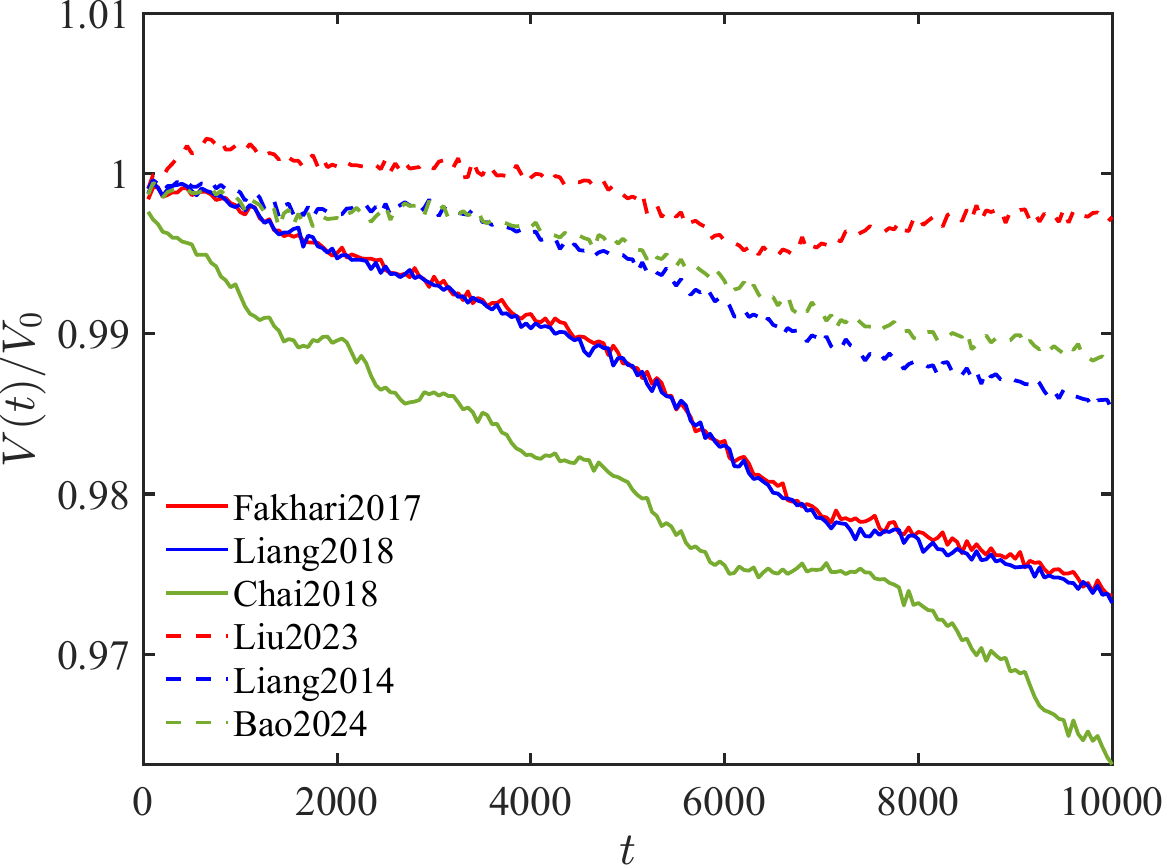}
        \caption{$\rho^*=2$}
        \label{fig:Droplet_volume_density2}
    \end{subfigure}
    \hfill
    % 子图 (b) ρ*=10
    \begin{subfigure}[b]{0.48\textwidth}
        \centering
        \includegraphics[width=\textwidth]{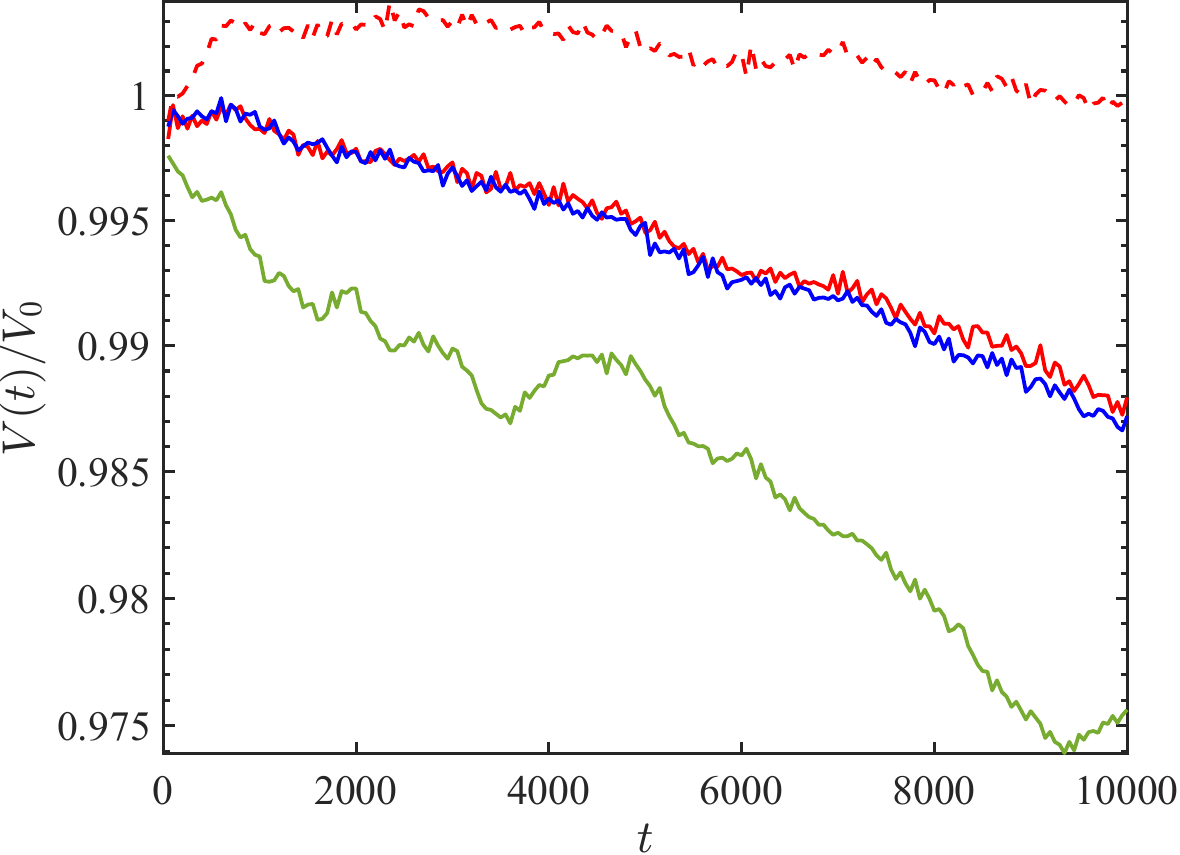}
        \caption{$\rho^*=10$}
        \label{fig:Droplet_volume_density10}
    \end{subfigure}

    \vspace{1em} % 调整上下行间距

    % 子图 (c) ρ*=50
    \begin{subfigure}[b]{0.48\textwidth}
        \centering
        \includegraphics[width=\textwidth]{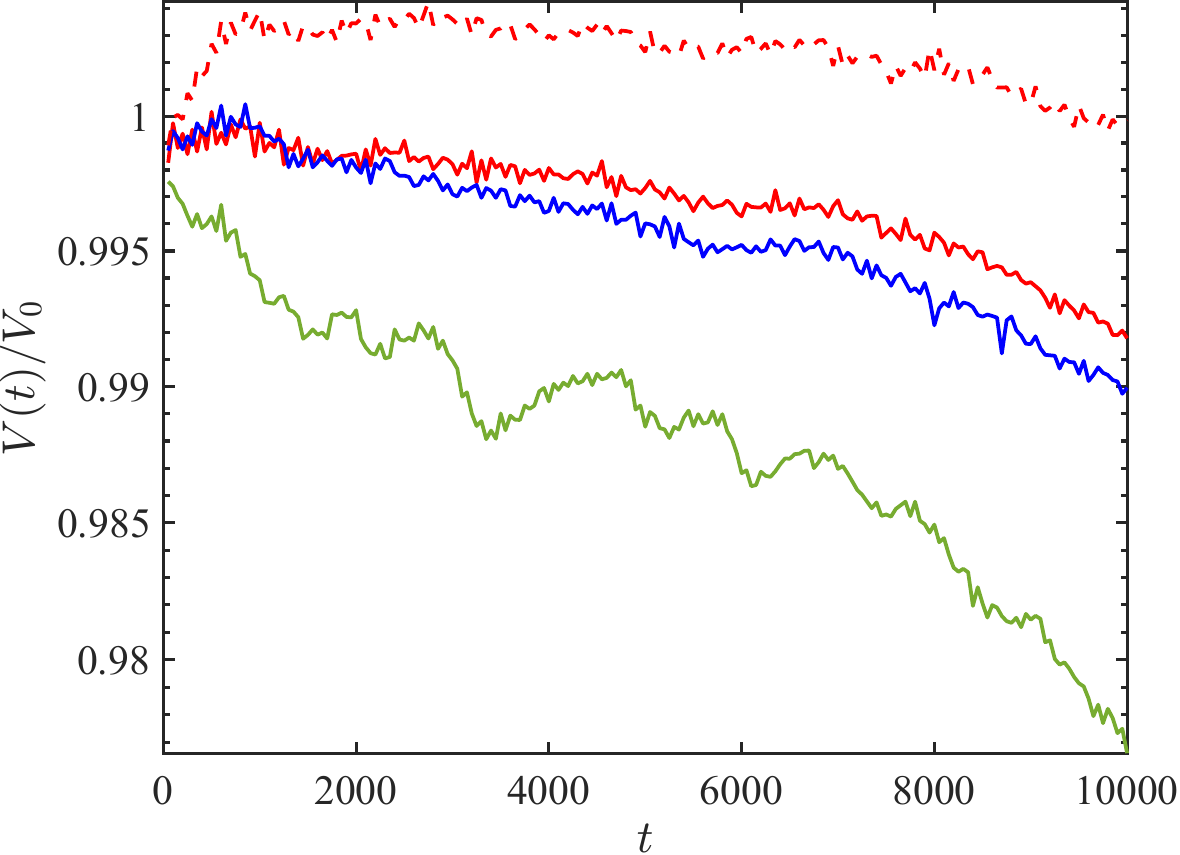}
        \caption{$\rho^*=50$}
        \label{fig:Droplet_volume_density50}
    \end{subfigure}
    \hfill
    % 子图 (d) ρ*=100
    \begin{subfigure}[b]{0.48\textwidth}
        \centering
        \includegraphics[width=\textwidth]{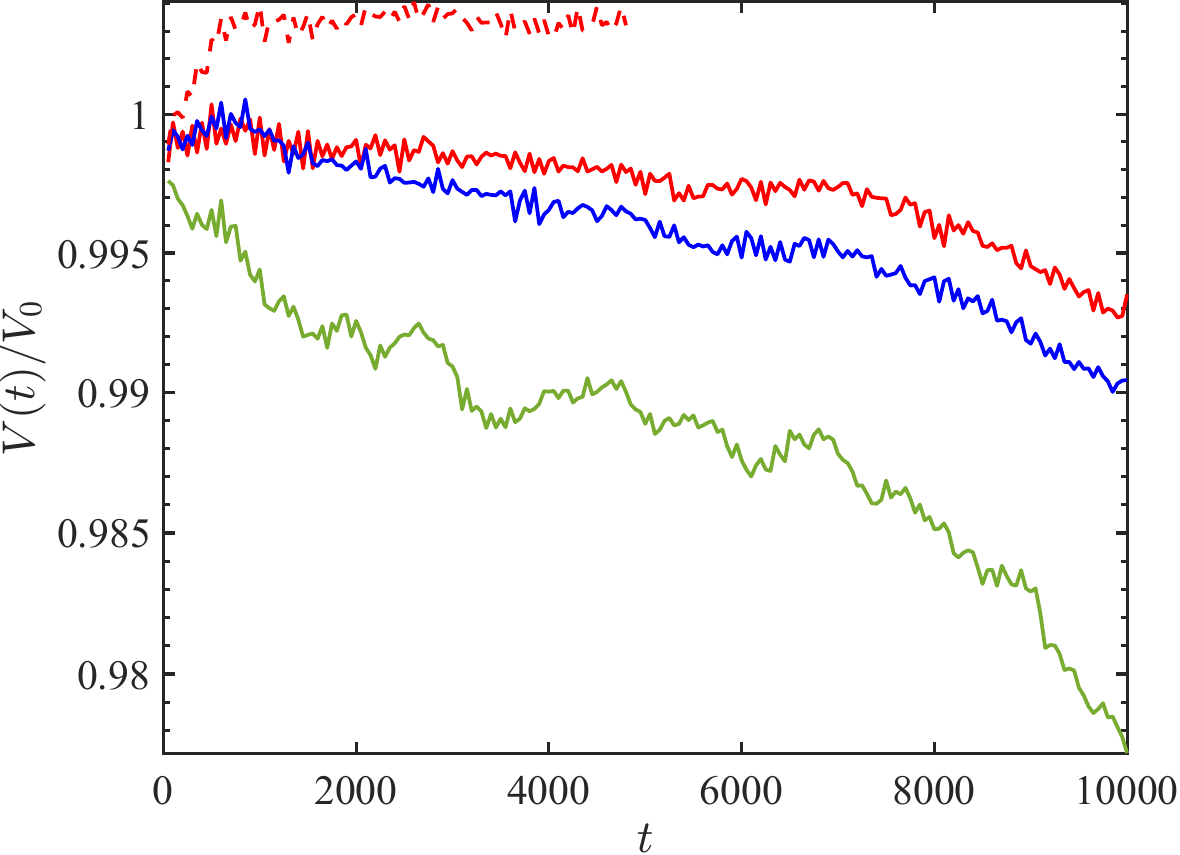}
        \caption{$\rho^*=100$}
        \label{fig:Droplet_volume_density100}
    \end{subfigure}

    \caption{Droplet volume evolution for different density ratios.}
    \label{fig:Droplet_volume_density}
\end{figure}

At low Weber numbers, the droplet is not expected to break up but only to deform under turbulent shear. Figure~\ref{fig:Droplet_volume_density} presents the time evolution of the total droplet volume, computed by counting grid points with $\phi > \phi_{\rm ave} = (\phi_H + \phi_L)/2$ and multiplying by the cell volume $\Delta x^3$. All models preserve the droplet volume reasonably well; however, except for the hybrid AC model of Liu2023, each model exhibits a certain degree of volume loss, revealing limitations in volume conservation. This issue has been widely reported in previous phase-field simulations of droplet-laden turbulent flows~\cite{elghobashi2019direct}.

The CH models (Liang2014 and Bao2024) exhibit less severe volume loss than the conservative AC models (Fakhari2017 and Liang2018), whereas the nonlocal AC model (Chai2018) performs the worst in this respect. However, the CH models lose numerical stability once the density ratio increases to 10. At even higher ratios ($\rho^* = 100$), the hybrid AC model (Liu2023) also becomes unstable. In contrast, the two conservative AC models and the nonlocal AC model remain comparatively more stable, although they too eventually fail at sufficiently high density ratios. The two conservative AC models yield nearly identical volume losses at low density ratios. As the density ratio increases, however, Fakhari2017—which recovers the non-conservative form of the NS equations—slightly outperforms Liang2018, which recovers the conservative form.

\begin{figure}
    \centering

    % 子图 (a) ρ*=2
    \begin{subfigure}[b]{0.48\textwidth}
        \centering
        \includegraphics[width=\textwidth]{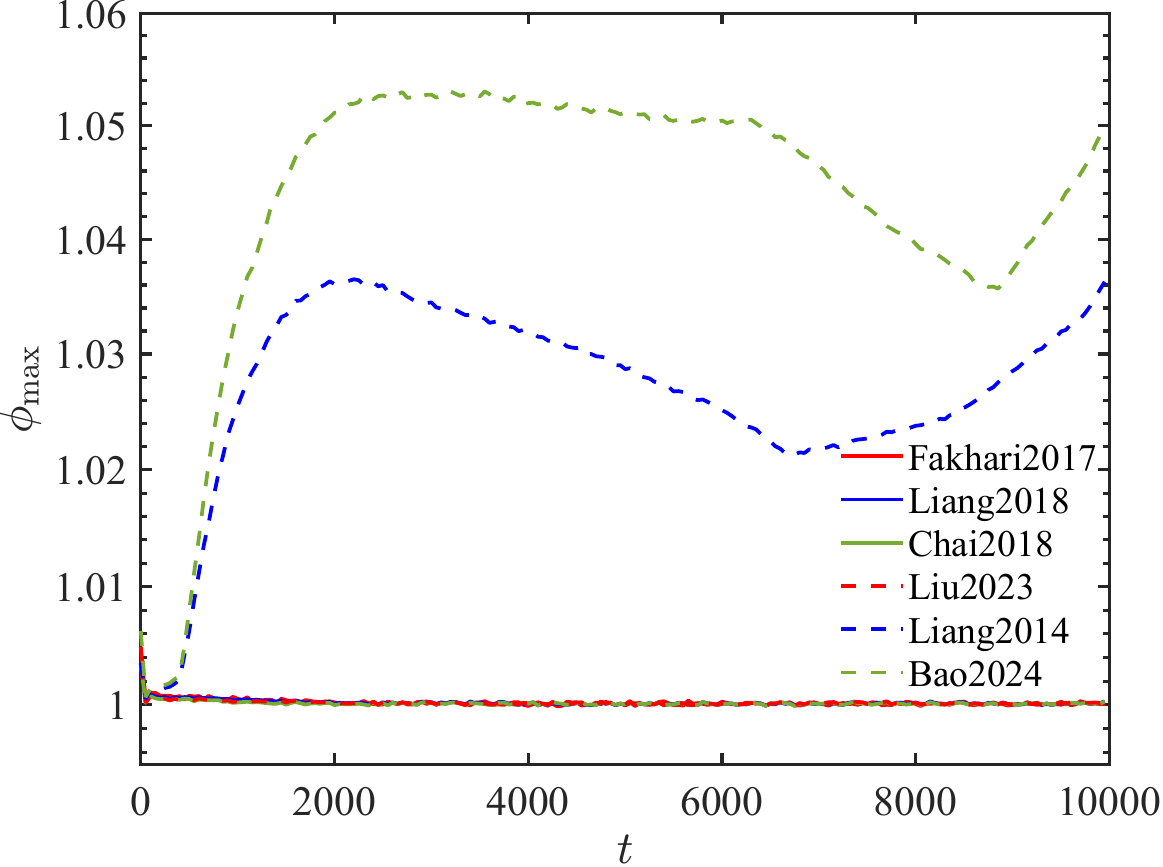}
        \caption{$\rho^*=2$}
        \label{fig:max_phi_density2}
    \end{subfigure}
    \hfill
    % 子图 (b) ρ*=10
    \begin{subfigure}[b]{0.48\textwidth}
        \centering
        \includegraphics[width=\textwidth]{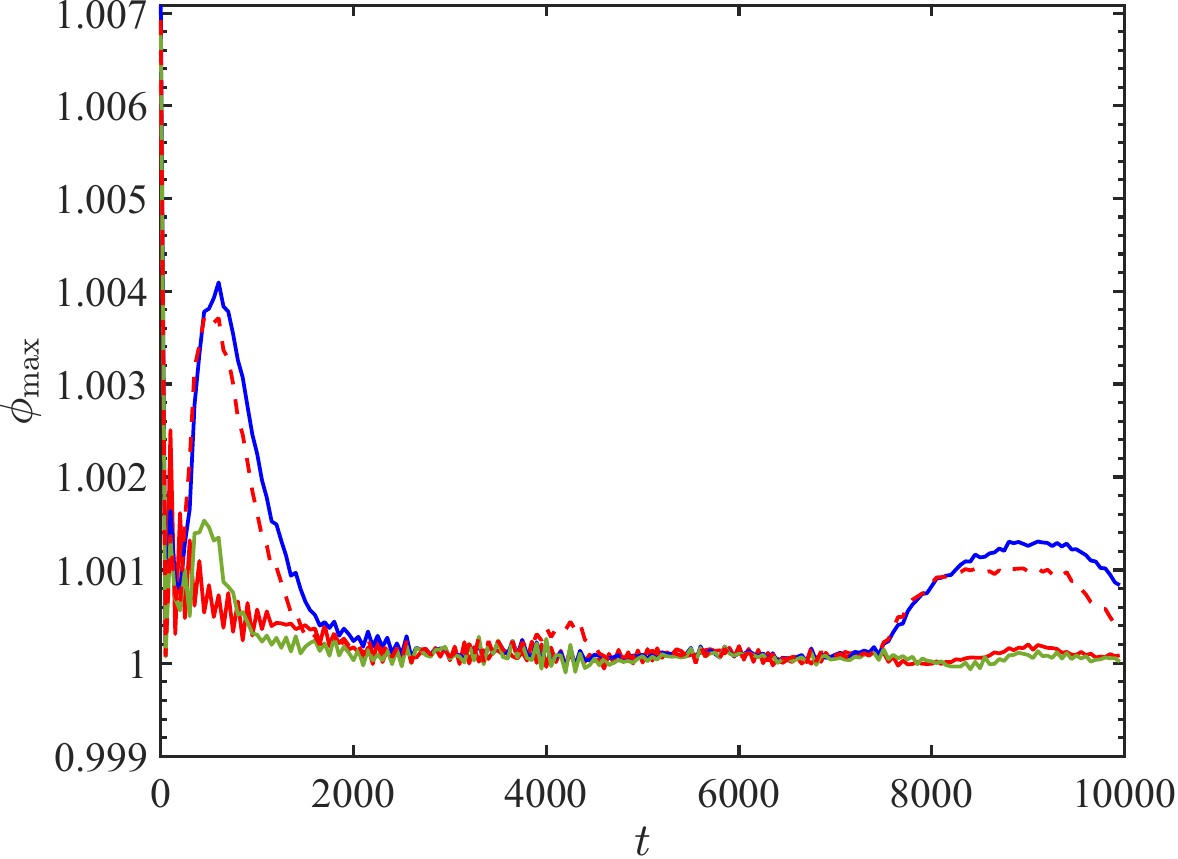}
        \caption{$\rho^*=10$}
        \label{fig:max_phi_density10}
    \end{subfigure}

    \vspace{1em} % 调整上下行间距

    % 子图 (c) ρ*=50
    \begin{subfigure}[b]{0.48\textwidth}
        \centering
        \includegraphics[width=\textwidth]{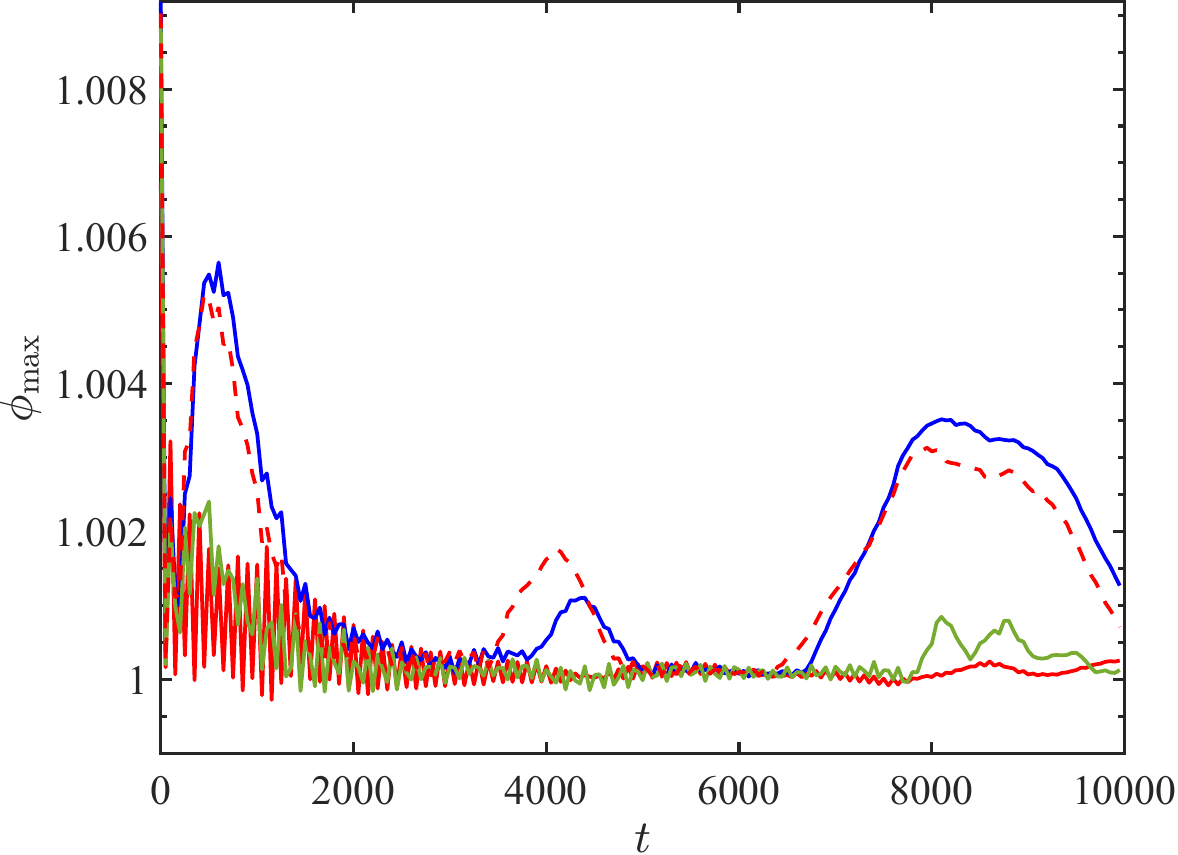}
        \caption{$\rho^*=50$}
        \label{fig:max_phi_density50}
    \end{subfigure}
    \hfill
    % 子图 (d) ρ*=100
    \begin{subfigure}[b]{0.48\textwidth}
        \centering
        \includegraphics[width=\textwidth]{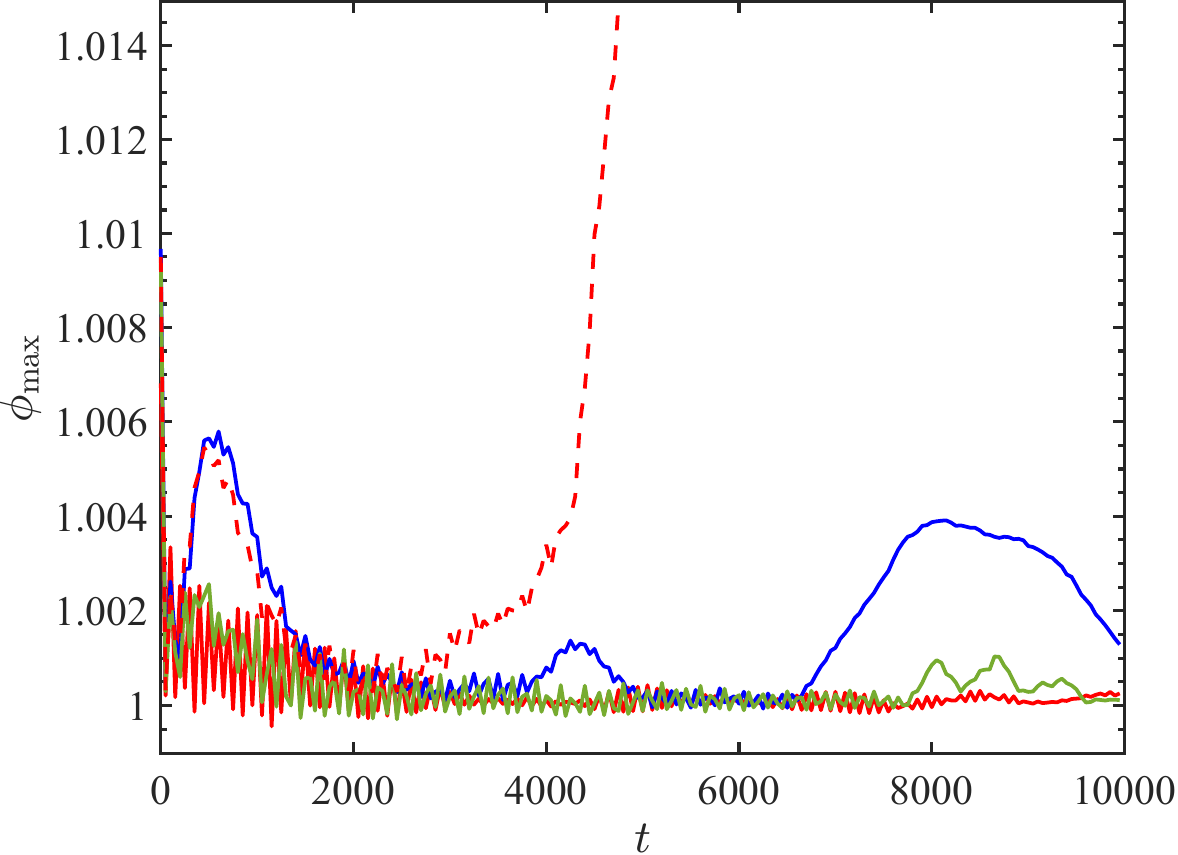}
        \caption{$\rho^*=100$}
        \label{fig:max_phi_density100}
    \end{subfigure}

    \caption{Evolution of the maximum order parameter for different density ratios.}
    \label{fig:max_phi_density}
\end{figure}

\begin{figure}
    \centering

    % 子图 (a) ρ*=2
    \begin{subfigure}[b]{0.48\textwidth}
        \centering
        \includegraphics[width=\textwidth]{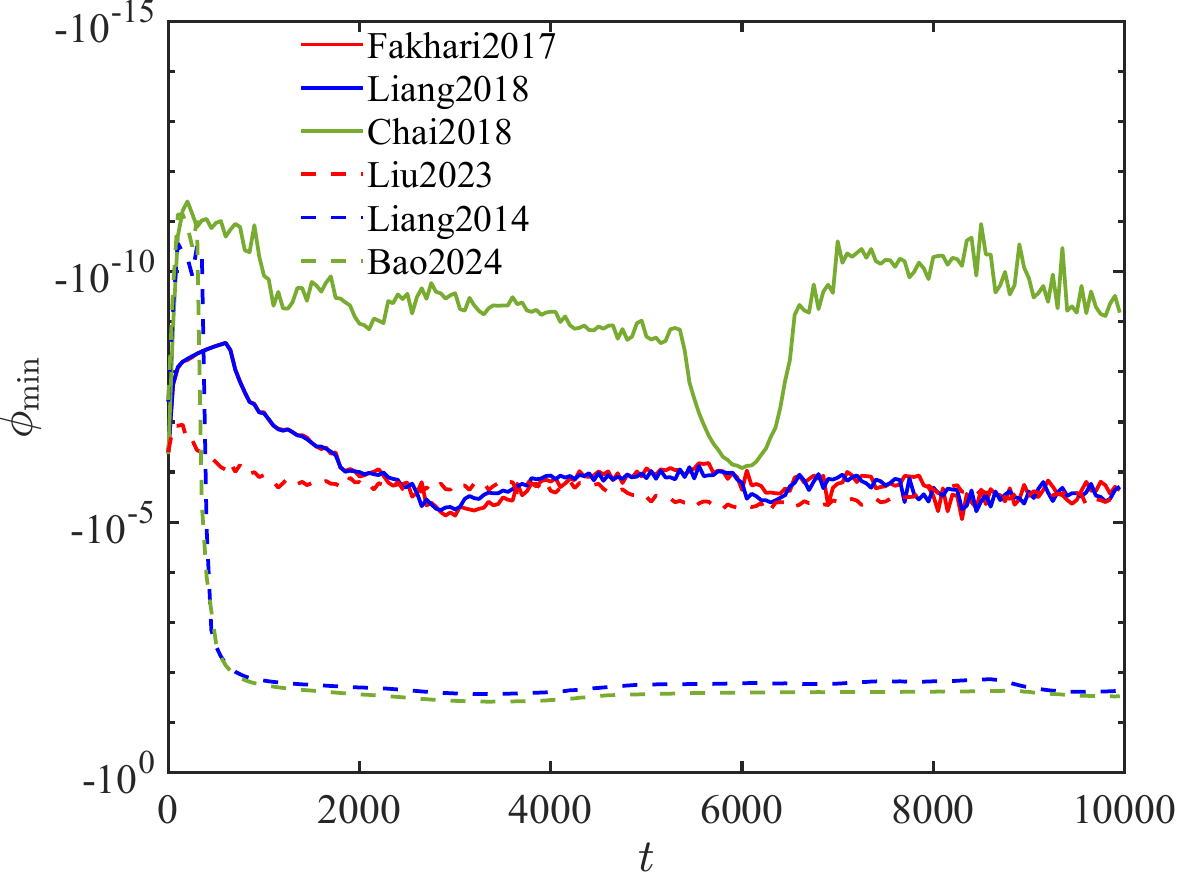}
        \caption{$\rho^*=2$}
        \label{fig:min_phi_density2}
    \end{subfigure}
    \hfill
    % 子图 (b) ρ*=10
    \begin{subfigure}[b]{0.48\textwidth}
        \centering
        \includegraphics[width=\textwidth]{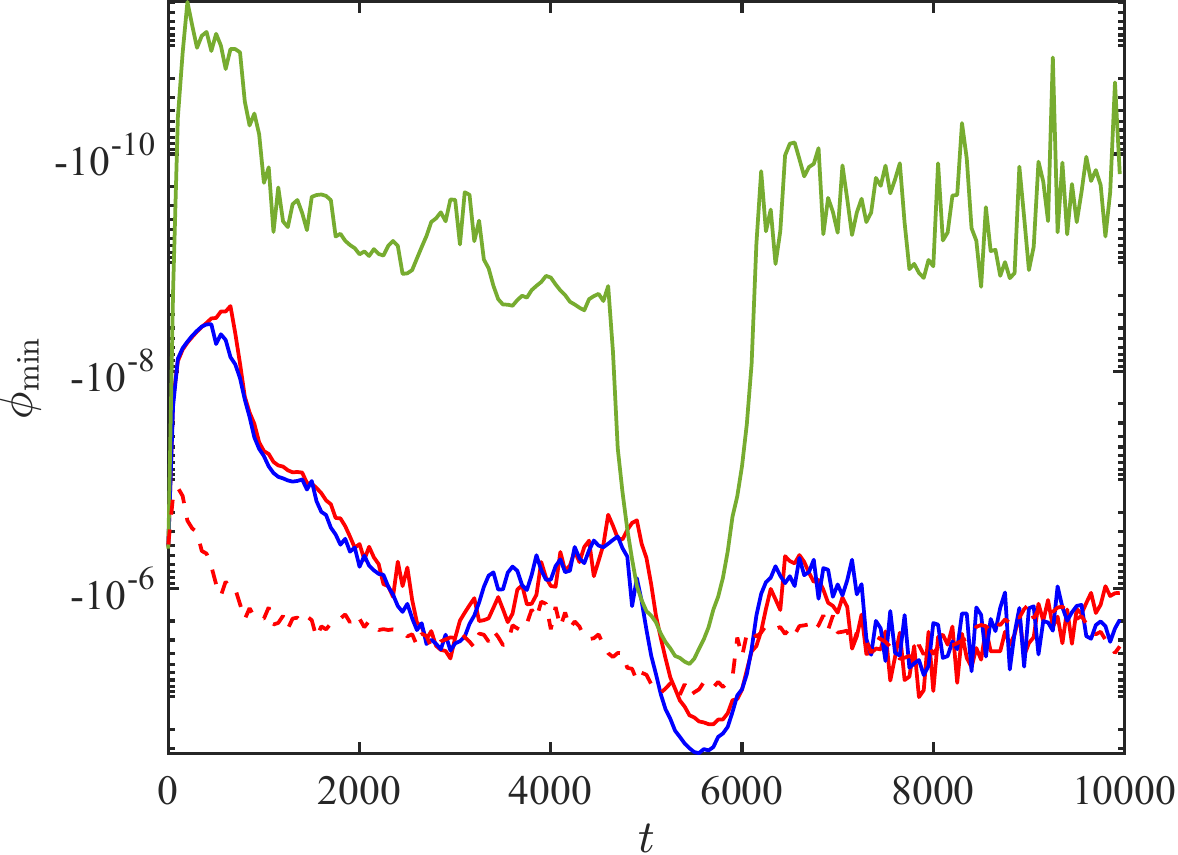}
        \caption{$\rho^*=10$}
        \label{fig:min_phi_density10}
    \end{subfigure}

    \vspace{1em} % 调整上下行间距

    % 子图 (c) ρ*=50
    \begin{subfigure}[b]{0.48\textwidth}
        \centering
        \includegraphics[width=\textwidth]{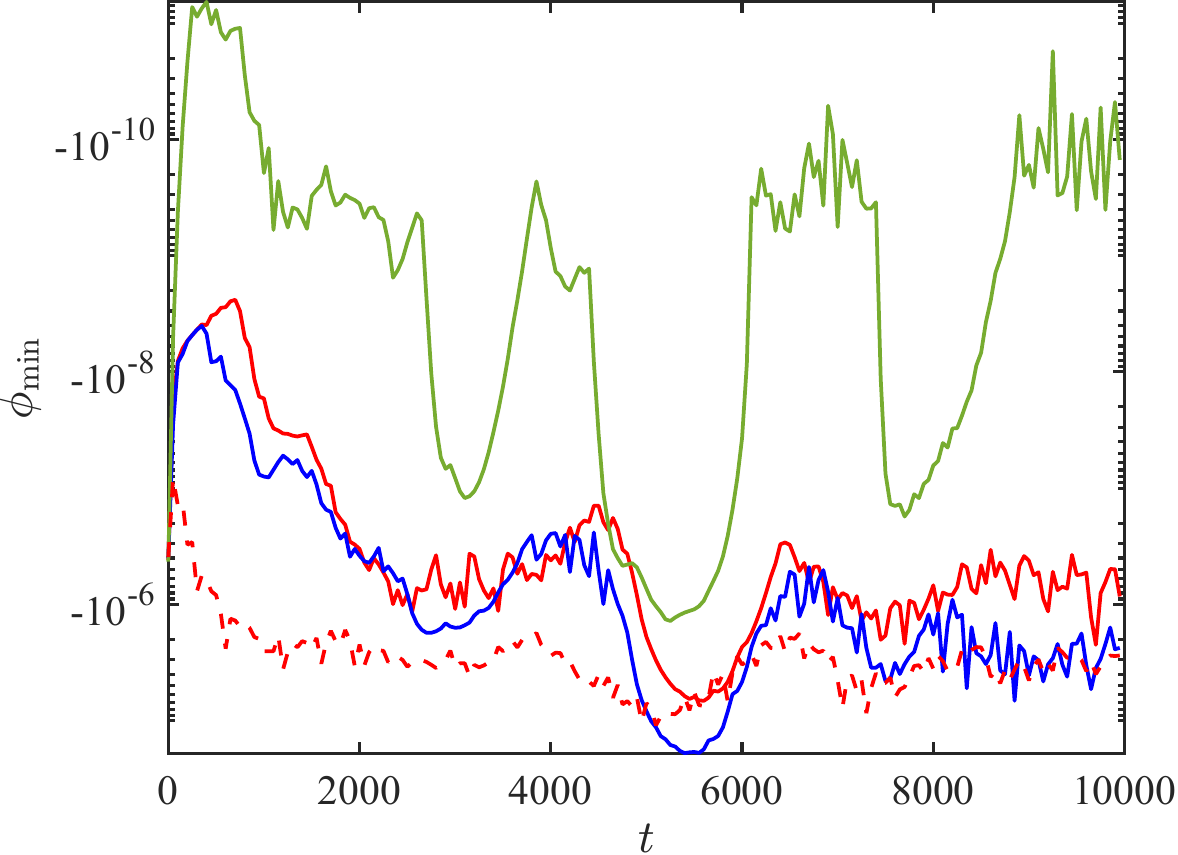}
        \caption{$\rho^*=50$}
        \label{fig:min_phi_density50}
    \end{subfigure}
    \hfill
    % 子图 (d) ρ*=100
    \begin{subfigure}[b]{0.48\textwidth}
        \centering
        \includegraphics[width=\textwidth]{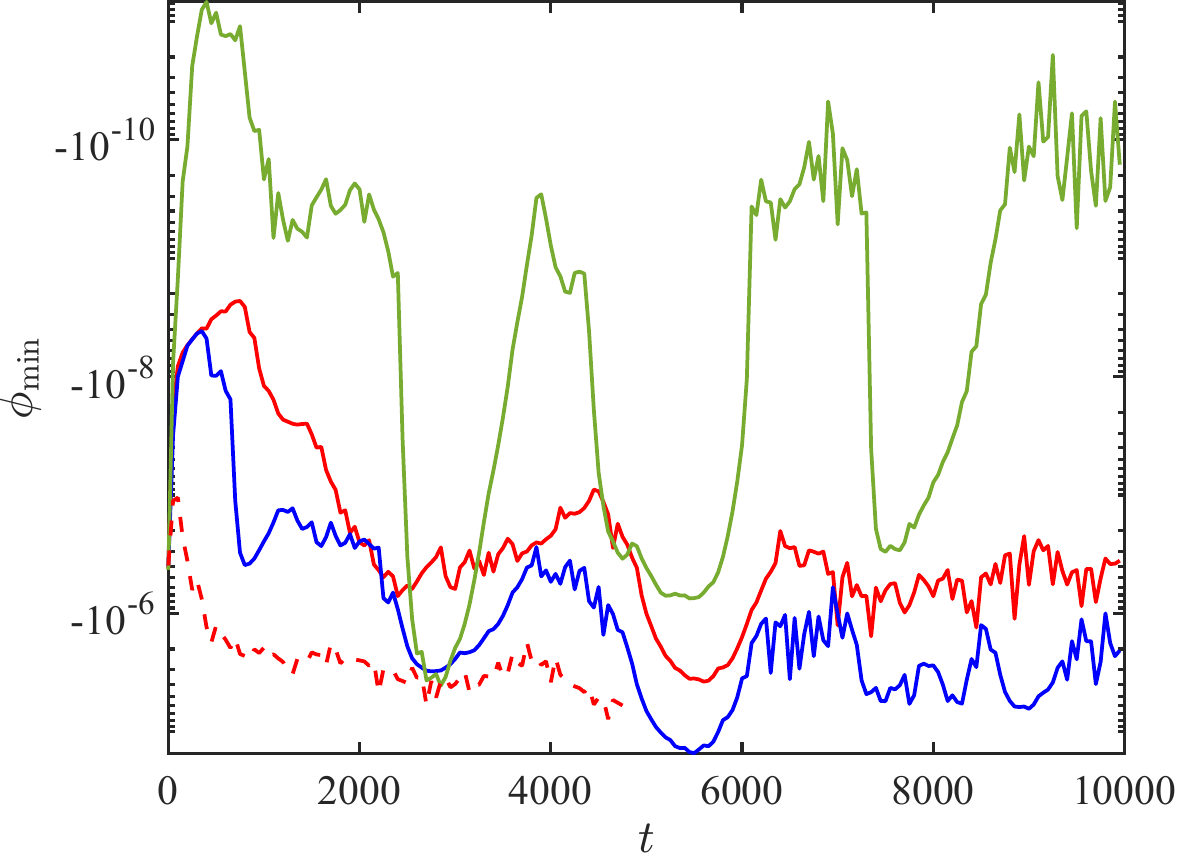}
        \caption{$\rho^*=100$}
        \label{fig:min_phi_density100}
    \end{subfigure}

    \caption{Evolution of the minimum order parameter for different density ratios.}
    \label{fig:min_phi_density}
\end{figure}

Figures~\ref{fig:max_phi_density} and \ref{fig:min_phi_density} show the temporal evolution of the maximum and minimum order parameters. The CH models clearly exhibit poorer boundedness than the AC models, consistent with the earlier two-dimensional results (see Table~\ref{tab:diagonal_translation_combined_8cols}). Because the density and viscosity are interpolated from the local $\phi$, any overshoot beyond the bulk values leads to extrapolation to unphysical density and viscosity, which in turn promotes numerical instability. This mechanism explains the reduced numerical stability of the CH models. Furthermore, the Fakhari2017 model generally performs better than Liang2018 in maintaining the upper bound of $\phi$, a result attributed to its recovery of the non-conservative form of the NS equations. This aspect will be discussed in greater detail in the next section.

In summary, the density-ratio tests indicate that the AC models maintain superior numerical stability and better boundedness of the order parameter than the CH models. At the tested low Weber number, all models exhibit some degree of droplet volume loss, except for the hybrid AC model, which preserves the volume well. As will be shown later, this issue of volume conservation becomes considerably more pronounced once droplet breakup occurs at higher Weber numbers.

\subsection{Effect of weber number}\label{subsubsection:WeberNumber}

Next, we conducted simulations to examine the effects of different Weber numbers on droplet dynamics. The kinematic viscosities of both phases were set to $\nu_H = \nu_L = 0.006$, and the density ratio was fixed at $\rho^* = 2$. Four Weber numbers were considered: ${\rm We} = 1.2, 12, 120,$ and $600$. The interface thickness was reduced to $\xi = 3$, which was sufficient to ensure numerical stability, and the mobility coefficient was fixed at $M_\phi = 0.01$ in all simulations. These settings allow for a consistent comparison of droplet behavior across different Weber numbers.

\begin{figure}
    \centering
    \includegraphics[width=1.0\textwidth]{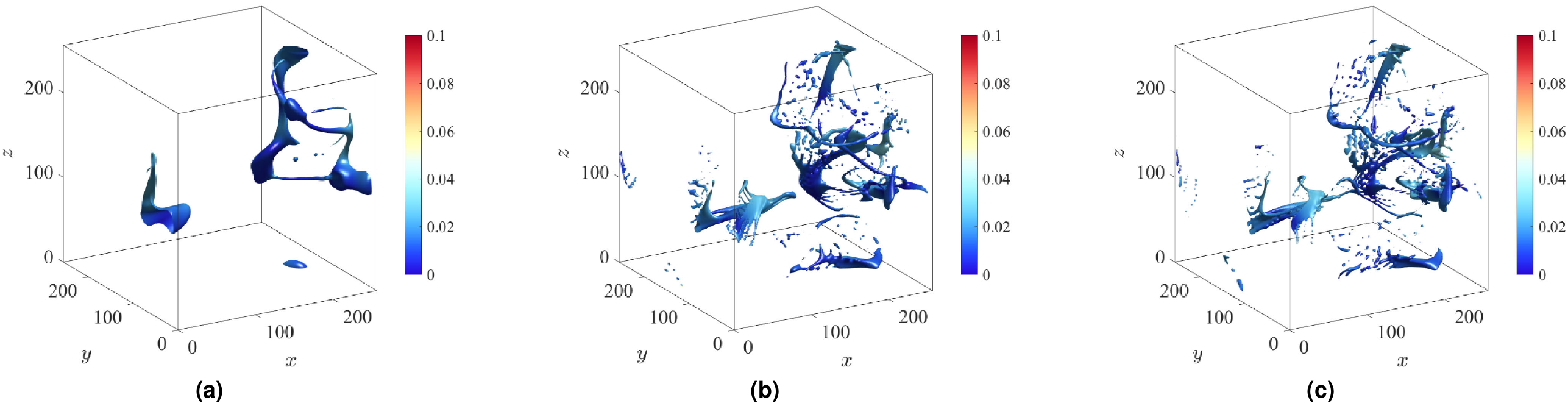}
    \caption{Snapshots of droplets for the local ACE model at $t=10000$: (a)${\rm We} = 12$, (b)${\rm We} = 120$, (c)${\rm We} = 600$.}
    \label{fig:Snapshots of droplets_weber}
\end{figure}

The Weber number strongly influences droplet breakup. Figure~\ref{fig:Snapshots of droplets_weber} shows droplet interfaces $\phi = 0.5(\phi_H + \phi_L)$ at $t = 10,000$ for ${\rm We} = 12, 120,$ and $600$. As ${\rm We}$ increases, the initial droplet breaks up into progressively smaller droplets.

\begin{figure}
    \centering

    % 子图 (a) We=1.2
    \begin{subfigure}[b]{0.48\textwidth}
        \centering
        \includegraphics[width=\textwidth]{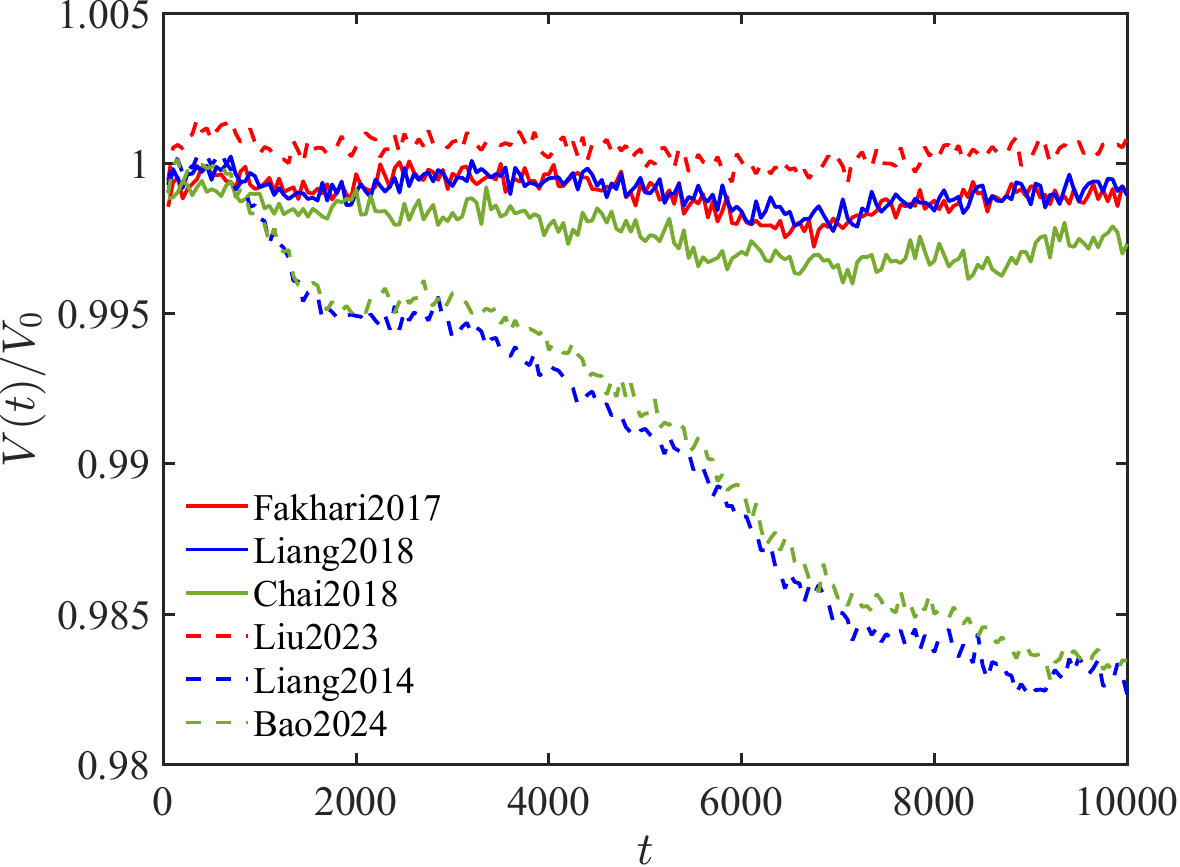}
        \caption{We = 1.2}
        \label{fig:Droplet_volume_We1_2}
    \end{subfigure}
    \hfill
    % 子图 (b) We=12
    \begin{subfigure}[b]{0.48\textwidth}
        \centering
        \includegraphics[width=\textwidth]{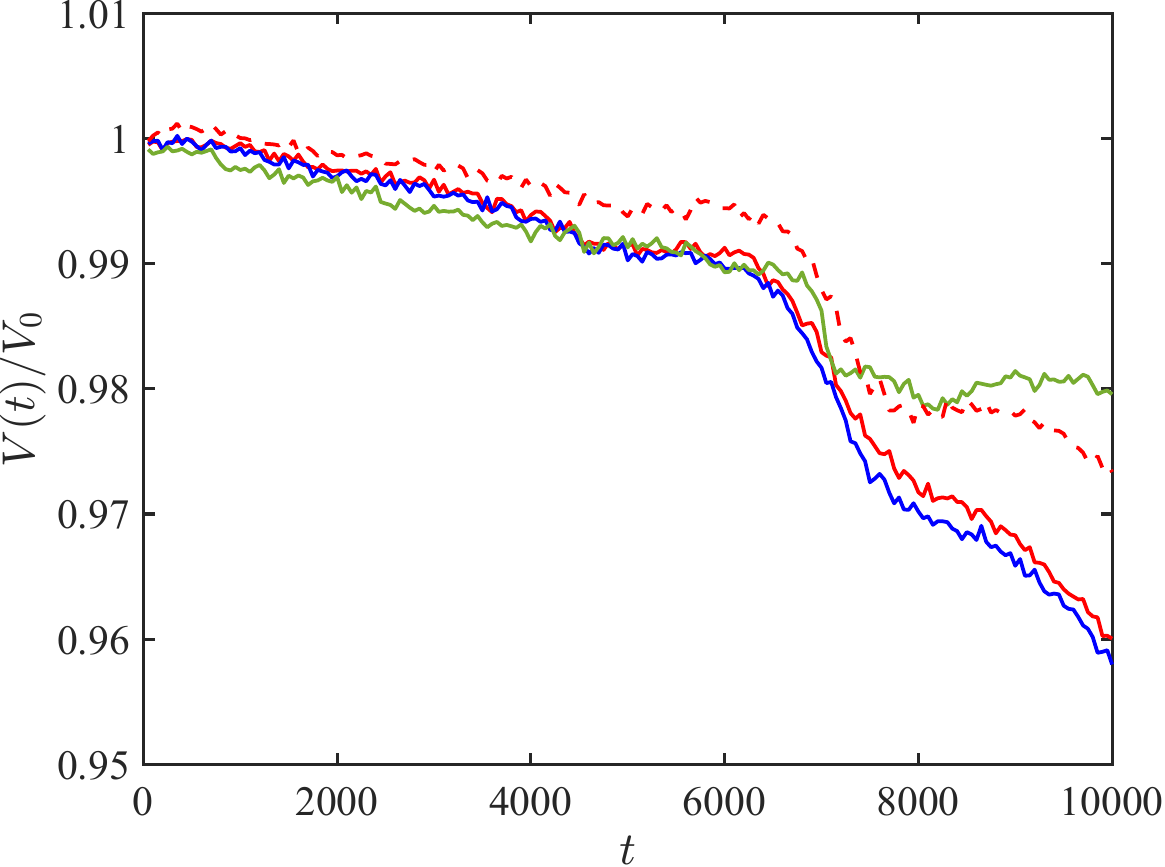}
        \caption{We = 12}
        \label{fig:Droplet_volume_We12}
    \end{subfigure}

    \vspace{1em} % 调整上下行间距

    % 子图 (c) We=120
    \begin{subfigure}[b]{0.48\textwidth}
        \centering
        \includegraphics[width=\textwidth]{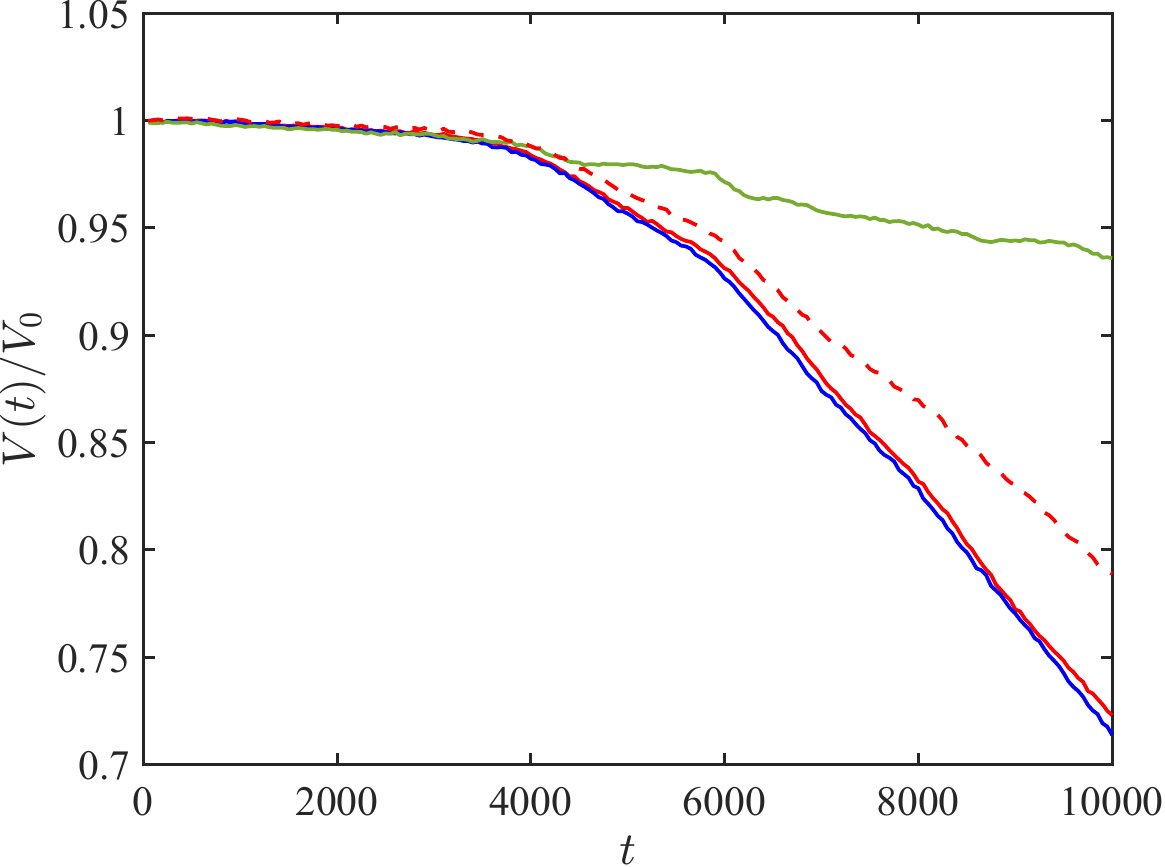}
        \caption{We = 120}
        \label{fig:Droplet_volume_We120}
    \end{subfigure}
    \hfill
    % 子图 (d) We=600
    \begin{subfigure}[b]{0.48\textwidth}
        \centering
        \includegraphics[width=\textwidth]{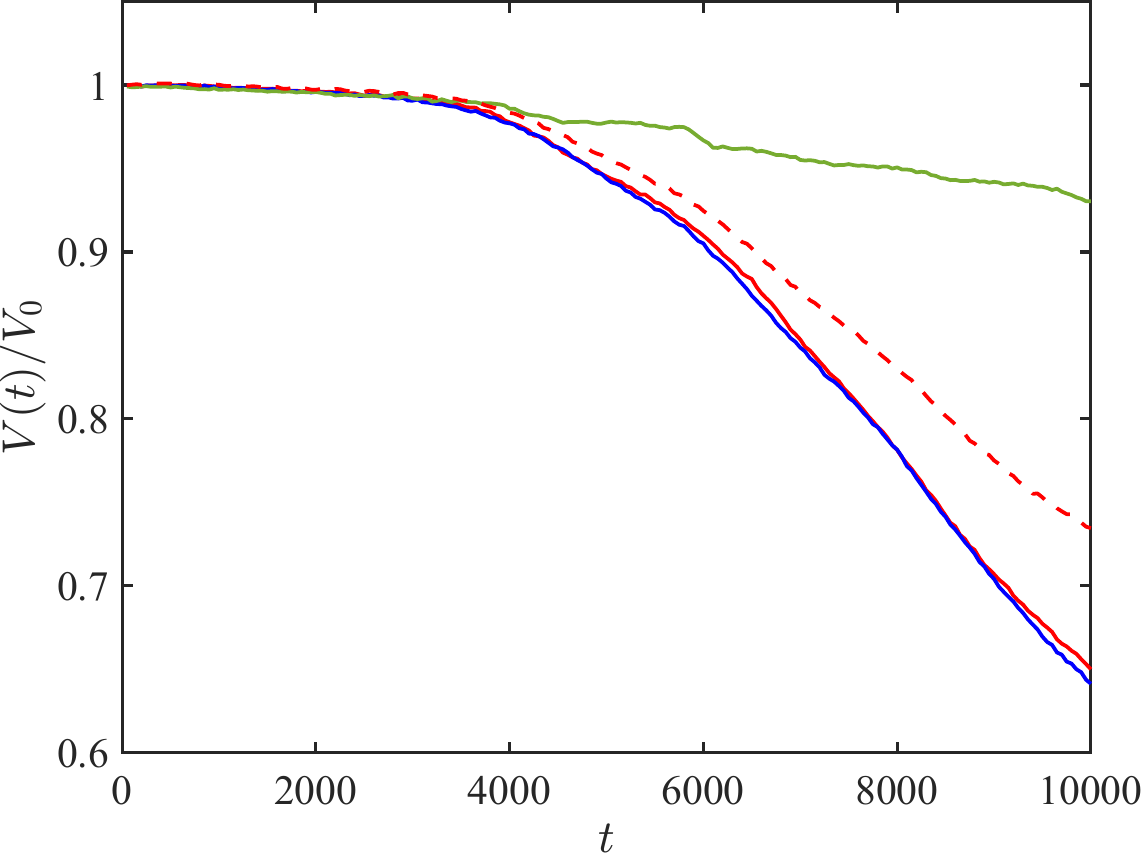}
        \caption{We = 600}
        \label{fig:Droplet_volume_We600}
    \end{subfigure}

    \caption{Droplet volume evolution for different Weber numbers.}
    \label{fig:Droplet_volume_We}
\end{figure}

Figure~\ref{fig:Droplet_volume_We} shows the time evolution of the total droplet volume, normalized by its initial value, for different Weber numbers. At ${\rm We} = 1.2$, where the initial large droplet remains intact without breakup, all four AC models preserve the droplet volume well. Compared with the results in Figure~\ref{fig:Droplet_volume_density2}, where the conservative AC models (Fakhari2017 and Liang2018) and the nonlocal AC model (Chai2018) exhibited slight volume loss, the only difference lies in the interface thickness: $\xi = 3$ in the present case versus $\xi = 6$ previously. This suggests that reducing the interface thickness improves volume preservation in the AC models.

In contrast, the two CH models show the opposite trend: the droplet volume decreased by about 1.7\% with $\xi = 3$, compared to only 1.3\% with $\xi = 6$, indicating that increasing the interface thickness mitigates volume loss in CH models under droplet-laden turbulent flow conditions. This behavior is opposite to that observed in laminar cases. For CH models, the spontaneous shrinkage of small droplets is widely recognized as the main cause of droplet volume loss~\cite{yue2007spontaneous}. As the interface thickness increases, the critical droplet radius below which spontaneous shrinkage occurs also increases, implying that droplets over a broader range of sizes tend to lose volume when the interface becomes thicker. However, this trend does not hold in the turbulent scenario considered here. The contrasting influence of interface thickness on droplet volume preservation will be further examined across a broader parameter range in the next section.

As ${\rm We}$ increases, the CH models again become unstable, showing that not only density ratio but also strong topological changes adversely affect their numerical stability. At high ${\rm We}$, the AC models also exhibit significant droplet volume loss within a short simulation time. Among them, the nonlocal AC model performs best, followed by the hybrid AC model, and then the conservative AC model.

\begin{figure}
    \centering
    \includegraphics[width=0.6\textwidth]{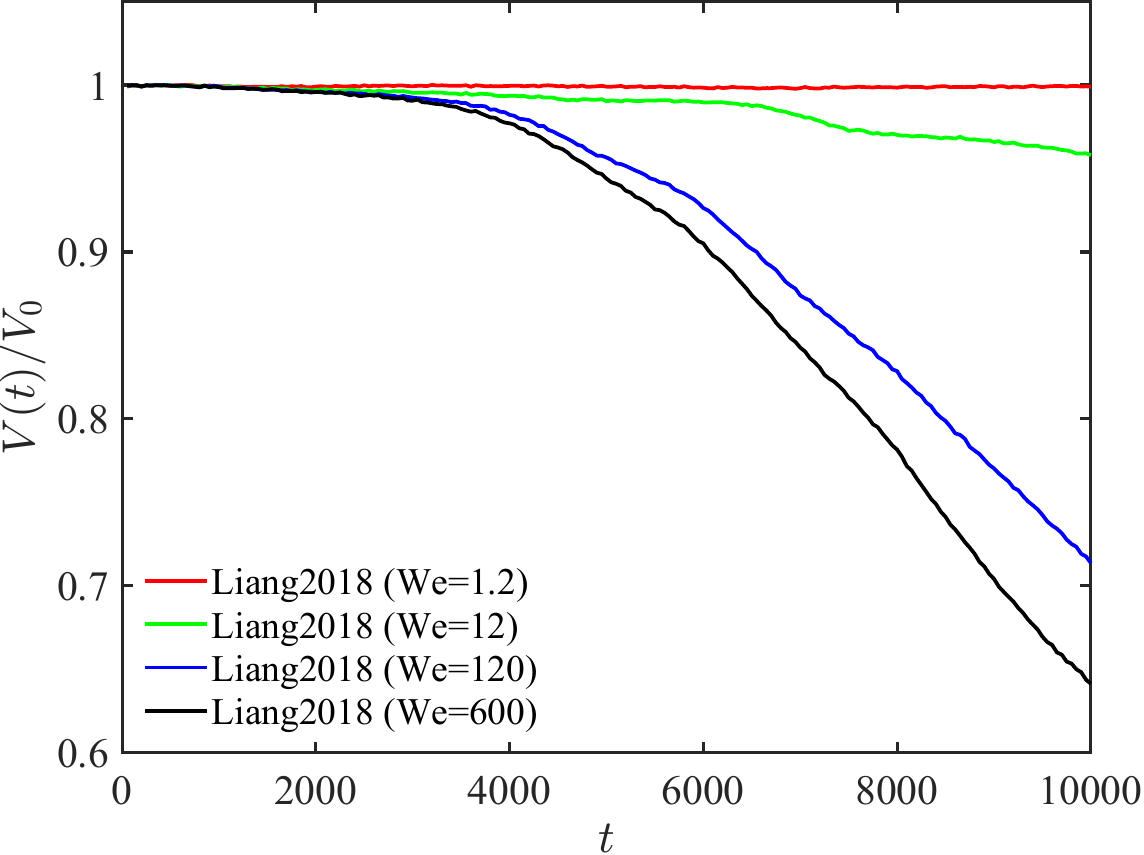}
    \caption{Volume evolution of droplets for the conservative AC model of Liang2018 at different Weber numbers.}
    \label{fig:liang_different_Weber_numbers}
\end{figure}

Figure~\ref{fig:liang_different_Weber_numbers} replots the volume evolution from Liang2018 at different ${\rm We}$. The volume loss grows more pronounced at larger ${\rm We}$, reflecting the dissolution of small droplets into the carrier phase. The higher the Weber number, the more small droplets are generated, and the faster the total droplet volume decays (see Figure~\ref{fig:Snapshots of droplets_weber}).

The nonlocal AC model exhibits the least volume loss because its coarsening effect transfers part of the volume from small droplets—which would otherwise dissolve—back into larger ones, thereby partially delaying volume decay. While this reduces overall volume loss, it comes at the expense of artificially eliminating small droplets, leading to significant deviations from physical reality. This is illustrated in Figure~\ref{fig:Snapshots of droplets_weber2}, which compares droplet distributions at $t = 10,000$ for ${\rm We} = 600$ between the nonlocal AC and conservative AC models. A substantial number of small droplets are absorbed into larger droplets in the nonlocal AC case.

\begin{figure}
    \centering
    \includegraphics[width=1.0\textwidth]{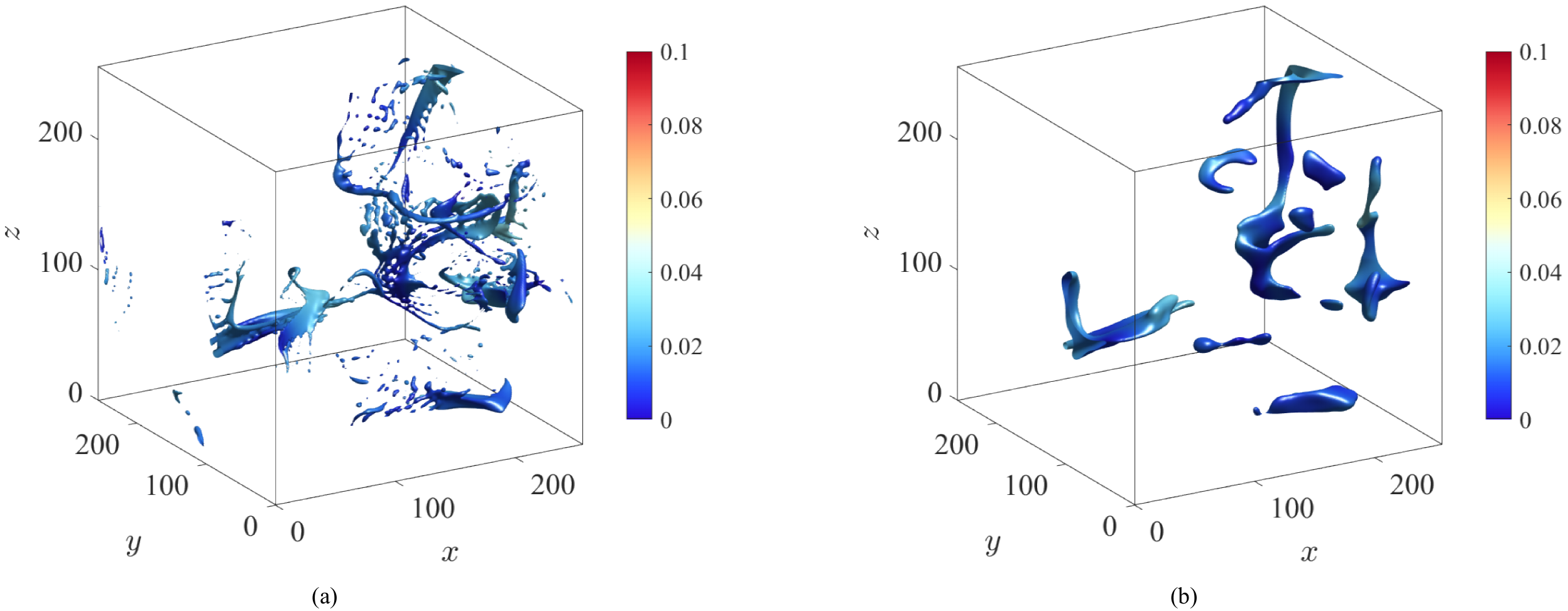}
    \caption{Snapshots of droplets for conservative AC and nonlocal AC models at $t=10,000$ with ${\rm We = 600}$: (a)Liang2018, (b)Chai2018}
    \label{fig:Snapshots of droplets_weber2}
\end{figure}

\begin{figure}[htbp]
    \centering

    \begin{subfigure}[b]{0.4\textwidth}
        \centering
        \includegraphics[width=\textwidth]{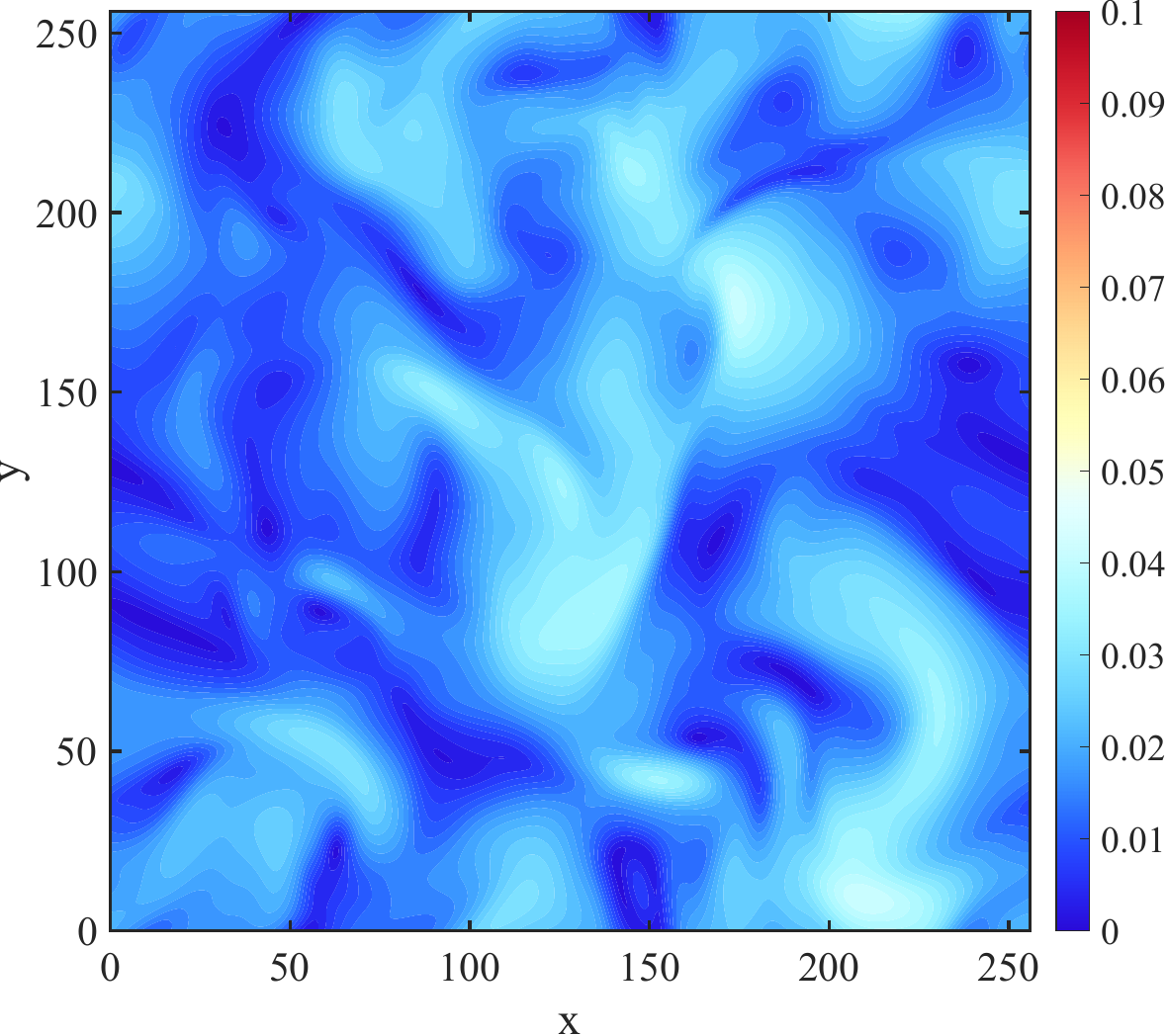}
        \caption{Fakhari2017}
        \label{fig:fakhari2017_cut}
    \end{subfigure}
 ~~~~\begin{subfigure}[b]{0.4\textwidth}
        \centering
        \includegraphics[width=\textwidth]{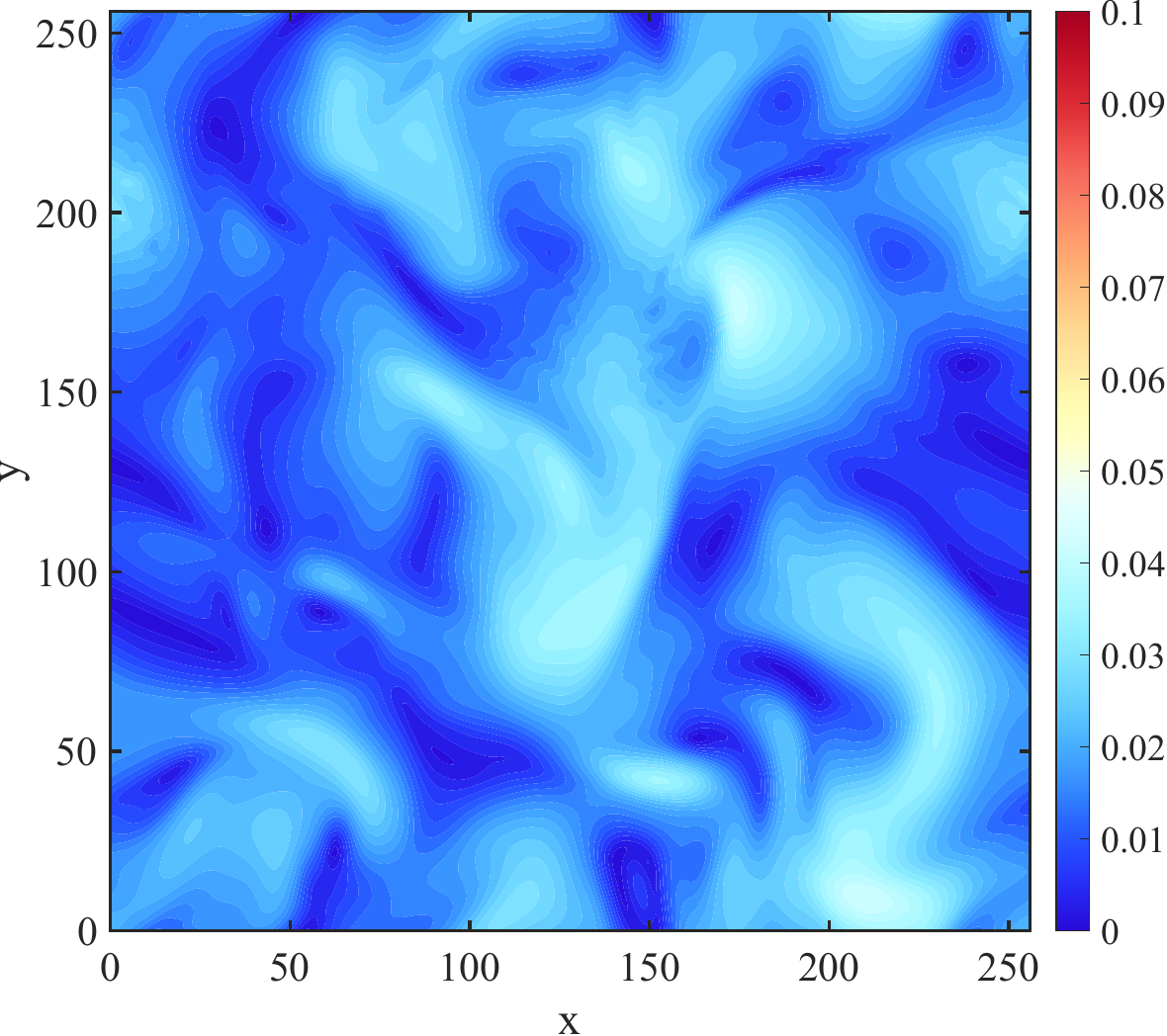}
        \caption{Liang2018}
        \label{fig:liang2018cut}
    \end{subfigure}

    \caption{Velocity distribution on a plane intersecting the droplet at 10,000 time steps .}
    \label{fig:vel_cut}
\end{figure}

In addition to droplet volume preservation, the boundedness of the order parameter is also examined. The overshoot of the order parameter generally becomes more pronounced as the Weber number increases. Figure~\ref{fig:Snapshots of droplets_weber2} presents the temporal evolution of the maximum order parameter at ${\rm We} = 120$ for the four AC models that remain numerically stable. It is evident that the maximum order parameter in the Fakhari2017 model is noticeably lower than in the other models. As discussed earlier, this advantage arises because Fakhari2017 recovers the non-conservative form of the NS equations, whereas the other tested models employ the conservative form. However, this improved performance of Fakhari2017 comes at the expense of computational efficiency (as will be discussed in Sec.~\ref{subsubsection:EfficiencyComparison}) due to its more complex definition of the force term. While the two conservative AC models exhibit distinct behavior regarding the boundedness of the order parameter, their velocity fields remain largely comparable. As shown in Figure~\ref{fig:vel_cut}, which displays the velocity magnitude distribution on a plane intersecting the domain center after 10,000 time steps, no appreciable difference is observed between the two models.

\begin{figure}
    \centering
    \includegraphics[width=0.6\textwidth]{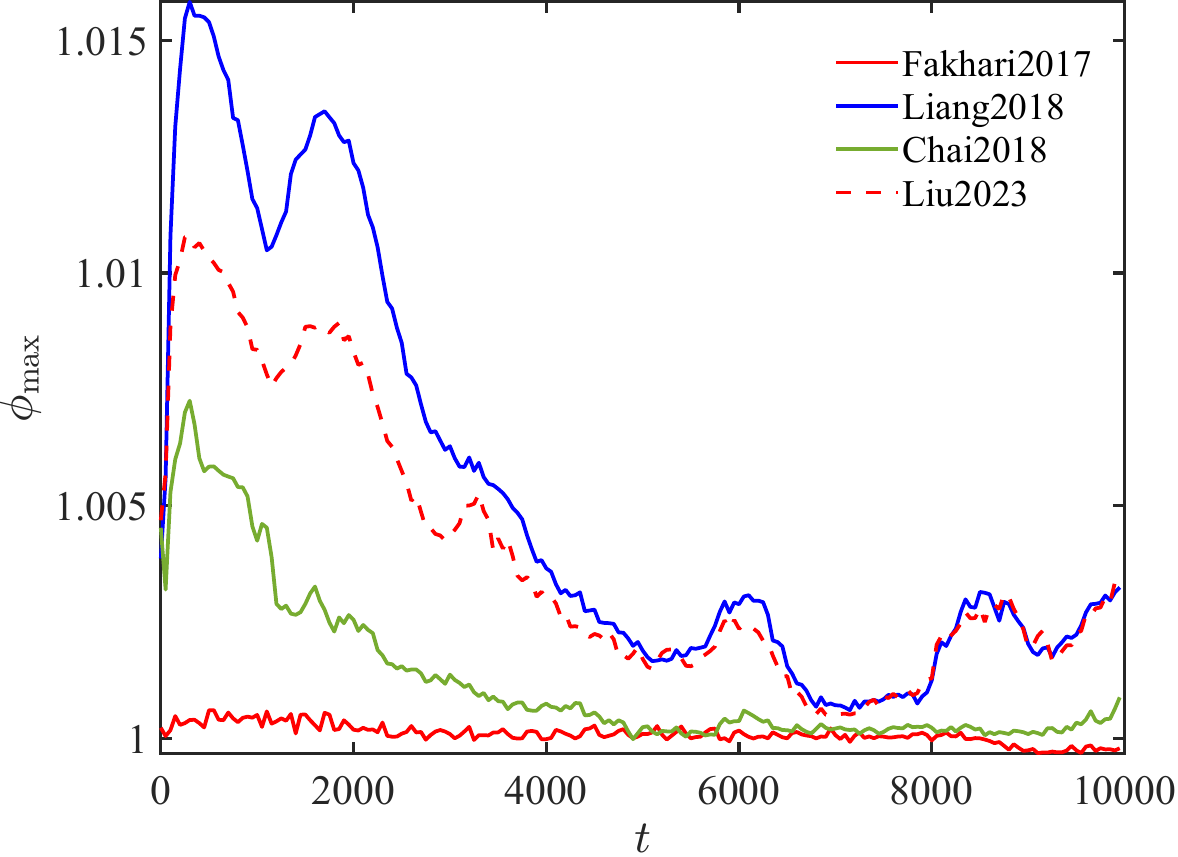}
    \caption{Evolution of the maximum order parameter for ${\rm We = 120}$}
    \label{fig:Snapshots of droplets_weber2}
\end{figure}

\begin{figure}
    \centering
    \includegraphics[width=0.6\textwidth]{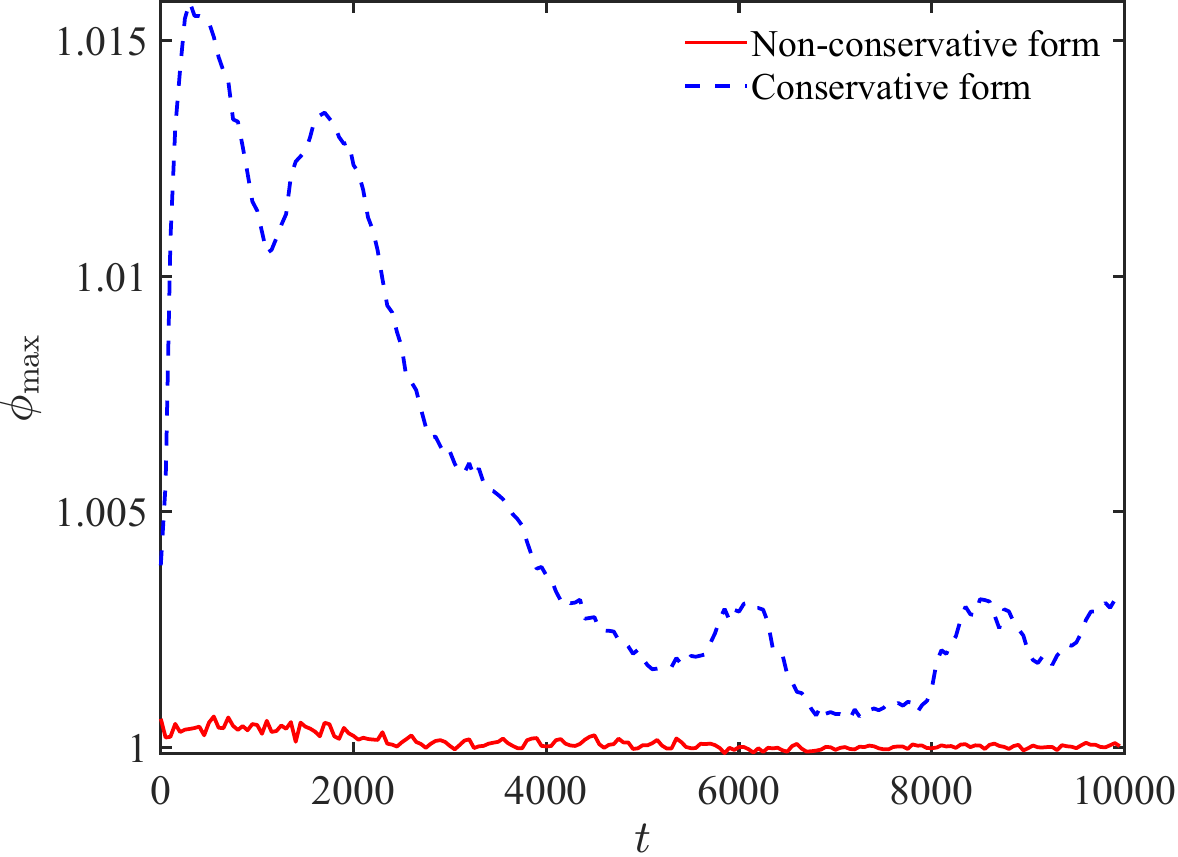}
    \caption{Evolution of the maximum order parameter for ${\rm We = 120}$}
    \label{fig:Evolution of the maximum_nc_c}
\end{figure}

To confirm that the differences in the boundedness of the order parameter arise from whether the NS equations are recovered in conservative or non-conservative form, rather than from differences in the reproduction of the AC equation itself, we modified the Liang2018 model to recover the non-conservative form and reran the above test. Figure~\ref{fig:Evolution of the maximum_nc_c} compares the temporal evolution of the maximum order parameter for the conservative and non-conservative forms of the NS equations within the same Liang2018 AC model. It is evident that the improved boundedness of the order parameter is attributable to the non-conservative form of the NS equations.

To summarize, the Weber number exerts a strong influence on droplet dynamics. As ${\rm We}$ increases, the initial droplet experiences greater shear and deformation, eventually breaking into multiple smaller droplets and accelerating droplet volume loss. Even the AC models, which exhibit only minor volume decreases at low ${\rm We}$, show substantial reductions at high ${\rm We}$; for example, at ${\rm We} = 600$, the droplet volume decreases to approximately 60\% of its initial value after 10,000 time steps. In contrast, the CH models become unstable at moderate Weber numbers (${\rm We} = 12$), highlighting their limitations in maintaining numerical stability under large topological deformations, particularly in turbulent flows. Furthermore, comparison between the two conservative AC models indicates that adopting the non-conservative form of the NS equations provides superior boundedness of the order parameter at high Weber numbers.

\subsection{Effect of interface thickness}\label{subsubsection:InterfaceThickness}

We further investigate the effects of interface thickness on droplet–turbulence interaction simulations. The density ratio was fixed at $\rho^* = 2$, and the Weber number was kept constant at ${\rm We} = 1.2$. Three interface thicknesses were considered: $\xi = 3, 6,$ and $9$, while the mobility coefficient was set to $M_\phi = 0.01$ in all cases. The first two values were already examined in Figures~\ref{fig:Droplet_volume_density2} and \ref{fig:Droplet_volume_We1_2}, and the results highlighted contrasting trends: for CH models, droplet volume preservation improves with increasing interface thickness, whereas for AC models, larger interface thickness leads to poorer volume conservation.

\begin{figure}
    \centering
    \includegraphics[width=0.6\textwidth]{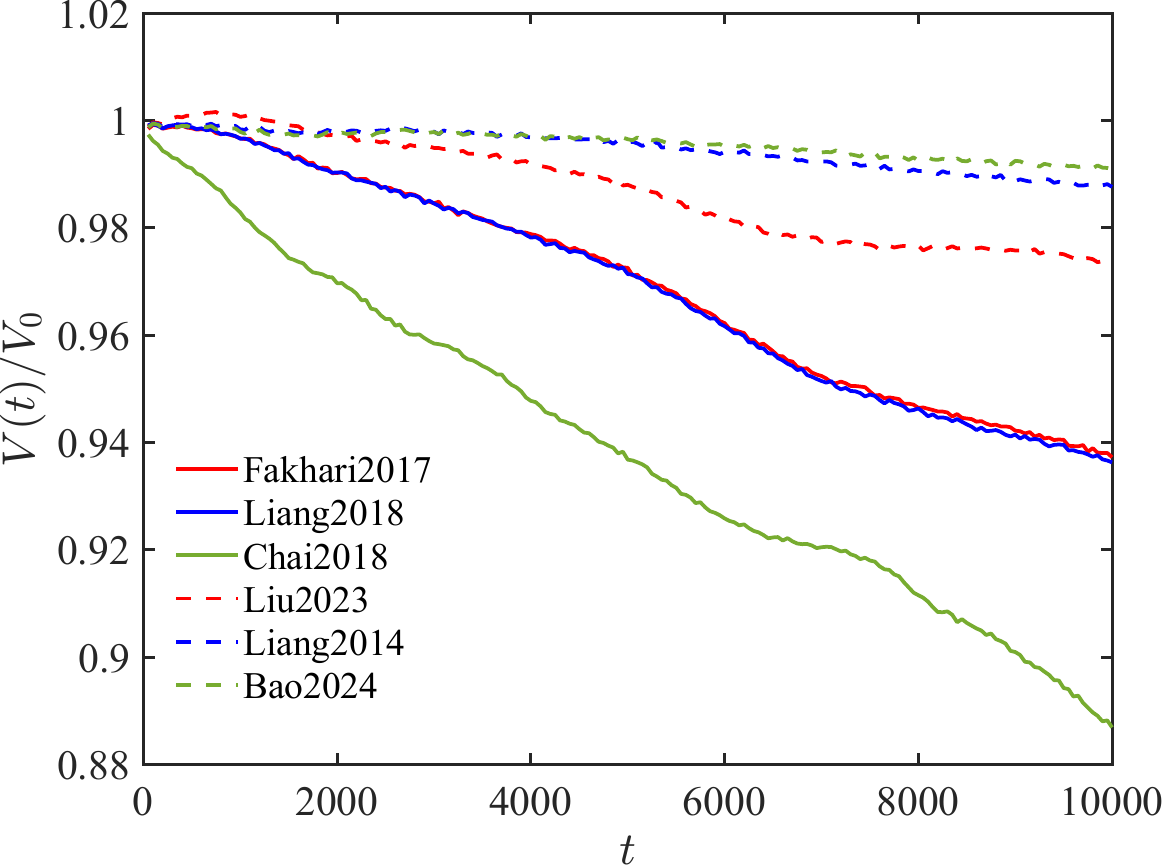}
    \caption{Droplet volume evolution under interface thickness $\xi=9$.}
    \label{fig:Droplet_Size_W9}
\end{figure}

\begin{figure}
    \centering

    % 子图 (a) ξ=3
    \begin{subfigure}[b]{0.48\textwidth}
        \centering
        \includegraphics[width=\textwidth]{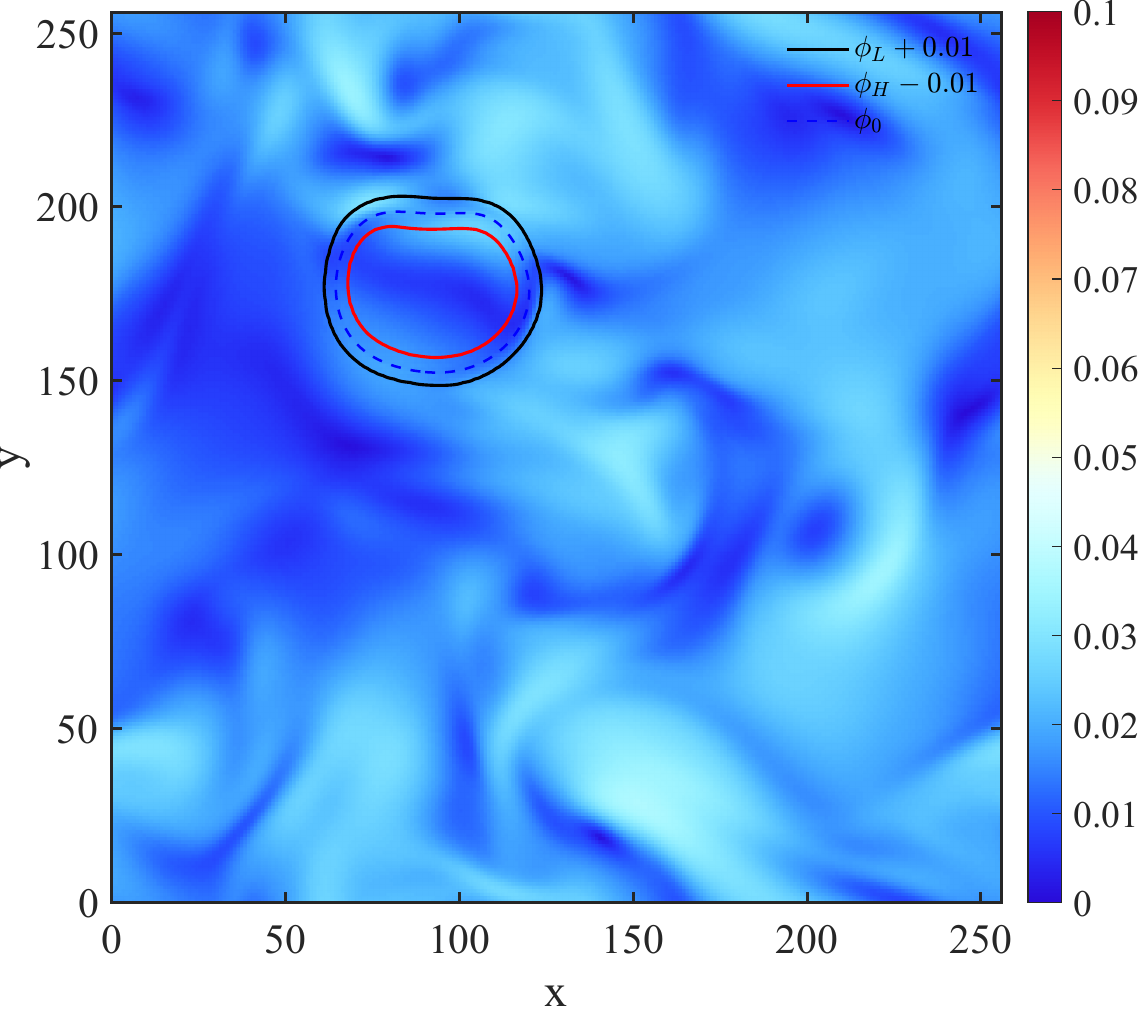}
        \caption{$\xi=3$}
        \label{fig:xi3_velocity}
    \end{subfigure}
    \hfill
    % 子图 (b) ξ=9
    \begin{subfigure}[b]{0.48\textwidth}
        \centering
        \includegraphics[width=\textwidth]{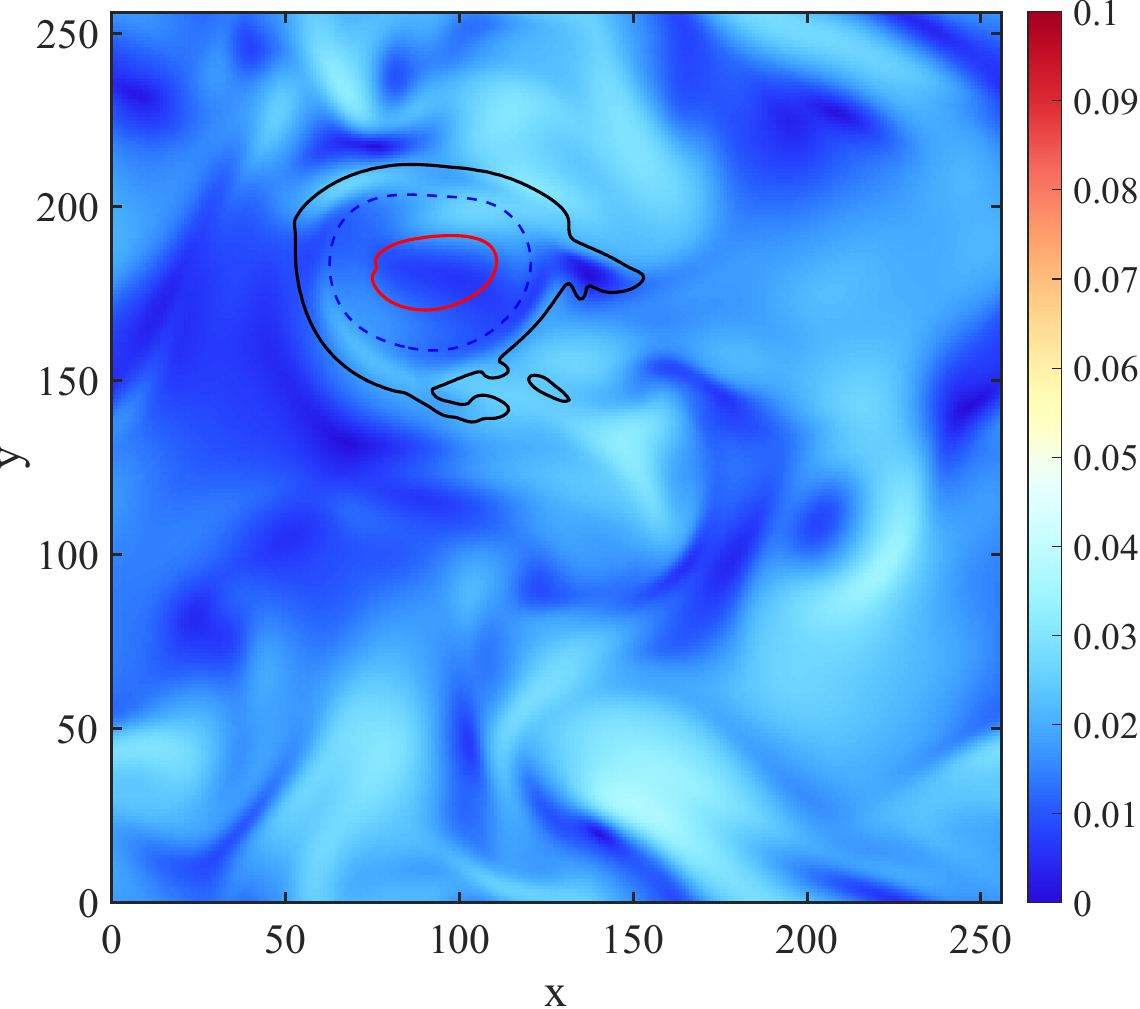}
        \caption{$\xi=9$}
        \label{fig:xi9_velocity}
    \end{subfigure}

    \caption{Velocity distribution on a plane intersecting the droplet at 5,000 time steps, computed using the conservative AC model of Liang2018.}
    \label{fig:bubble_velocity_comparison}
\end{figure}

To further confirm this trend, we tested a third value, $\xi = 9$, and the corresponding droplet volume evolution is presented in Figure~\ref{fig:Droplet_Size_W9}. As the interface thickness increases, CH models continue to exhibit reduced volume decay, with loss rates dropping below 1.2\% after 10,000 time steps. In contrast, volume loss for AC models increases, with the nonlocal AC model exceeding 11\% and even the best-performing hybrid AC model reaching 2.6\%. However, excessively wide interfaces should be avoided, as they inevitably smooth out small structures in turbulent flows. As illustrated in Figure~\ref{fig:bubble_velocity_comparison}, while the overall velocity structures are preserved, the widely spread interface region ($\phi_L + 0.01(\phi_H-\phi_L) < \phi < \phi_H - 0.01(\phi_H-\phi_L)$) modifies the nearby velocity distribution.

\begin{figure}
    \centering

    % 子图 (a) PHImax
    \begin{subfigure}[b]{0.48\textwidth}
        \centering
        \includegraphics[width=\textwidth]{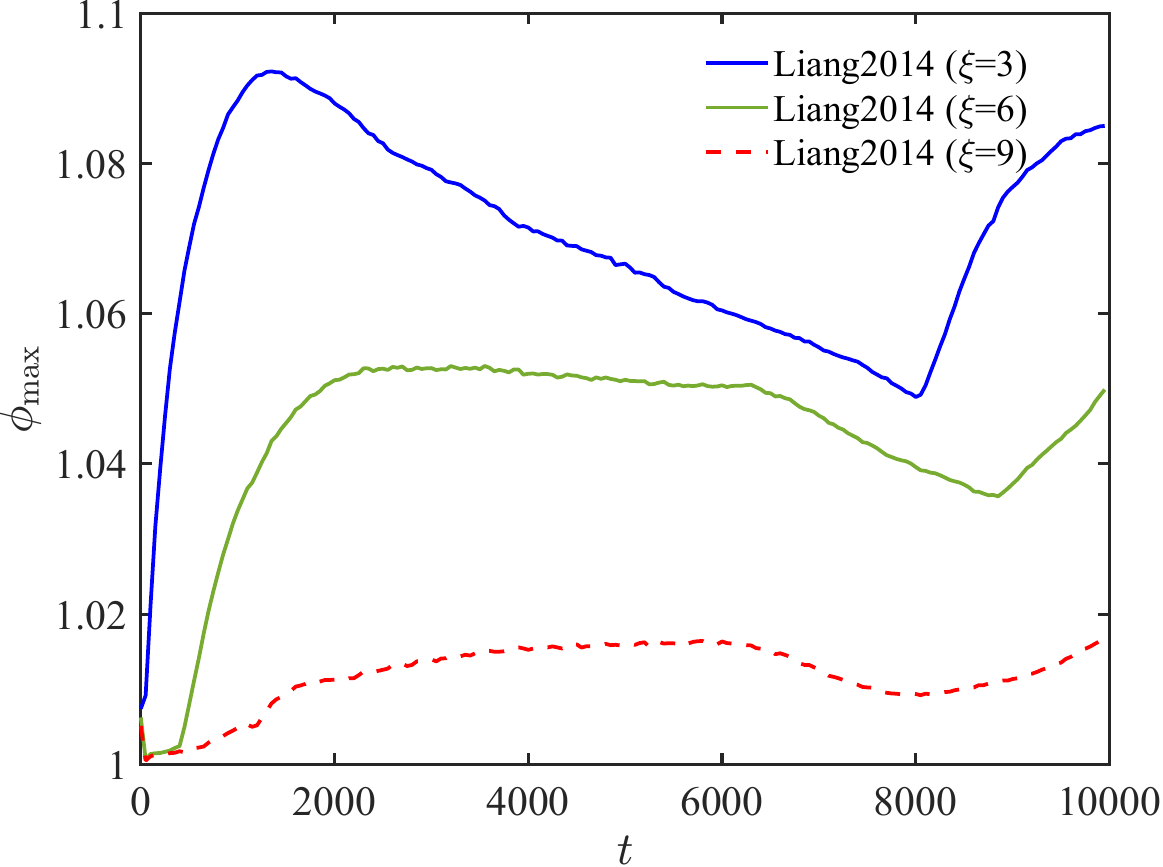}
        \caption{Temporal evolution of the maximum order parameter.}
        \label{fig:Phi_Max_Liang2014}
    \end{subfigure}
    \hfill
    % 子图 (b) PHImin
    \begin{subfigure}[b]{0.48\textwidth}
        \centering
        \includegraphics[width=\textwidth]{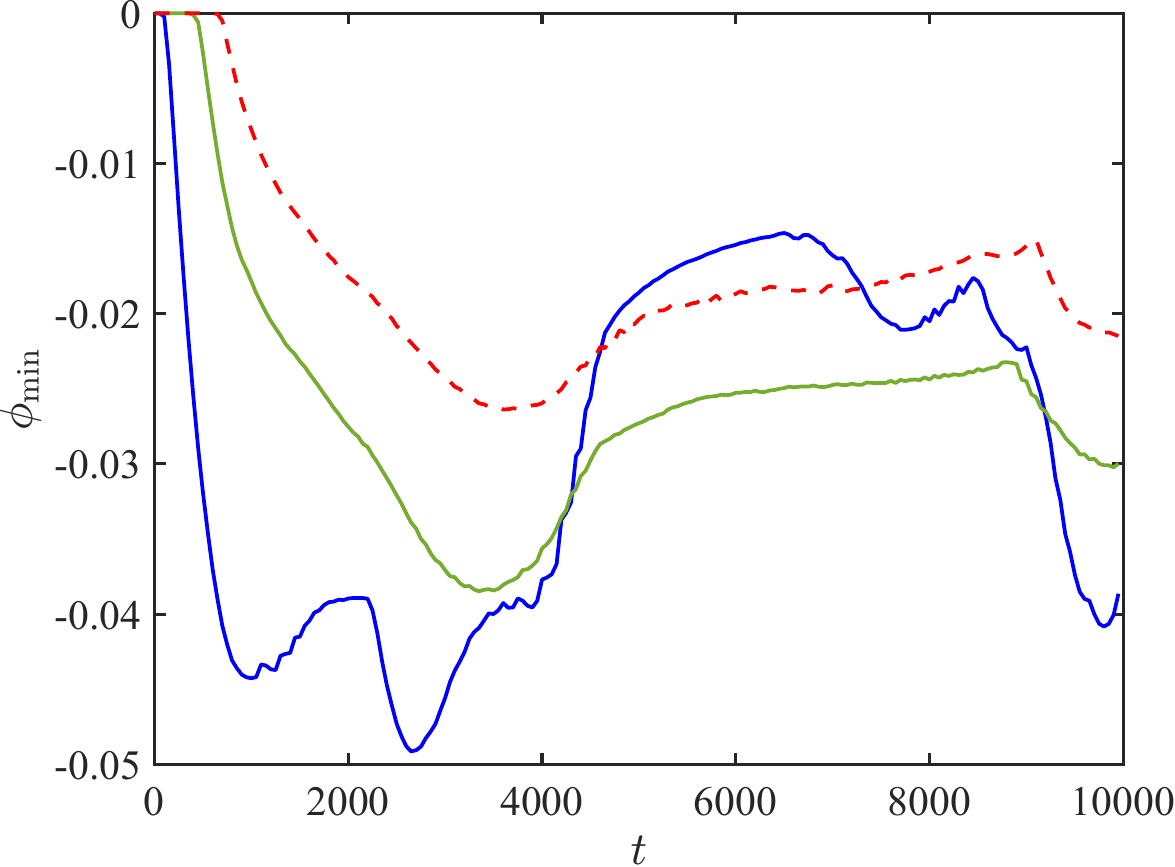}
        \caption{Temporal evolution of the minimum order parameter.}
        \label{fig:Phi_Min_Liang2014}
    \end{subfigure}

    \caption{Temporal evolution of the maximum and minimum order parameters under different interface thicknesses for the incompressible CH model of Liang2014.}
    \label{fig:order27}
\end{figure}

Figure~\ref{fig:order27} shows the temporal evolution of the maximum and minimum order parameters in the incompressible CH model of Liang2014 for different interface thicknesses. As noted previously, the boundedness of the order parameter in CH models is relatively poor, but this behavior improves as the interface thickness increases.

\subsection{Comparison of computational efficiency}\label{subsubsection:EfficiencyComparison}

Finally, we evaluate the computational performance of the six models in three-dimensional simulations of a large droplet evolving in homogeneous isotropic turbulence. The computational domain is $L_x \times L_y \times L_z = 256 \times 256 \times 256$, and the simulations employ 256 MPI processes arranged in a two-dimensional decomposition with 16 processes in both the $y$- and $z$- directions. This setup ensures a balanced workload and efficient inter-process communication. All models are implemented using the same code structure and run on an identical platform and environment. Table~\ref{tab:time_comparison} reports the wall-clock time for each model over 2,000 steps.

\begin{table}
\centering
\caption{Wall clock time comparison of different models.}
\label{tab:time_comparison}
\begin{tabular}{c c c c  c c c}
\hline
Model &Fakhari2017  &Liang2018  & Chai2018  & Liu2023  & Liang2014 & Bao2024  \\
\hline
Time (s) & $139.07$ &$124.47$ & $120.83$ & $124.36$ & $132.56$ & $144.24$ \\
\hline
\end{tabular}%
\end{table}

As shown in Table~\ref{tab:time_comparison}, the AC models (Liang2018, Chai2018, Liu2023) are generally more computationally efficient than the CH models (Liang2014, Bao2024), although the differences are not decisive. The lower efficiency of CH models is mainly due to the need to solve higher-order partial differential equations, which involve additional gradient and Laplacian operations, increasing both computational complexity and inter-process communication overhead. In Bao2024, the use of a lattice kinetic scheme to maintain stability with singular mobility further increases computational cost, as strain-rate computations are required when updating the hydrodynamic quantities. Among the AC models, Fakhari2017 shows slightly lower computational efficiency compared to others, primarily due to the more complex calculation of viscous forces, which also involves the strain-rate computations. Nevertheless, the overall differences in computational efficiency remain modest, suggesting that computational cost should not be a decisive factor when selecting among these models.

\section{Summary}\label{sec:summary}

In this work, we systematically evaluated eight representative phase-field lattice Boltzmann models, classified into five categories: the conservative AC model, the nonlocal AC model, the hybrid AC model, the incompressible CH model, and the quasi-incompressible CH model. The evaluation was conducted through a series of benchmark tests covering both laminar and turbulent flows to comprehensively assess their numerical performance.

From the laminar flow tests, we found that conservative AC models generally exhibit the highest accuracy in predicting droplet shape and interface deformation.
The nonlocal AC model, while providing better boundedness of the order parameter, suffers from coarsening effects that continuously transfer mass from small to large structures, leading to the unphysical absorption of small droplets.
Hybrid AC models, which linearly combine conservative and nonlocal components, inherit both the advantages and drawbacks of their constituents: although they improve boundedness, they still display mild coarsening and fail to preserve delicate small-scale structures under extreme conditions.
The CH models, in contrast, preserve the richest small-scale structures during severe interface deformation and breakup, but they yield the poorest boundedness of the order parameter and allow small droplets to disappear spontaneously. The introduction of singular mobility can slow down such droplet shrinkage but cannot fully eliminate it.

In turbulent simulations with dispersed droplets, all tested models encounter significant challenges in preserving droplet volume, particularly at high Weber numbers, where large droplets experience strong deformation and breakup under turbulence-induced shear. Among them, the nonlocal AC model exhibits the best droplet volume conservation, primarily due to its pronounced coarsening effect, which transfers mass from smaller droplets to larger ones before the latter eventually dissolve into the surrounding fluid. Overall, the AC models outperform the CH models in both numerical stability and droplet volume conservation, especially under high density ratio and high Weber number conditions. The numerical instability observed in CH models is closely associated with their poor boundedness of the order parameter, which leads to extrapolated density and viscosity values reaching unphysical extremes. The influence of interface thickness also differs between model types: increasing the interface thickness improves volume conservation and boundedness in CH models, whereas reducing it benefits the AC models. Furthermore, adopting the non-conservative form of the NS equations effectively maintains the boundedness of the order parameter, which in turn potentially enhances numerical stability.

Finally, although the computational efficiency of all models is generally comparable, AC models are typically more efficient than CH models because they involve only up to second-order derivative computations. Among the AC models, those that recover the non-conservative form of the NS equations exhibit slightly reduced efficiency due to the more complex definition of the force term.

To summarize, while most examined models can handle simple laminar flow simulations with satisfactory performance, our systematic comparison suggests that conservative AC models represent the most reliable choice for three-dimensional droplet-laden turbulence simulations under critical parameter settings with large interface deformation. This study also underscores the need for further improvements to PF-LB models, particularly in mitigating the dissolution of small droplets, which remains a major limitation for practical applications.

\textbf{Statement for Data Availability}: The source codes (written in FORTRAN 90) and the data for selected test cases reported in this work are available from the corresponding author upon reasonable request.

\begin{acknowledgments}
This work has been supported by the National Natural Science Foundation of China
(grant number 12472254),  the Natural Science Foundation of Shandong Province, China (grant number ZR2024QA083), and the Fundamental Research Program of Shanxi Province(grant number 202303021211160).
The financial support from Shandong University through the Qilu Young Scholar Program is also acknowledged. Computing resources are provided by the National Supercomputing Center at Jinan.
\end{acknowledgments}

\bibliography{apssamp.bib}% Produces the bibliography via BibTeX.

\end{document}